\documentclass[%
 reprint,
superscriptaddress,
onecolumn,
notitlepage,
 amsmath,amssymb,
 aps,
prc,
]{revtex4-1}

\usepackage{graphicx}
\usepackage{dcolumn}
\usepackage{bm}

\usepackage{nicefrac}

\usepackage{silence}
\usepackage[american]{babel}
\usepackage{amsmath}
\usepackage[version=3]{mhchem} 
\usepackage{mathtools}

\usepackage{xfrac}
\usepackage{lmodern}
\usepackage[hidelinks]{hyperref}
\usepackage{microtype}
\usepackage{listings}
\usepackage [autostyle, english = american]{csquotes}
\usepackage{booktabs,siunitx}
\usepackage{gensymb}
\usepackage[normalem]{ulem}

\newcommand{\etal}{\emph{et~al.}}

\usepackage[colorinlistoftodos]{todonotes}
 
\newcommand{\cmmnt}[1]{}

\usepackage{siunitx}
\newcommand{\mmicro}{\si\micro}

\usepackage{subfig}
\newcommand{\subfigimg}[4][,]{%
  \setbox1=\hbox{\noindent\includegraphics[#1]{#3}}
  \leavevmode\rlap{\usebox1}
  \rlap{\hspace*{#4pt}\raisebox{\dimexpr\ht1-2\baselineskip}{#2}}
  \phantom{\usebox1}
}

\usepackage{makecell}

\usepackage{adjustbox}

\begin{document}


\title{Proton-induced reactions on Fe, Cu, \& Ti from threshold to 55 MeV}

\author{Andrew S. Voyles}
 \email{andrew.voyles@berkeley.edu}
\affiliation{%
 Department of Nuclear Engineering, University of California, Berkeley,  Berkeley, CA 94720, USA
}%
\author{Amanda M. Lewis}%
\affiliation{%
 Department of Nuclear Engineering, University of California, Berkeley, Berkeley, CA 94720, USA
}%
\author{Jonathan T. Morrell}%
\affiliation{%
 Department of Nuclear Engineering, University of California, Berkeley, Berkeley, CA 94720, USA
}%

\author{M. Shamsuzzoha Basunia}
\affiliation{Nuclear Science Division, Lawrence Berkeley National Laboratory,  Berkeley, CA 94720, USA}
\author{Lee A. Bernstein}
\affiliation{%
 Department of Nuclear Engineering, University of California, Berkeley, Berkeley, CA 94720, USA
}%
\affiliation{Nuclear Science Division, Lawrence Berkeley National Laboratory,  Berkeley, CA 94720, USA}
\author{Jonathan W. Engle}
\affiliation{Department of Medical Physics, University of Wisconsin -- Madison,  Madison, WI 53705, USA}
\author{Stephen A. Graves}
\affiliation{Department of Radiology, University of Iowa,  Iowa City, IA 52242, USA}
\author{Eric F. Matthews}%
\affiliation{%
 Department of Nuclear Engineering, University of California, Berkeley, Berkeley, CA 94720, USA
}%

%
%

\date{\today}

\begin{abstract}

Theoretical models often differ significantly from measured data in their predictions of the magnitude of nuclear reactions that produce radionuclides for medical, research, and national security applications.
In this paper, we compare \emph{a priori} predictions from several state-of-the-art reaction modeling packages (CoH, EMPIRE, TALYS, and ALICE) to cross sections measured using the stacked-target activation method.  
The experiment was performed using the LBNL 88-Inch Cyclotron with beams of 25 and 55\,MeV protons on a stack of 
iron, copper, and titanium foils.  
34 excitation functions were measured for 4~\textless~$E_p$~\textless~55\,MeV, including the first measurement of the independent cross sections for \ce{^{nat}Fe}(p,x)\ce{^{49,51}Cr}, \ce{^{51,52m,52g,56}Mn}, and \ce{^{58m,58g}Co}.  
All of the models failed to reproduce the isomer-to-ground state ratio for 
reaction channels at 
compound and pre-compound energies, suggesting issues in modeling the deposition or distribution of angular momentum in these residual nuclei.


\end{abstract}

\maketitle



\section{\label{sec:intro_fe}Introduction}

Clinical practice of nuclear medicine is rapidly growing with the inclusion of a broader array of radiopharmaceuticals. 
Future growth is anticipated, given the pre-clinical success of many  new and emerging radionuclides. 
Although the physical and chemical properties of these novel radionuclides tend to be
well-established, their broad-scale  clinical applications are reliant upon well-characterized nuclear data to facilitate 
production. 
%
%
%
One particular set of  emerging radionuclides are 
positron-emitting isotopes of manganese, which have been identified as having potential for a range of diagnostic applications\,\cite{J.2013,Graves2015,Lewis2015,PhysRevC.96.014613,Wooten2017,Hernandez2017}.
In particular, a significant interest has been expressed in producing the emerging radionuclides \ce{^{51}Mn} for clinical use in quantitative positron emission tomography (PET) studies, as well as  \ce{^{52g}Mn} for pre-clinical imaging of neural and immune processes via PET\,\cite{Graves2016}.

Manganese radionuclides are desirable for radiopharmaceutical applications, as 
they possess well-established biochemistry, and have been 
chelated 
by the complexing agent DOTA for tracking monoclonal antibodies with high biostability at neutral pH\,\cite{Graves2015}.
\ce{^{52}Mn} ($t_{1/2}$ = 5.591$~\pm~$0.003\,d, $I_{\beta^+}$ = 29.4\%, $E_{\beta\text{, avg}}$ = 0.242\,MeV\,\cite{Dong2015}) has been shown to be useful for immuno-PET applications, 
offering the possibility for imaging within minutes of injection,
making it highly suitable for pre-clinical imaging as a longer-lived complement   to the more established immuno-PET agents \ce{^{89}Zr} and   \ce{^{64}Cu}.  
However, its  long half-life and unfavorable high-energy decay gamma-rays 
make it undesirable for clinical applications.
The short half-life of the  \ce{^{52m}Mn} isomer ($t_{1/2}$ = 21.1$~\pm~$0.2\,min)
makes production and handling difficult, and with a  high-intensity gamma emission (1434.06\,keV, $I_\gamma = 98.2~\pm~0.5\%$),  
\ce{^{52m}Mn} is similarly undesirable for clinical applications\,\cite{Dong2015}.
In contrast, \ce{^{51}Mn} ($t_{1/2}$ = 46.2$~\pm~$0.1\,min, $I_{\beta^+}$ = 96.86\%, $E_{\beta\text{, avg}}$ = 0.964\,MeV\,\cite{Wang2017}), 
is more clinically suitable for rapid metabolic studies.
\ce{^{51}Mn} lacks any strong decay gamma-rays (its longer-lived daughter \ce{^{51}Cr} [$t_{1/2}$ = 27.704$~\pm~$0.003\,d] has only a single  320.0284\,keV [$I_\gamma = 9.910~\pm~0.010\%$] line), making it the best choice of these radionuclides for clinical imaging.

Developing production of these radionuclides requires well-characterized cross section data, or predictive models when such data have not been measured.   
Modeling of nuclear reactions in 
the A=40--70 mass region presents numerous challenges including uncertainties in nuclear level densities as a function of spin due to the opening of the $f_{7/2}$ orbital and the presence of enhanced $\gamma$-strength at low energies\,\cite{Voinov2004a,Algin2008,Algin2007}.  
Therefore, as part of a larger campaign to address deficiencies in cross-cutting nuclear data needs, our group has
measured the   nuclear excitation functions of the radionuclides \ce{^{51}Mn},   \ce{^{52m}Mn}, and \ce{^{52g}Mn} from proton-induced reactions on Fe.
We used the thin-foil stacked-target technique to study proton-induced reactions on 
Fe foils of natural isotopic abundance with 
Ti and 
Cu monitor foils.
This work 
complements 
earlier measurements using 40--100\,MeV protons  and extends them down to reaction thresholds,  to investigate the feasibility of production using the international network   of low-energy medical cyclotrons\,\cite{Graves2016}. 
Furthermore, 
we used our measured data to probe the role of angular momentum in the transitional energy region between compound and direct reactions since both the ground and long-lived isomeric states in \ce{^{52}Mn} were populated.



In addition to their interest for PET studies, the \ce{^{51,52g,52m}Mn} excitation functions 
offer an opportunity to study the distribution of angular momentum in compound nuclear and direct pre-equilibrium reactions via observation of the \ce{^{52m}Mn} ($t_{1/2}$ = 21.1$~\pm~$0.2\,min; J$^\pi=2^+$) to \ce{^{52g}Mn} ($t_{1/2}$ = 5.591$~\pm~$0.003\,d; J$^\pi=6^+$)   ratio\,\cite{Dong2015,Wang2017}.
Measurements of isomer-to-ground state ratios have been used for over 20\,years to probe the spin distribution of excited nuclear states in the A\,$\approx$\,190 region\,\cite{PhysRevC.73.034613,PhysRevC.45.1171}.
These measurements also provide an opportunity to benchmark 
the predictive capabilities of  reaction modeling codes used for nuclear reaction evaluations and the way in which they implement the underlying physical reaction mechanisms.  

\section{\label{sec:experimental_fe}Experimental Methods and Materials}

The work described herein follows the  methods utilized in our recent work and established by Graves \etal\ 
for monitor reaction characterization of beam energy and fluence in stacked target irradiations\,\cite{Voyles2018a,Graves2016}.
%
%
%
Preliminary results  were reported in a Master's thesis \cite{springer2017investigation}; here we report the final analysis of that work.
Unless otherwise stated, all values are presented herein as mean $\pm$ SD, or as the calculated result $\pm$ half the width of a 68\% confidence interval.

\subsection{\label{sec:target_design_fe}Stacked-target design}

\begin{figure}[ht]
 \centering
 \includegraphics[width=0.45\textwidth]{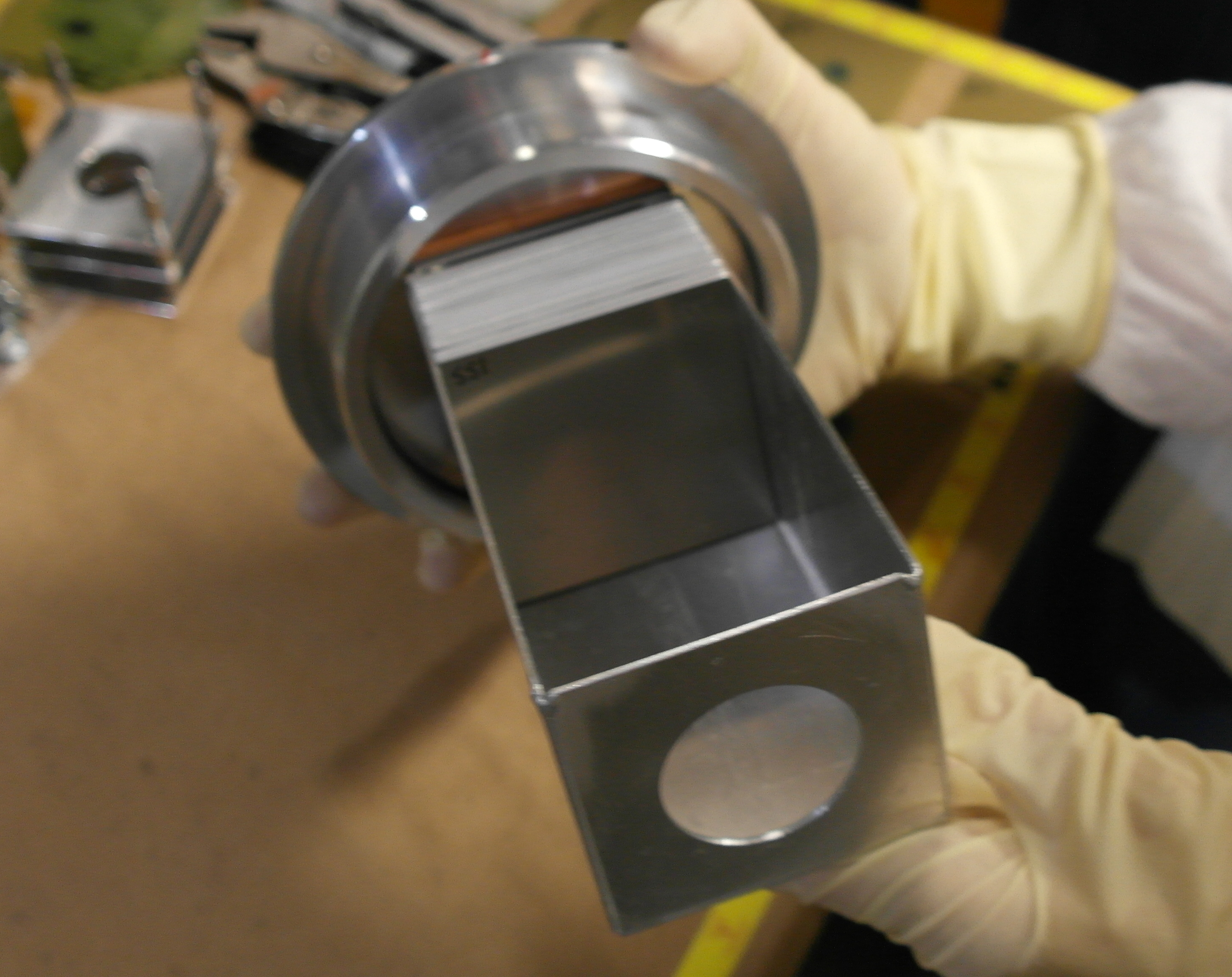}
 \caption{\label{fig:fe_target_stack}Photograph of the assembled 25\,MeV target stack.
 The proton beam enters through the circular entrance in the foreground, and the upstream stainless steel profile monitor (SS-5) is visible at the front of the stack.}
\end{figure}

We constructed a pair of target stacks 
for this work,
one stack covering the 55--20\,MeV range and the other  25--0\,MeV.
This minimized the systematic uncertainties associated with significant degradation of beam energy, and
included multiple overlapping measurements between 20--25\,MeV as a consistency check between the stacks.
A series of nominal 25\,\mmicro m \ce{^{nat}Fe} foils (99.5\%, lot \#LS470411), 25\,\mmicro m \ce{^{nat}Ti} foils (99.6\%, lot \#LS471698), and 25\,\mmicro m \ce{^{nat}Cu} foils (99.95\%, lot \#LS471698) were used (all from Goodfellow Corporation, Coraopolis, PA 15108, USA) as targets.
In each stack, seven foils 
were cut down to 2.5$\times$2.5\,cm squares and spatially characterized 
at four different locations using a digital caliper and micrometer (Mitutoyo America Corp.).
Four mass measurements were performed using an analytical balance 
in order to determine their areal density. 
The foils were  sealed into \enquote{packets} using two pieces of  3M 5413-Series Kapton polyimide film tape 
consisting of 43.2 \mmicro m of a silicone adhesive (nominal 4.79\,mg/cm$^2$) on 25.4 \mmicro m of a polyimide backing (nominal 3.61\,mg/cm$^2$).
The sealed foils were mounted over the hollow center of  1.5875 mm-thick aluminum frames.
Plates of 6061 aluminum alloy  served as proton energy degraders  between energy positions.
The target box, seen in \autoref{fig:fe_target_stack}, is machined from 6061 aluminum alloy and mounts on the end of an electrically-isolated beamline.
The specifications of both target stacks 
are in \autoref{tab:fe_stack_table} of Appendix \ref{sec:fe_stack_design}.

Both target stacks were 
separately irradiated at the  Lawrence Berkeley National Laboratory  (LBNL)
88-Inch Cyclotron, a  K=140 sector-focused cyclotron \cite{7999622}.
The 25\,MeV stack was irradiated for 
20\,minutes at a nominal current of 100\,nA,  for an  integral current of 31.61\,nAh, measured using a current integrator on the electrically-isolated beamline. 
The 55\,MeV stack was irradiated for 
10\,minutes at a nominal current of 120\,nA,  for an  integral current of 20.78\,nAh. 
The beam current
remained stable under these conditions for the duration of each irradiation.
The approximately 1\,cm-diameter proton beam incident upon each stack's upstream stainless steel profile monitor had a maximum energy of either 25 or 55\,MeV, with an approximately 2\% energy width due to multi-turn extraction --- these energy profiles were used for all later analysis.
Following end-of-bombardment (EoB), each stack was removed from the beamline and disassembled.
All activated foils were transported to a counting lab for gamma spectrometry, which started approximately 20\,minutes following the end of each irradiation.


\subsection{\label{sec:spectroscopy_fe}Quantification of induced activities}

\begin{figure*}
 \centering
 \includegraphics[width=0.75\textwidth]{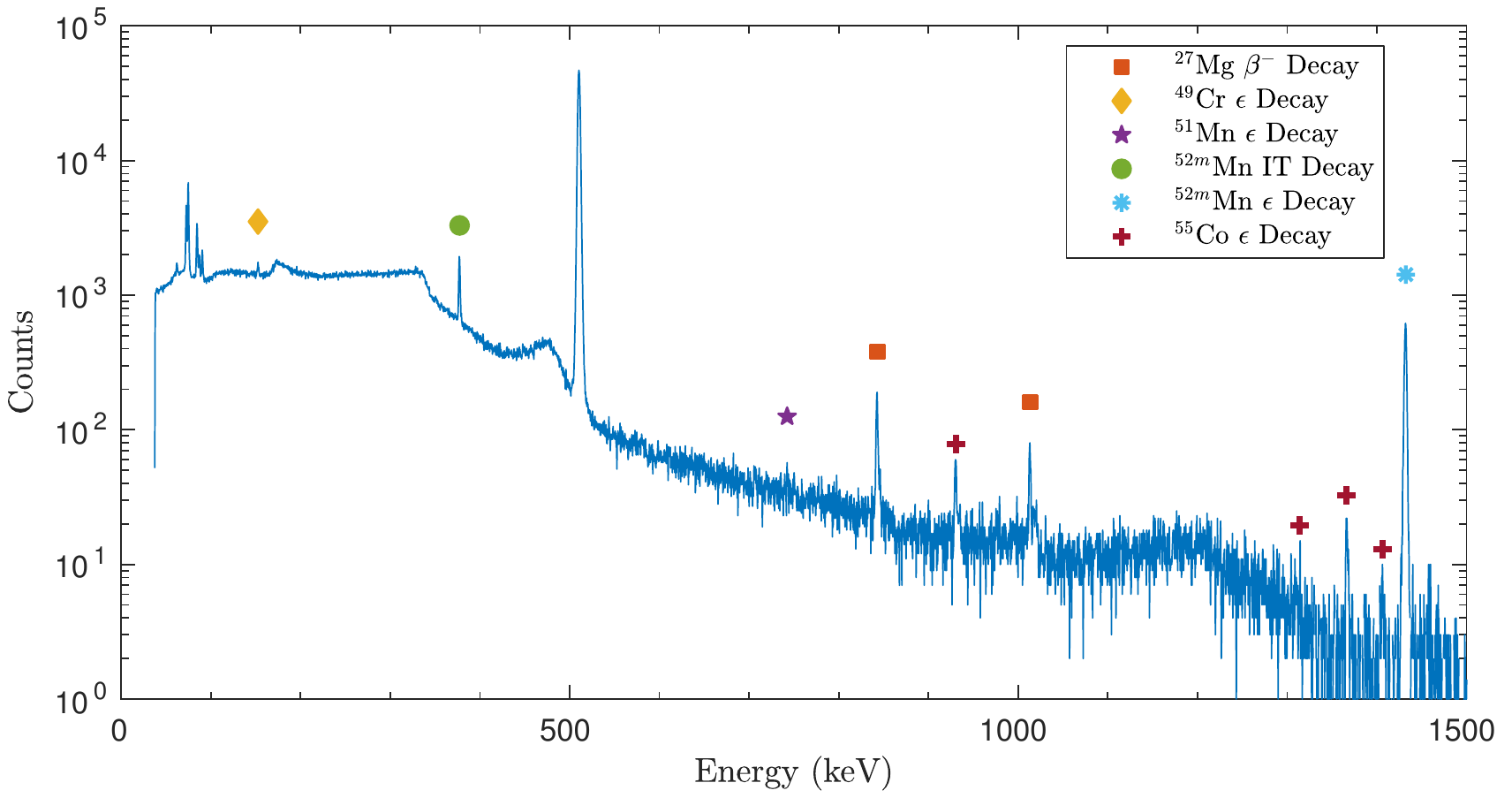}
 \caption{A gamma spectrum  from an activated Fe foil at approximately 55\,MeV (the maximum incident proton energy), collected 25 minutes after end-of-bombardment. Several observed reaction products are visible in this spectrum, and  the \ce{^{51}Cr} and \ce{^{52m}Mn} decay lines, which form two of the   primary reaction channels of interest, are  clearly isolated from surrounding peaks. }
 \label{fig:gspec_femn}
\end{figure*}

A single 
ORTEC GMX Series (model \#GMX-50220-S)  High-Purity Germanium (HPGe) detector was used to determine the activities in each target.
Samples were counted at fixed positions ranging 5--60  cm (5\% maximum permissible dead-time) from the front face of the detector.
The foils were counted  for 
4 weeks following 
(EoB).
An example of one of the gamma-ray spectra collected 
is shown in \autoref{fig:gspec_femn}.
Net peak areas were fitted using the 
code FitzPeaks\,\cite{fitzgerald2009fitzpeaks}, which utilizes the SAMPO fitting algorithms for gamma-ray spectra\,\cite{Aarnio2001}.

The net  counts in each fitted gamma-ray photopeak were converted into  activities for the decaying  activation products.
%
%
%
The   half-lives and gamma-ray branching ratios  used for all calculations of measured cross sections reported in this work  have been taken from the most recent edition of  Nuclear Data Sheets for each  mass chain
\cite{Singh2015a,Chen2011a,Wu2000,Burrows2007,Burrows2006,Burrows2008,Wang2017,Dong2015,Dong2014,JUNDE2008787,Junde2011,Bhat1998,Nesaraja2010,Browne2013,Zuber20151,NICHOLS2012973,ERJUN2001,Singh2007}.
Corrections (typically \textless0.2\%) for gamma-ray attenuation within each foil packet were made, using  photon attenuation coefficients from the XCOM photon cross sections database\,\cite{berger2011xcom}.
EoB activities were determined by $\chi^2$-fitting of all observed decay gammas for a product to the decay curve.
The total  uncertainty in  activity is the propagated sum of the uncertainty in  fitted peak areas, uncertainty in detector efficiency calibration,  uncertainty in the gamma-ray branching ratio data, and uncertainty in photon attenuation coefficients (taken as~5\%).

As in our previous work, these activities were used to calculate 
cumulative and independent cross sections\,\cite{Voyles2018a}.
For the first product nuclide in a mass chain with observable decay gammas, its (p,x) cross section is reported as a cumulative cross section ($\sigma_c$), which is the sum of direct production of that nucleus, as well as decay of its  precursors and any other independent cross sections leading to that nucleus. 
Cumulative cross sections are reported whenever it is impossible to use decay spectrometry to distinguish independent production of a nucleus from decay feeding.
For all remaining observed reaction products in the mass chain, and cases where no decay precursors exist, independent cross sections  ($\sigma_i$) corresponding to a single 
residual product are reported, 
facilitating comparison to reaction model calculations.  
Solutions to the first- and higher-order  Bateman equations are used for separation of  feeding contributions from decay precursors, so that  independent cross sections may be reported\,\cite{bateman1910solution,Cetnar2006}.

\subsection{\label{sec:dosimetry_fe}Proton fluence determination}

Thin \ce{^{nat}Ti} and \ce{^{nat}Cu} foils  
were co-irradiated to measure beam current at each position within the stack.
The IAEA-recommended \ce{^{nat}Ti}(p,x)\ce{^{46}Sc}, \ce{^{nat}Ti}(p,x)\ce{^{48}V},  \ce{^{nat}Cu}(p,x)\ce{^{62}Zn}, and \ce{^{nat}Cu}(p,x)\ce{^{63}Zn} monitor reactions were used 
\cite{Hermanne2018}.
Systematically enhanced fluence from \ce{^{nat}Ti}(p,x)\ce{^{48}Sc} co-production was avoided by only using the 928.327, 944.130, and 2240.396\,keV decay gammas  from \ce{^{48}V}.
Using the formalism outlined in our previous work, the integral form of the well-known activation equation was used to  determine proton fluence ($I \Delta t $),
in order to account for energy loss across each monitor foil \cite{Voyles2018a}.
The propagated uncertainty in proton fluence is calculated as the quadrature sum of (1) the uncertainty in quantified EoB activity, (2) uncertainty in the duration of irradiation (conservatively estimated at 10 s, to account for any transient changes in beam current), (3) uncertainty in foil areal density, (4) uncertainty in monitor product half-life (included, but normally negligible), (5) uncertainty in IAEA recommended cross section (using values  from the 2017 IAEA re-evaluation\,\cite{Hermanne2018}), and (6) uncertainty in differential proton fluence (from transport simulations).


\subsection{\label{sec:proton_transport_fe}Proton transport calculations}

Estimates of the proton beam energy for preliminary stack designs were calculated using the Anderson \& Ziegler (A\&Z) stopping power formalism\,\cite{Andersen_Ziegler_1977,Ziegler1985,Ziegler1999}.
However, the 
transport code FLUKA-2011.2x.3 was used for simulation of the full 3-D target stack and to determine the full proton energy and fluence distribution for each foil\,\cite{Bohlen2014a}. 
$10^8$ source protons were used for all FLUKA simulations, 
yielding a statistical uncertainty 
of less than 0.01\%.
As with the determination of proton fluence in the monitor foils, the progressively increasing energy straggle towards the rear of each stack is accounted for using 
FLUKA.
These energy distributions $\frac{d\phi}{dE}$ were used to calculate a flux-weighted average proton  energy $\langle E \rangle$, which accounts for the slowing-down of protons within a foil (particularly in the low-energy stack) and reports the effective  energy centroid for each foil.
To report a complete description of the representative energy for each foil, a bin width is provided through the  energy uncertainty, calculated as the full width at half maximum (FWHM) of the FLUKA-modeled energy distribution for each foil.

The \enquote{variance minimization} techniques utilized in our recent work and established by Graves \etal\ have been used to reduce uncertainty in proton fluence assignments due to poorly-characterized stopping power\,\cite{Voyles2018a,Graves2016}.
This method is based on the assumption that the independent measurements of proton fluence from the different monitor reactions 
should all be consistent at each 
position.
This disagreement is minor at the front of the stack, but gets progressively worse as the beam travels through the stack, due to the compounded effect of systematic uncertainties.

When performing  a variance minimization, it is important to apply this variation of effective areal density  to the stack components which  have the most significant impact on beam energy loss.
Therefore, the aluminum degraders are used for variance minimization for the 55\,MeV stack, as they make up more than 80\% of the areal density of the stack.
For the 25\,MeV stack, the Kapton tape was chosen for variance minimization, as the foil packets themselves are responsible for the majority of beam degradation.
While it only makes up approximately 20\% of the low-energy stack's areal density, the Kapton surrounding each foil packet has a greater areal density than the foil itself.
In addition, it is far easier to directly characterize the areal density of the metallic foils than it is for the Kapton, resulting in only an approximate value for the latter.
The contributions to the slowing of the beam due to the adhesive have often been neglected in much work performed to date. 
This is of relatively minor consequence for higher-energy irradiations (especially relative to any beam degraders), but 
becomes increasingly important for proton energies below 
25\,MeV, causing as much as an additional loss of 8\,MeV  by the time it reaches the end of the stack.

\begin{figure*}
    \centering
    \subfloat{
        \centering
        \hspace{-5pt}\subfigimg[width=0.485\textwidth]{a)}{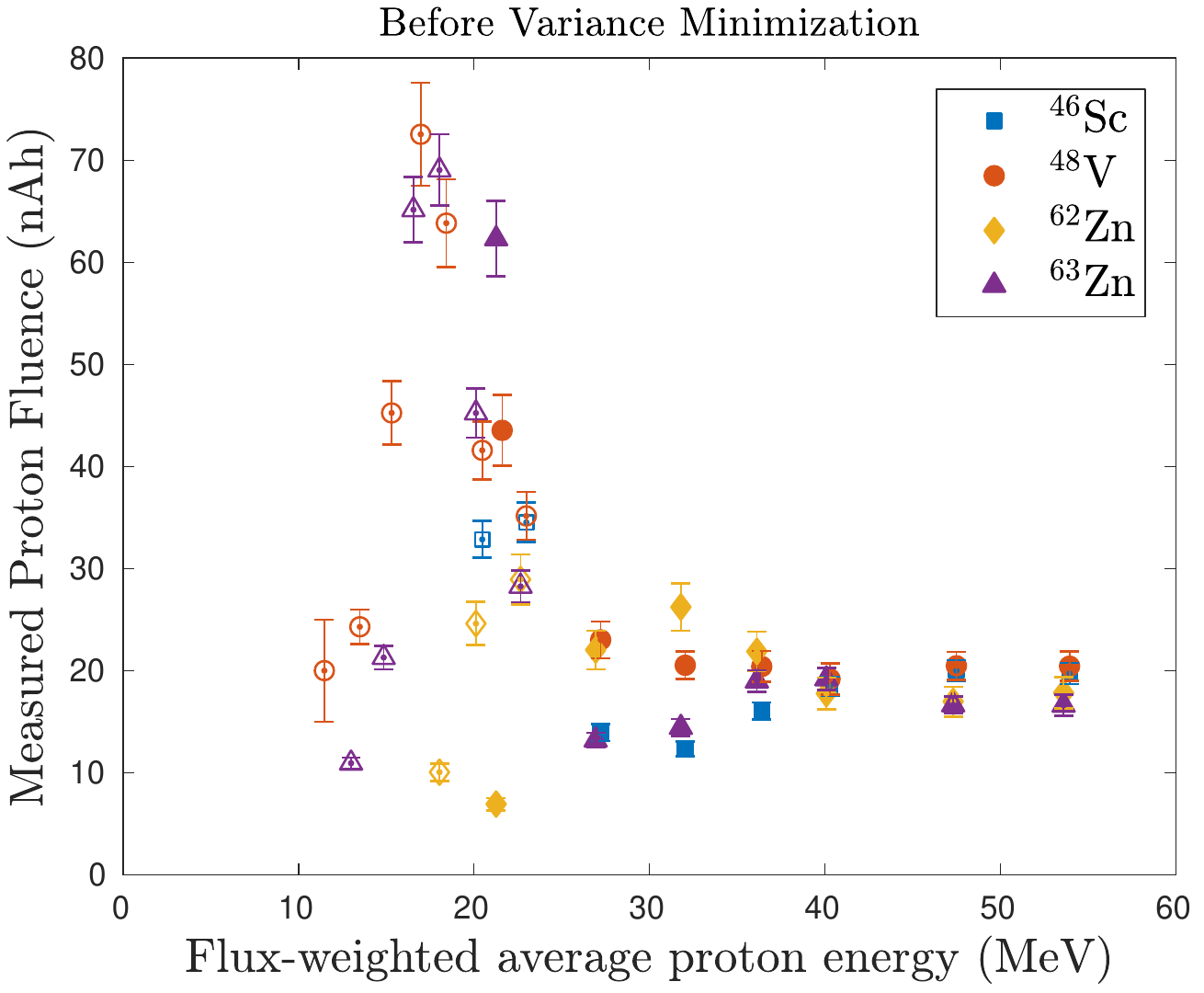}{50}
         \label{fig:fe_before_minimization}
   \hspace{-5pt}}%
     \subfloat{
        \centering

        \subfigimg[width=0.485\textwidth]{b)}{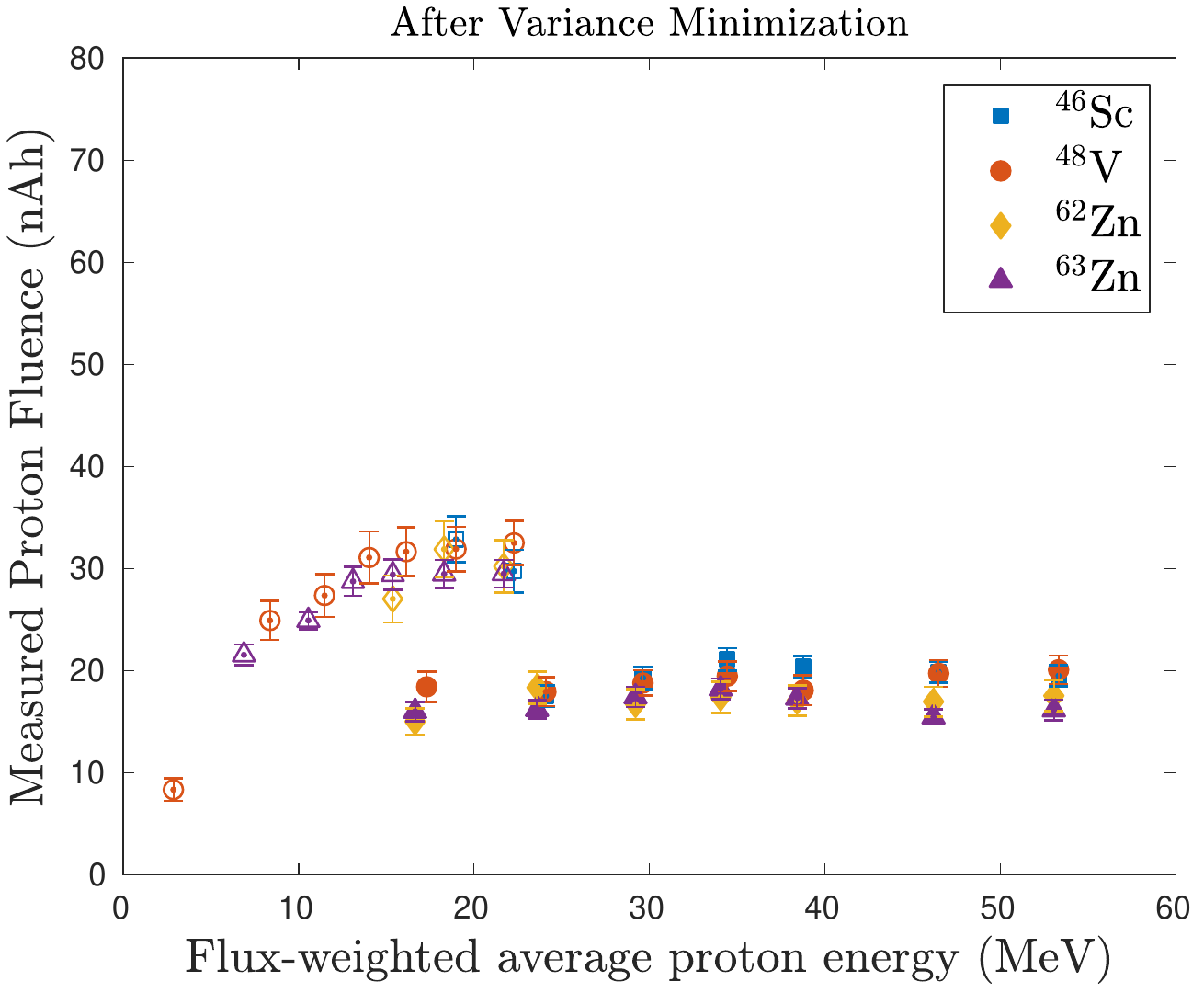}{50}
         \label{fig:fe_after_minimization}
   \hspace{-5pt}}%
    \caption{Results of variance minimization through increasing the effective areal density of the  aluminum degraders by 4.43\% (55\,MeV stack) and Kapton tape by 5.69\% (25\,MeV stack). A noticeable reduction of variance in measured proton fluence is seen,  particularly at the  rear stack positions.  Open data points represent the 25\,MeV stack, and closed data points represent the 55\,MeV stack.} 
     \label{fig:fe_variance_mins}
\end{figure*}

In performing the  minimization, the areal density of each of the  aluminum degraders (for the 55\,MeV stack)  was varied uniformly in FLUKA simulations  by a factor of up to $\pm$25\% of nominal values, to find the effective density which minimized variance in the measured proton fluence at the lowest energy position (Ti-07, Cu-07).
For the 25\,MeV stack, the areal density reached in the minimization of the 55\,MeV stack was used for  the E-09 and H-01 aluminum degraders
and the areal density of each of the  Kapton tape  layers  was 
varied 
by 
$\pm$25\%, 
to find the effective density which minimized variance in the measured proton fluence at the next-to-lowest energy position (Ti-19, Cu-19).
These 
positions were chosen as  minimization candidates as they are the most sensitive to systematic uncertainties in stack design.
In the 25\,MeV stack, activity was not seen in gamma spectrometry for the lowest-energy (Cu-20) monitor foil, implying that the beam was stopped at some point in between Ti-20 and Cu-20.
This observation 
indicates that  the true areal densities of the stack components differ from nominally measured values (primarily for the difficult-to-characterize Kapton tape), as transport calculations using nominal areal densities predict that the beam should exit the stack 
with an energy of approximately 7\,MeV.
As a result, this position was not used for minimization, with the Ti-19 and Cu-19 position being the lowest-energy reliable monitor foils in the stack.
%
%
%
The results of the minimization technique indicate a clear minimum in proton fluence variance for flux-weighted average 22.71\,MeV protons entering the last energy position of the 55\,MeV stack.
This is approximately 2\,MeV lower than the 
FLUKA simulations, and approximately 2\,MeV lower than 
A\&Z calculations, both of which used the nominal 2.80\,g/cm$^3$ measured density of the  aluminum degraders.
This energy corresponds to an aluminum areal density  4.43\% greater than 
measurements and 
corrects for other minor systematic uncertainties in stack design.
Similarly, for the 25\,MeV stack, variance minimization converges on  flux-weighted average 9.23\,MeV protons entering the Fe-13/Ti-19/Cu-19 energy position, which is approximately 4\,MeV lower than the nominal FLUKA simulations, and approximately 5\,MeV lower than nominal A\&Z calculations.
This energy corresponds to a Kapton tape areal density of 5.69\% greater than nominal measurements, which is completely reasonable given the lack of areal density data from the manufacturer.
The impact of this variance minimization for improving disagreement in proton fluence is  clearly  seen in   \autoref{fig:fe_variance_mins}.

\begin{figure}
 \centering
 \includegraphics[width=0.495\textwidth]{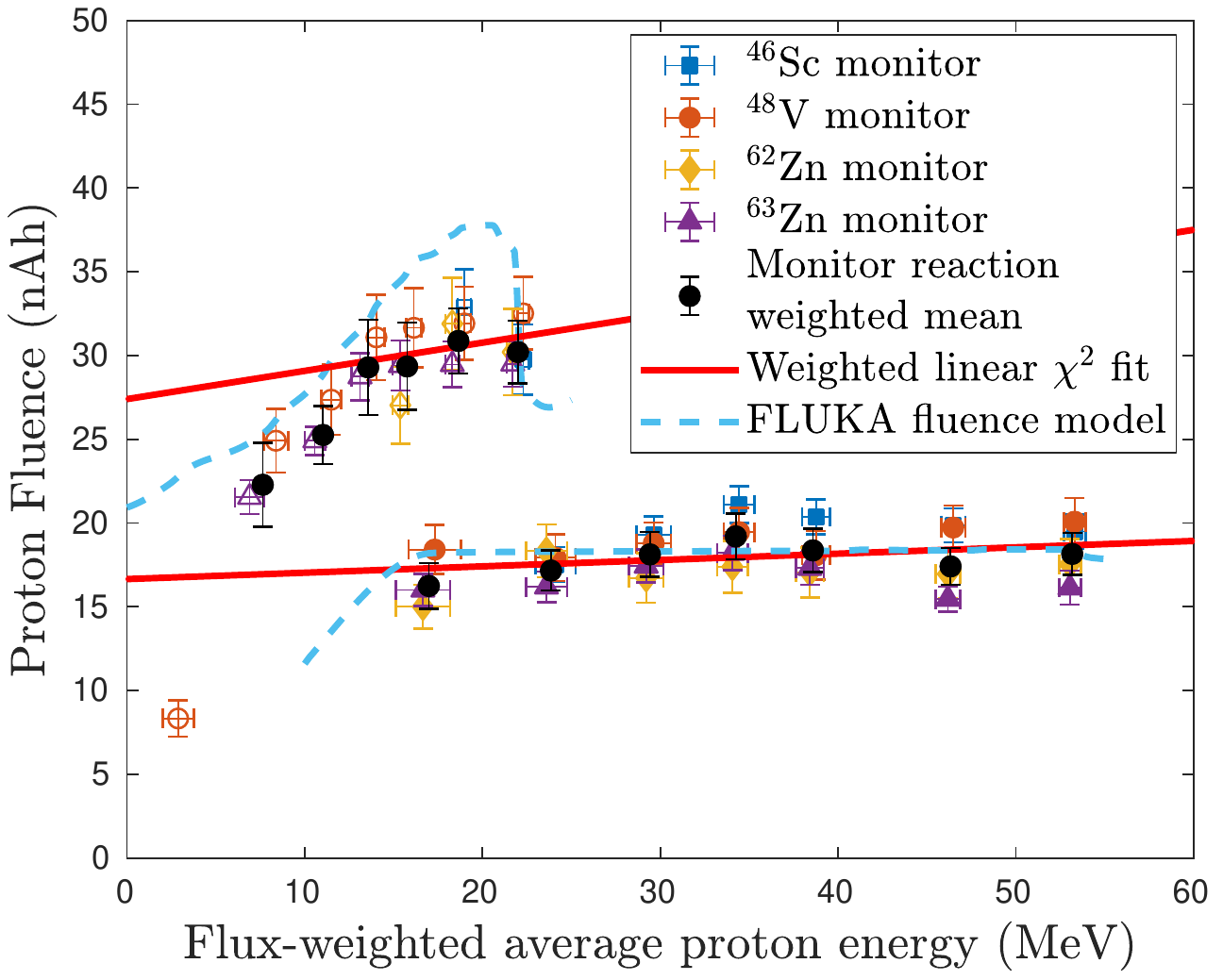}
 \caption{Final uncertainty-weighted mean proton fluences throughout the target stack, based on the variance-minimized observed fluence from the the  \ce{^{nat}Ti}(p,x)\ce{^{46}Sc}, \ce{^{nat}Ti}(p,x)\ce{^{48}V}, \ce{^{nat}Cu}(p,x)\ce{^{62}Zn}, and \ce{^{nat}Cu}(p,x)\ce{^{63}Zn} monitor reactions. Open data points represent the 25\,MeV stack, and closed data points represent the 55\,MeV stack.} 
 \label{fig:fe_fluence_plot}
\end{figure}

%
An enhanced version of the final \ce{^{nat}Ti}(p,x)\ce{^{46}Sc}, \ce{^{nat}Ti}(p,x)\ce{^{48}V}, \ce{^{nat}Cu}(p,x)\ce{^{62}Zn}, and \ce{^{nat}Cu}(p,x)\ce{^{63}Zn} monitor reaction fluences is shown in \autoref{fig:fe_fluence_plot}.
The uncertainty-weighted mean  for the two \ce{^{nat}Cu}(p,x) and two \ce{^{nat}Ti}(p,x) monitor channels was calculated at each energy position, to determine the final fluence assignments for the Cu and Ti foils, respectively, and the uncertainty-weighted mean  for all four monitor channels was used to determine the final fluence assignments for the Fe foils.
Uncertainty in each final proton fluence  is  calculated by error propagation of the individual monitor channel fluence values  at each energy position.
These weighted-mean fluences are  plotted  in \autoref{fig:fe_fluence_plot}, along with the estimated fluence according to both  FLUKA transport 
and an uncertainty-weighted linear $\chi^2$ fit to the individual monitor channel fluence measurements.
Both models reproduce the observed fluence data consistently within uncertainty for the 55\,MeV stack, with the FLUKA model predicting a slightly greater fluence loss throughout the stack.
However, neither model is capable of accurately modeling the rapid decrease in apparent fluence at the rear of the 25\,MeV stack.   
These models are used purely to provide an extrapolation from the highest-energy position back to the \enquote{front} of each stack, to compare with the nominal fluence measured by  the beamline current integrators.
%
%
%
%
%


\subsection{\label{sec:calcs_sec_fe}Calculation of measured cross sections}

Using the quantified EoB activities along with the variance-minimized proton fluence, it is possible to calculate 
cross sections for 
observed (p,x) reactions.
While thin ($\approx$ 10--20\,mg/cm$^2$)  foils were irradiated to minimize the energy bins of these cross section measurements, 
all cross sections reported here are flux-averaged  
over the energy distribution subtended by each foil.
The beam current, measured using a current integrator connected to the electrically-isolated target box, remained stable for the duration of the irradiation.
The propagated uncertainty in cross section is calculated as the quadrature sum of the uncertainty in quantified EoB activity (which includes uncertainty in detector efficiencies), uncertainty in the duration of irradiation (conservatively estimated at 10\,s, to account for any minor transient changes in beam current), uncertainty in foil areal density, uncertainty in monitor product half-life (included, but normally negligible),  and uncertainty in proton current (quantified by error propagation of the monitor reaction fluence values  at each energy position).


\section{\label{sec:results_fe}Results and Discussion}

\subsection{Measurement of nuclear excitation functions}

After irradiation, all foils were 
still sealed in their Kapton packets, verifying that no activation products were lost due to packet failure.
With the exception of a single foil (Cu-20, in the 25\,MeV stack), each activated foil had a small \enquote{blister} under the Kapton tape layer, caused by a combination of 
off-gassing of oxides and the formation of gaseous short-lived beta activities in the tape.
This blister   verifies that the primary proton beam was incident upon the foil, and provides additional 
evidence 
that the beam was stopped in the stack between Ti-20 and Cu-20.
Using the \ce{^{nat}Ti}(p,x)\ce{^{46}Sc}, \ce{^{nat}Ti}(p,x)\ce{^{48}V}, \ce{^{nat}Cu}(p,x)\ce{^{62}Zn}, and \ce{^{nat}Cu}(p,x)\ce{^{63}Zn} monitor reactions, as discussed in \autoref{sec:proton_transport_fe}, a fluence of 17.9$~\pm~$1.0\,nAh 
was calculated to be incident upon the 55\,MeV target stack using the FLUKA fluence model, and a  fluence of 19.0$~\pm~$1.3\,nAh using the linear fit model.
Similarly, for the 25\,MeV stack, a fluence of 27.5$~\pm~$8.3\,nAh 
was calculated to be incident upon the 
target stack using the FLUKA fluence model, and a  fluence of 31.7$~\pm~$3.7\,nAh using the linear fit model to the four frontmost compartments (before the fluence loss becomes strongly nonlinear).
Both linear models are consistent with the nominal fluence of 20.78\,nAh (for the 55\,MeV stack) and 31.61\,nAh (for the 25\,MeV stack) measured using the 
current integrators.
However, for both target stacks, the FLUKA transport model predicts a significant increase in proton fluence, in particular for the 25\,MeV stack.
This model fails to reproduce the fluence loss seen in monitor foils, and predicts a significantly higher production of lower-energy secondary protons not seen in the activation data.
As fluence loss scales with $\sigma_{\mathrm{tot}}\rho\Delta r$, it is expected that an extrapolation back to the stack entrance (through the SS-3/SS-5 profile monitors) will underestimate the nominal fluence incident upon the box.
This incident fluence dropped by approximately 
8.9\% to  17.3$~\pm~$1.5\,nAh using the linear fit model over the length of the 55\,MeV stack, which is consistent with similar measurements at the Los Alamos National Laboratory's Isotope Production Facility in the past\,\cite{Voyles2018a,Graves2016}.
This loss of fluence is due to a combination of 
(p,x) reactions throughout the target stack, as well as large-angle deflections (primarily in the aluminum degraders) from scattering of the beam.

Using the final proton fluence at each energy position, cross sections for \ce{^{48,49,51}Cr}, \ce{^{48}V},  \ce{^{51,52m,52g,52,54,56}Mn},   \ce{^{52}Fe}, and \ce{^{55,56,57,58m,58g,58}Co}
 were extracted for (p,x) reactions  on \ce{^{nat}Fe} foils up to 55\,MeV, presented in \autoref{tab:fe_rp_table}.
For  (p,x) reactions on \ce{^{nat}Cu}, the (p,x) cross sections for  \ce{^{54}Mn}, \ce{^{57}Ni}, \ce{^{57,60,61}Co},   and \ce{^{60,61,64}Cu}  were extracted, 
presented in \autoref{tab:cufe_rp_table}.
For  (p,x) reactions on \ce{^{nat}Ti}, the (p,x) cross sections for  \ce{^{43}K} and \ce{^{44g,44m,44,47,48}Sc}  were extracted, 
presented in \autoref{tab:ti_rp_table}.
In addition, as there exist a number of isomers with radioactive ground states in these mass regions,  independent measurements of isomer-to-ground-state branching ratios for \ce{^{nat}Fe}(p,x)\ce{^{52m/g}Mn}, \ce{^{nat}Fe}(p,x)\ce{^{58m/g}Co}, and \ce{^{nat}Ti}(p,x)\ce{^{44m/g}Sc} were  extracted and are presented in \autoref{tab:fe_ibr_table}.
Comparisons  of the measured cross sections and isomer branching ratios with literature data (retrieved from EXFOR\,\cite{Otuka2014272}) are seen in the figures of Appendices \ref{sec:fe_xs_figures} and \ref{sec:fe_ibr_figures}.
The propagated uncertainty in these cross sections varies widely based on the reaction product in question, with the major components  arising from uncertainty in EoB activity ($\pm$3--10\%), proton fluence ($\pm$5--13\%), and foil areal density ($\pm$0.1--0.3\%).

\begin{table}
\centering
\caption{Measured cross sections for the various \ce{^{nat}Fe}(p,x) reaction products observed in this work. Cumulative cross sections are designated as $\sigma_c$, independent cross sections are designated as $\sigma_i$.}
\label{tab:fe_rp_table}
\small
\resizebox{\textwidth}{!}{%
\begin{tabular}{@{}ccccccccccccccc@{}}
\toprule\toprule
                            & \multicolumn{14}{c}{Production cross section (mb)}                                                                                                         \\ \cmidrule(l){2-15} 
E$_\text{p}$ (MeV)&	\makecell{53.45(61)} &	\makecell{46.63(68)} &	\makecell{38.93(78)} &	\makecell{34.62(85)} &	\makecell{29.84(96)} &	\makecell{24.4(11)} &	\makecell{22.71(43)} &	\makecell{19.36(56)} &	\makecell{17.6(15)} &	\makecell{16.54(48)} &	\makecell{14.52(49)} &	\makecell{12.11(53)} &	\makecell{9.23(61)} &	\makecell{4.10(73)}\\  \midrule
\ce{^{48}Cr}\,($\sigma_c$)&	\makecell{0.172(11)} &	\makecell{0.01376(80)} &	\makecell{0.00435(29)} &	-- &	-- &	-- &	-- &	-- &	-- &	-- &	-- &	-- &	-- &	--\\
\ce{^{48}V}\,($\sigma_i$)&	\makecell{6.38(43)} &	\makecell{4.64(29)} &	\makecell{0.513(35)} &	\makecell{0.1096(80)} &	-- &	-- &	-- &	-- &	-- &	-- &	-- &	-- &	-- &	--\\
\ce{^{48}V}\,($\sigma_c$)&	\makecell{6.55(43)} &	\makecell{4.65(29)} &	\makecell{0.517(35)} &	\makecell{0.1096(80)} &	-- &	-- &	-- &	-- &	-- &	-- &	-- &	-- &	-- &	--\\
\ce{^{49}Cr}\,($\sigma_c$)&	\makecell{1.83(12)} &	\makecell{2.25(18)} &	\makecell{1.20(11)} &	\makecell{0.315(36)} &	-- &	-- &	-- &	-- &	-- &	-- &	-- &	-- &	-- &	--\\
\ce{^{51}Mn}\,($\sigma_c$)&	\makecell{11.77(74)} &	\makecell{15.69(94)} &	\makecell{11.78(77)} &	\makecell{6.15(41)} &	\makecell{0.475(39)} &	\makecell{0.679(47)} &	\makecell{1.060(63)} &	\makecell{1.97(15)} &	\makecell{2.10(17)} &	\makecell{1.98(19)} &	\makecell{1.46(13)} &	\makecell{0.584(41)} &	-- &	--\\
\ce{^{51}Cr}\,($\sigma_i$)&	\makecell{65.9(58)} &	\makecell{81.0(59)} &	\makecell{56.5(45)} &	\makecell{27.6(23)} &	\makecell{3.83(35)} &	\makecell{0.90(12)} &	\makecell{0.70(13)} &	\makecell{0.150(49)} &	\makecell{0.110(24)} &	-- &	-- &	-- &	-- &	--\\
\ce{^{51}Cr}\,($\sigma_c$)&	\makecell{77.6(57)} &	\makecell{96.7(58)} &	\makecell{68.3(44)} &	\makecell{33.7(23)} &	\makecell{4.30(35)} &	\makecell{1.58(11)} &	\makecell{1.76(11)} &	\makecell{2.12(13)} &	\makecell{2.21(17)} &	\makecell{1.98(19)} &	\makecell{1.46(13)} &	\makecell{0.584(41)} &	-- &	--\\
\ce{^{52}Fe}\,($\sigma_c$)&	\makecell{2.74(17)} &	\makecell{1.82(11)} &	\makecell{1.60(10)} &	\makecell{2.25(15)} &	\makecell{0.770(52)} &	\makecell{0.206(15)} &	\makecell{0.192(13)} &	\makecell{0.01297(75)} &	\makecell{0.00242(21)} &	-- &	-- &	-- &	-- &	--\\
\ce{^{52m}Mn}\,($\sigma_i$)&	\makecell{8.29(52)} &	\makecell{9.49(54)} &	\makecell{13.69(88)} &	\makecell{17.9(12)} &	\makecell{23.3(15)} &	\makecell{11.82(74)} &	\makecell{5.78(33)} &	\makecell{0.0763(44)} &	\makecell{0.0763(57)} &	\makecell{0.0754(61)} &	\makecell{0.0585(52)} &	-- &	-- &	--\\
\ce{^{52g}Mn}\,($\sigma_i$)&	\makecell{11.58(72)} &	\makecell{13.46(76)} &	\makecell{20.8(13)} &	\makecell{28.2(18)} &	\makecell{21.8(15)} &	\makecell{16.3(10)} &	\makecell{10.37(64)} &	\makecell{0.300(17)} &	\makecell{0.1124(85)} &	\makecell{0.0428(38)} &	\makecell{0.00540(50)} &	-- &	-- &	--\\
\ce{^{52g}Mn}\,($\sigma_c$)&	\makecell{13.66(90)} &	\makecell{15.6(11)} &	\makecell{23.0(18)} &	\makecell{30.9(21)} &	\makecell{31.7(22)} &	\makecell{18.5(13)} &	\makecell{10.57(62)} &	\makecell{0.313(19)} &	\makecell{0.0393(45)} &	\makecell{0.0556(48)} &	\makecell{0.0185(19)} &	-- &	-- &	--\\
\ce{^{54}Mn}\,($\sigma_i$)&	\makecell{131.0(85)} &	\makecell{162(10)} &	\makecell{167(12)} &	\makecell{129(10)} &	\makecell{42.2(34)} &	\makecell{2.46(25)} &	\makecell{1.10(13)} &	\makecell{1.09(14)} &	\makecell{1.47(17)} &	\makecell{1.53(16)} &	\makecell{1.36(12)} &	\makecell{1.133(80)} &	\makecell{0.747(75)} &	\makecell{0.0600(80)}\\
\ce{^{55}Co}\,($\sigma_i$)&	\makecell{9.43(63)} &	\makecell{12.5(10)} &	\makecell{15.7(12)} &	\makecell{21.5(15)} &	\makecell{48.4(36)} &	\makecell{64.7(56)} &	\makecell{61.0(45)} &	\makecell{43.6(30)} &	\makecell{33.6(34)} &	\makecell{13.4(12)} &	\makecell{0.377(35)} &	\makecell{0.0421(29)} &	-- &	--\\
\ce{^{56}Mn}\,($\sigma_c$)&	-- &	\makecell{0.518(39)} &	\makecell{0.610(44)} &	\makecell{0.462(45)} &	\makecell{0.506(54)} &	\makecell{0.405(33)} &	\makecell{0.223(13)} &	\makecell{0.0962(56)} &	\makecell{0.0329(43)} &	\makecell{0.0253(21)} &	\makecell{0.0132(14)} &	-- &	-- &	--\\
\ce{^{56}Co}\,($\sigma_i$)&	\makecell{13.0(11)} &	\makecell{16.3(14)} &	\makecell{18.9(16)} &	\makecell{23.6(17)} &	\makecell{29.2(26)} &	\makecell{47.8(32)} &	\makecell{51.6(30)} &	\makecell{82.6(50)} &	\makecell{176(13)} &	\makecell{197(16)} &	\makecell{344(30)} &	\makecell{376(24)} &	\makecell{288(29)} &	\makecell{3.54(47)}\\
\ce{^{57}Co}\,($\sigma_i$)&	-- &	\makecell{0.476(47)} &	\makecell{0.539(60)} &	\makecell{0.648(44)} &	\makecell{1.170(90)} &	\makecell{1.84(12)} &	\makecell{2.36(14)} &	\makecell{2.50(16)} &	\makecell{3.20(25)} &	\makecell{3.40(28)} &	\makecell{5.14(47)} &	\makecell{8.18(52)} &	\makecell{11.5(12)} &	\makecell{5.49(73)}\\
\ce{^{58m}Co}\,($\sigma_i$)&	-- &	-- &	-- &	\makecell{0.0427(28)} &	\makecell{0.0619(42)} &	\makecell{0.1054(69)} &	\makecell{0.172(11)} &	\makecell{0.236(15)} &	\makecell{0.241(19)} &	\makecell{0.300(28)} &	\makecell{0.475(44)} &	\makecell{0.545(35)} &	\makecell{0.477(49)} &	\makecell{0.170(25)}\\
\ce{^{58g}Co}\,($\sigma_i$)&	-- &	-- &	-- &	\makecell{0.0884(66)} &	\makecell{0.0980(74)} &	\makecell{0.1118(82)} &	\makecell{0.1229(84)} &	\makecell{0.1484(90)} &	\makecell{0.333(28)} &	\makecell{0.318(27)} &	\makecell{0.919(90)} &	\makecell{1.276(96)} &	\makecell{1.56(17)} &	\makecell{0.623(83)}\\
\ce{^{58g}Co}\,($\sigma_c$)&	-- &	-- &	-- &	\makecell{0.1311(72)} &	\makecell{0.1599(86)} &	\makecell{0.217(11)} &	\makecell{0.295(14)} &	\makecell{0.384(18)} &	\makecell{0.574(34)} &	\makecell{0.618(39)} &	\makecell{1.39(10)} &	\makecell{1.82(10)} &	\makecell{2.04(18)} &	\makecell{0.792(86)}\\ \bottomrule\bottomrule
\end{tabular}
}
\end{table}

\begin{table}
\centering
\caption{Measured cross sections for the various \ce{^{nat}Cu}(p,x) reaction products observed in this work. Cumulative cross sections are designated as $\sigma_c$, independent cross sections are designated as $\sigma_i$.}
\label{tab:cufe_rp_table}
\small
\resizebox{\textwidth}{!}{%
\begin{tabular}{@{}ccccccccccccccc@{}}
\toprule\toprule
                            & \multicolumn{14}{c}{Production cross section (mb)}                                                                                                         \\ \cmidrule(l){2-15} 
E$_\text{p}$ (MeV)&	\makecell{53.04(61)} &	\makecell{46.18(68)} &	\makecell{38.42(79)} &	\makecell{34.06(86)} &	\makecell{29.21(97)} &	\makecell{23.6(12)} &	\makecell{21.70(33)} &	\makecell{18.30(38)} &	\makecell{16.6(15)} &	\makecell{15.38(44)} &	\makecell{13.11(49)} &	\makecell{10.57(57)} &	\makecell{6.90(82)} &	\makecell{1.4(13)}\\ \midrule
\ce{^{54}Mn}\,($\sigma_i$)&	\makecell{2.09(13)} &	\makecell{0.428(24)} &	\makecell{0.0931(61)} &	\makecell{0.0517(33)} &	\makecell{0.0223(15)} &	\makecell{0.0185(12)} &	-- &	-- &	-- &	-- &	-- &	-- &	-- &	--\\
\ce{^{57}Ni}\,($\sigma_c$)&	\makecell{2.15(15)} &	\makecell{0.775(45)} &	\makecell{0.0530(44)} &	-- &	-- &	-- &	-- &	-- &	-- &	-- &	-- &	-- &	-- &	--\\
\ce{^{57}Co}\,($\sigma_i$)&	\makecell{49.8(34)} &	\makecell{33.8(23)} &	\makecell{3.79(42)} &	\makecell{1.206(80)} &	\makecell{1.67(11)} &	\makecell{1.053(72)} &	\makecell{0.707(46)} &	\makecell{0.264(27)} &	-- &	-- &	-- &	-- &	-- &	--\\
\ce{^{57}Co}\,($\sigma_c$)&	\makecell{51.9(34)} &	\makecell{34.6(23)} &	\makecell{3.84(42)} &	\makecell{1.206(80)} &	\makecell{1.67(11)} &	\makecell{1.053(72)} &	\makecell{0.707(46)} &	\makecell{0.264(27)} &	-- &	-- &	-- &	-- &	-- &	--\\
\ce{^{60}Co}\,($\sigma_c$)&	\makecell{9.41(59)} &	\makecell{8.08(47)} &	\makecell{6.14(40)} &	\makecell{3.12(22)} &	\makecell{0.794(63)} &	\makecell{0.201(17)} &	\makecell{0.125(10)} &	\makecell{0.0199(22)} &	-- &	-- &	-- &	-- &	-- &	--\\
\ce{^{60}Cu}\,($\sigma_c$)&	\makecell{25.3(16)} &	\makecell{16.86(97)} &	\makecell{1.46(12)} &	\makecell{0.578(38)} &	-- &	-- &	-- &	-- &	-- &	-- &	-- &	-- &	-- &	--\\
\ce{^{61}Co}\,($\sigma_c$)&	\makecell{4.26(54)} &	\makecell{5.98(66)} &	\makecell{6.94(62)} &	\makecell{6.61(70)} &	\makecell{5.94(83)} &	\makecell{0.872(81)} &	\makecell{0.253(14)} &	\makecell{0.1178(69)} &	\makecell{0.0415(35)} &	-- &	-- &	-- &	-- &	--\\
\ce{^{61}Cu}\,($\sigma_c$)&	\makecell{79.7(51)} &	\makecell{106.4(64)} &	\makecell{161(11)} &	\makecell{155(10)} &	\makecell{104.1(72)} &	\makecell{6.97(52)} &	\makecell{1.84(13)} &	\makecell{1.195(80)} &	\makecell{0.809(69)} &	-- &	-- &	-- &	-- &	--\\
\ce{^{64}Cu}\,($\sigma_i$)&	\makecell{50.9(32)} &	\makecell{55.4(32)} &	\makecell{58.4(38)} &	\makecell{62.4(41)} &	\makecell{101.8(68)} &	\makecell{145(14)} &	\makecell{83.1(47)} &	\makecell{57.0(33)} &	\makecell{46.4(35)} &	\makecell{22.2(18)} &	-- &	-- &	-- &	--\\ \bottomrule\bottomrule
\end{tabular}
}
\end{table}

\begin{table}
\centering
\caption{Measured cross sections for the various \ce{^{nat}Ti}(p,x) reaction products observed in this work. Cumulative cross sections are designated as $\sigma_c$, independent cross sections are designated as $\sigma_i$.}
\label{tab:ti_rp_table}
\small
\resizebox{\textwidth}{!}{%
\begin{tabular}{@{}ccccccccccccccc@{}}
\toprule\toprule
                            & \multicolumn{14}{c}{Production cross section (mb)}                                                                                                         \\ \cmidrule(l){2-15} 
E$_\text{p}$ (MeV)&	\makecell{53.31(61)} &	\makecell{46.48(68)} &	\makecell{38.76(78)} &	\makecell{34.44(86)} &	\makecell{29.63(96)} &	\makecell{24.1(11)} &	\makecell{22.29(32)} &	\makecell{18.98(37)} &	\makecell{17.3(15)} &	\makecell{16.14(42)} &	\makecell{14.03(47)} &	\makecell{11.49(55)} &	\makecell{8.38(70)} &	\makecell{2.88(88)}\\  \midrule
\ce{^{43}K}\,($\sigma_c$)&	\makecell{1.55(11)} &	\makecell{0.889(67)} &	\makecell{0.189(17)} &	\makecell{0.0462(42)} &	-- &	-- &	-- &	-- &	-- &	-- &	-- &	-- &	-- &	-- \\
\ce{^{44m}Sc}\,($\sigma_i$)&	\makecell{12.60(72)} &	\makecell{12.37(63)} &	\makecell{15.20(93)} &	\makecell{17.8(11)} &	\makecell{16.8(11)} &	\makecell{8.63(53)} &	\makecell{4.26(24)} &	\makecell{1.451(82)} &	\makecell{1.317(88)} &	\makecell{1.269(95)} &	\makecell{0.839(70)} &	--  &	-- &	-- \\
\ce{^{44g}Sc}\,($\sigma_i$)&	\makecell{25.0(24)} &	\makecell{27.0(22)} &	\makecell{37.0(29)} &	\makecell{52.6(48)} &	\makecell{47.8(39)} &	\makecell{29.9(30)} &	\makecell{8.09(45)} &	\makecell{3.49(20)} &	\makecell{2.70(31)} &	\makecell{3.02(23)} &	\makecell{2.49(21)} &	-- &	-- &	-- \\
\ce{^{44g}Sc}\,($\sigma_c$)&	\makecell{37.6(25)} &	\makecell{39.4(23)} &	\makecell{52.2(31)} &	\makecell{70.4(49)} &	\makecell{64.7(40)} &	\makecell{38.6(31)} &	\makecell{12.35(51)} &	\makecell{4.94(22)} &	\makecell{4.02(32)} &	\makecell{4.29(25)} &	\makecell{3.33(22)} &	-- &	-- &	-- \\
\ce{^{47}Sc}\,($\sigma_c$)&	\makecell{21.2(12)} &	\makecell{20.5(10)} &	\makecell{21.7(13)} &	\makecell{23.5(14)} &	\makecell{25.1(16)} &	\makecell{15.63(97)} &	\makecell{11.53(70)} &	\makecell{5.50(32)} &	\makecell{2.75(18)} &	\makecell{1.57(12)} &	\makecell{0.810(67)} &	\makecell{0.361(21)} &	\makecell{0.218(23)} &	-- \\
\ce{^{48}Sc}\,($\sigma_i$)&	\makecell{1.66(13)} &	\makecell{1.68(19)} &	\makecell{1.29(15)} &	\makecell{0.772(65)} &	\makecell{0.700(64)} &	\makecell{0.339(26)} &	\makecell{0.318(18)} &	\makecell{0.185(13)} &	\makecell{0.135(14)} &	\makecell{0.0625(55)} &	-- &	-- &	-- &	-- \\ \bottomrule\bottomrule
\end{tabular}
}
\end{table}

\begin{table}
\centering
\caption{Measured isomer-to-ground-state branching ratios for the various \ce{^{nat}Fe}(p,x) and \ce{^{nat}Ti}(p,x) reaction products observed in this work.}
\label{tab:fe_ibr_table}
\small
\resizebox{\textwidth}{!}{%
\begin{tabular}{@{}ccccccccccccccc@{}}
\toprule\toprule
                               & \multicolumn{14}{c}{Isomer branching ratio}                                                                                                                \\ \cmidrule(l){2-15} 
E$_\text{p}$ (MeV)&	\makecell{53.45(61)} &	\makecell{46.63(68)} &	\makecell{38.93(78)} &	\makecell{34.62(85)} &	\makecell{29.84(96)} &	\makecell{24.4(11)} &	\makecell{22.71(43)} &	\makecell{19.36(56)} &	\makecell{17.6(15)} &	\makecell{16.54(48)} &	\makecell{14.52(49)} &	\makecell{12.11(53)} &	\makecell{9.23(61)} &	\makecell{4.10(73)}\\ \midrule
\ce{^{nat}Fe}(p,x)\ce{^{52}Mn} &	\makecell{0.417(38)} &	\makecell{0.414(37)} &	\makecell{0.396(40)} &	\makecell{0.388(37)} &	\makecell{0.517(49)} &	\makecell{0.420(40)} &	\makecell{0.358(29)} &	\makecell{0.202(17)} &	\makecell{0.404(56)} &	\makecell{0.638(76)} &	\makecell{0.92(12)} &	-- &	-- &	--\\
\ce{^{nat}Fe}(p,x)\ce{^{58}Co} &	-- &	-- &	-- &	\makecell{0.326(28)} &	\makecell{0.387(34)} &	\makecell{0.485(40)} &	\makecell{0.583(47)} &	\makecell{0.614(49)} &	\makecell{0.420(41)} &	\makecell{0.486(55)} &	\makecell{0.341(40)} &	\makecell{0.299(26)} &	\makecell{0.234(31)} &	\makecell{0.214(39)}\\         \vspace{1em}     \\ 
E$_\text{p}$ (MeV)&	\makecell{53.31(61)} &	\makecell{46.48(68)} &	\makecell{38.76(78)} &	\makecell{34.44(86)} &	\makecell{29.63(96)} &	\makecell{24.1(11)} &	\makecell{22.29(32)} &	\makecell{18.98(37)} &	\makecell{17.3(15)} &	\makecell{16.14(42)} &	\makecell{14.03(47)} &	\makecell{11.49(55)} &	\makecell{8.38(70)} &	\makecell{2.88(88)}\\ \midrule
\ce{^{nat}Ti}(p,x)\ce{^{44}Sc} &	\makecell{0.335(29)} &	\makecell{0.314(24)} &	\makecell{0.291(25)} &	\makecell{0.253(23)} &	\makecell{0.260(23)} &	\makecell{0.224(23)} &	\makecell{0.345(24)} &	\makecell{0.294(21)} &	\makecell{0.328(34)} &	\makecell{0.296(28)} &	\makecell{0.252(27)} &	-- &	-- &	--\\ \bottomrule\bottomrule
\end{tabular}
}
\end{table}


%

These results have several notable features.
The 
\ce{^{nat}Fe}, \ce{^{nat}Cu}, and \ce{^{nat}Ti}(p,x) cross sections measured here are in excellent agreement with 
literature,  but have been measured nearly exclusively with the highest precision to date.
While (p,x) reactions below 70\,MeV on these elements are well-characterized overall, measurements of several reaction channels are somewhat sparse in comparison.
Indeed, fewer than four existing measurements have been performed for the \ce{^{nat}Fe}(p,x)\ce{^{48}Cr},\ce{^{52}Fe}, \ce{^{nat}Cu}(p,x)\ce{^{60}Cu},\ce{^{61}Co}, and \ce{^{nat}Ti}(p,x)\ce{^{44}Sc}
reactions presented here.
Additionally, \ce{^{22,24}Na} activity is seen in all foils, consistent with proton activation of the trace  silicon in the Kapton tape used for foil encapsulation, as described in our previous work\,\cite{Voyles2018a}.
No cross sections for \ce{^{nat}Si}(p,x)\ce{^{22,24}Na} are reported due to the significant uncertainty in characterizing the layer of silicone adhesive, but this serves as another example of how the use of silicone-based adhesives may systematically enhance the apparent fluence when using the \ce{^{nat}Al}(p,x)\ce{^{22,24}Na} monitor reactions.

This work presents the first measurements of several observables in 
this mass region, including the \ce{^{nat}Fe}(p,x)\ce{^{49}Cr}, \ce{^{nat}Fe}(p,x)\ce{^{51}Mn}, \ce{^{nat}Fe}(p,x)\ce{^{52m}Mn},  \ce{^{nat}Fe}(p,x)\ce{^{56}Mn}, and \ce{^{nat}Fe}(p,x)\ce{^{58m}Co} reactions in the 0--70\,MeV region, 
the independent cross sections for       \ce{^{nat}Fe}(p,x)\ce{^{51}Cr}, \ce{^{nat}Fe}(p,x)\ce{^{52g}Mn}, \ce{^{nat}Fe}(p,x)\ce{^{58g}Co}, and the \ce{^{52\text{m}}Mn} ($2^+$) / \ce{^{52\text{g}}Mn}  ($6^+$) and \ce{^{58\text{m}}Co} ($5^+$) / \ce{^{58\text{g}}Co}  ($2^+$)  isomer branching ratios via \ce{^{nat}Fe}(p,x).  
The cumulative cross sections from these data are also consistent with existing measurements of the cumulative \ce{^{nat}Fe}(p,x)\ce{^{51}Cr},\ce{^{52g}Mn},\ce{^{58g}Co} cross sections.


Several of the activities produced via \ce{^{nat}Fe}(p,x)  could be useful as experimental monitor reactions. 
In particular, \ce{^{nat}Fe}(p,x)\ce{^{56}Co} ($t_{1/2}=77.236~\pm~0.026$ d, $\epsilon$=100\% to \ce{^{56}Fe}\,\cite{Junde2011})  is a strongly-fed reaction channel (peak cross section of approximately 400\,mb near 12\,MeV), with a sufficiently-long half-life permitting offline gamma spectrometry,  and  cannot be populated via secondary neutrons incident upon the monitor target. 
It possesses a number of intense gamma-rays, which are distinct from those populated by daughter states in the decay of  \ce{^{56}Mn}, and the gammas which are produced both in the decay of  \ce{^{56}Co} and  \ce{^{56}Mn} may be easily resolved based on differences in intensity and half-life. 
Similarly,  \ce{^{nat}Fe}(p,x)\ce{^{54}Mn}  ($t_{1/2}=312.20~\pm~0.020$ d, $\epsilon$=100\% to \ce{^{54}Cr}\,\cite{Dong2014}) has a convenient half-life and strong cross section (peak of approximately 160\,mb near 40\,MeV), and is immune from two different reactions on the same monitor foil leading to states in the same daughter nuclide.
This reaction could be useful for intermediate- to high-energy protons ($E_p \geq$30\,MeV), but would be susceptible to production via the high-energy secondary neutrons  (threshold 12.1\,MeV off of \ce{^{56}Fe}) produced in these facilities, though (p,x) production rates should dominate (n,x) through both particle flux and cross section. 
Additionally, the \ce{^{nat}Fe}(p,x)\ce{^{54}Mn} and \ce{^{nat}Fe}(p,x)\ce{^{56}Co} channels have a difference in apparent energetic thresholds of nearly 20\,MeV.
This provides some energy discrimination sensitivity in the same iron foil, particularly in the 20--50\,MeV region, which could be useful for determining incident proton energy near 40\,MeV, where fast neutrons are less significant than deeper in the stack.


Notably, this work is the most well-characterized measurement of the \ce{^{nat}Fe}(p,x)\ce{^{51,52}Mn} reactions below 70\,MeV to date, with cross sections measured  at the 6--10\% uncertainty level.
This is important, as it presents the first measurement of the \ce{^{nat}Fe}(p,x)\ce{^{51}Mn} reaction,  the first measurement of the independent \ce{^{nat}Fe}(p,x)\ce{^{52m,52g}Mn} cross sections, and extends the \ce{^{nat}Fe}(p,x)\ce{^{52}Mn} excitation function down to the lowest energy to date. 
\ce{^{nat}Fe}(p,x)\ce{^{51}Mn} appears to offer a compelling alternative to the more established \ce{^{50}Cr}(d,x)\ce{^{51}Mn} pathway, which necessitates an enriched \ce{^{50}Cr} target to avoid radio-manganese contamination from reactions on stable isotopes of Cr\,\cite{Klein2000}.
\ce{^{nat}Fe}(p,x)\ce{^{51}Mn} could be used for 
production of  \ce{^{51}Mn} ($\geq$98.8\% radioisotopic purity) below 20\,MeV using the \ce{^{54}Fe}(p,$\alpha$)\ce{^{51}Mn} channel.
In addition, this low-energy production is accessible using the international network of small medical and research cyclotrons, enabling in-house production of this short-lived ($t_{1/2}=46.2~\pm~0.1$ m\,\cite{Wang2017}) radionuclide.
To increase yields over \ce{^{nat}Fe}(p,x), an enriched \ce{^{54}Fe} target could be used to take advantage of the eight-fold increase in reaction cross section for production using 40--50\,MeV protons, without opening the additional manganese exit channels accessible on a natural target.

Likewise, \ce{Fe}(p,x) offers an interesting production pathway for  \ce{^{52}Mn}.
Conventional production uses the low-energy \ce{^{nat}Cr}/\ce{^{52}Cr}(p,n)\ce{^{52}Mn} pathways, which offer high radioisotopic purity (approximately 99.6\%)\,\cite{Graves2015,Wooten2015}.
\ce{^{nat}Fe}(p,x) offers a nearly threefold increase in production yield, but the low radioisotopic purity (99.1\% for 20--30\,MeV, decreasing to 60.8\% by 40\,MeV) at higher energies due to the opening of \ce{^{54}Mn} makes this route seem impractical.
Much like \ce{^{51}Mn}, the use of an enriched \ce{^{54}Fe} target would prevent production of \ce{^{54}Mn}, providing a higher-yield production route over   \ce{^{nat}Cr}(p,x), with the tradeoff of necessitating higher-energy protons ($\leq$35\,MeV) for production.
It is important to note that the \ce{Fe}(p,x) route provides $\geq$60\% feeding of \ce{^{52g}Mn} ($t_{1/2}$ = 5.591$~\pm~$0.003\,d\,\cite{Dong2015}), implying that the short-lived  \ce{^{52m}Mn} ($t_{1/2}$ = 21.1$~\pm~$0.2\,m\,\cite{Dong2015}) can be easily separated through the difference in half-life, to avoid the unfavorable gammas produced by the isomer. 
However, if nearly pure \ce{^{52m}Mn} is desired for preclinical imaging applications, the feeding of \ce{^{52}Mn} through the $\epsilon$ decay of \ce{^{52}Fe} ($t_{1/2}$ = 8.725$~\pm~$0.008\,h\,\cite{Dong2015}) exclusively populates the isomer, making this potentially suitable for production through \enquote{milking} of a \ce{^{52}Fe} generator\,\cite{Blauenstein1997}.
Clearly, the use of \ce{Fe}(p,x)\ce{^{51,52}Mn} has significant untapped potential, and additional work is needed to further characterize these reaction channels for $E_p \leq$60\,MeV.



In addition to the \ce{^{nat}Fe}(p,x)\ce{^{51,52}Mn} measurements, this experiment has also yielded measurements of  a number of additional  emerging radionuclides with medical applications.
These include the non-standard positron emitters 
\ce{^{44}Sc}\,\cite{Muller2013,Filosofov2010,Qaim2011}
\ce{^{55}Co}\,\cite{Thisgaard2011,Zaman1996,Hermanne2000a}
\ce{^{61}Ni}\,\cite{PMID:7632762,zweit1996medium,Graves2016,Rosch2014}, 
\ce{^{61}Cu}\,\cite{Szelecsenyi2005a,Fukumura2004}
\ce{^{64}Cu}\,\cite{Lewis2003,Bandari2014,mp500671j,Szelecsenyi1993,Aslam2009,Hilgers2003,Szelecsenyi2005,Voyles2017},
and the $\beta^-$-therapeutic agent 
\ce{^{47}Sc}\,\cite{Muller2014,Deilami-nezhad2016}.
Production of these radionuclides offers no major advantages over established pathways, with the generally lower yields and radioisotopic purities failing to justify the convenience of natural targets  via   \ce{^{nat}Fe}(p,x), \ce{^{nat}Cu}(p,x), and  \ce{^{nat}Ti}(p,x). 
The one potential exception to this is the production of \ce{^{58m}Co}, a potent agent for Auger electron-based targeted therapy\,\cite{Thisgaard2011a,Valdovinos2017b,Thisgaard2014a}.
While ingrowth of the long-lived \ce{^{58g}Co} ($t_{1/2}$ = 70.86(6) d\,\cite{Nesaraja2010}) is unavoidable, minimizing co-production of \ce{^{58g}Co} is necessary to minimize patient dose.
The \ce{^{nat}Fe}(p,x)\ce{^{58m}Co} pathway shows a clear \enquote{peak} in the \ce{^{58m}Co}/\ce{^{58g}Co} branching ratio for approximately 15--30\,MeV protons, which might have translational implications if Auger electron therapy becomes more clinically prevalent.

\subsection{Comparison of reaction modeling with experimental results}

\begin{table}
 \caption{Default settings for the reactions codes}
 \label{tab:fe_defaults}
 \resizebox{\textwidth}{!}{%
\begin{tabular}{ c c c c}

 \underline{Code Version} & \underline{Proton/Neutron Optical Model} & \underline{Alpha Optical Model}  & \underline{E1 $\gamma$SF Model}  \\ 
 EMPIRE-3.2.3\cite{Herman2007}       & Koning-Delaroche\cite{Koning2003} & Avrigeanu(2009)\cite{Avrigeanu2009}       & Modified Lorentzian\cite{belgya2006handbook}  \\  
 TALYS-1.8\cite{Koning2012}          & Koning-Delaroche & Specific folded potential\cite{Koning2012}      & Brink-Axel Lorentzian\cite{Koning2012}   \\
  CoH-3.5.3\cite{kawano2003coh,KAWANO2010} & Koning-Delaroche     & Avrigeanu(1994)\cite{Avrigeanu1994} & Generalized Lorentzian\cite{kawano2003coh,KAWANO2010} \\
  ALICE-2017\cite{Blann1996}     &   Nadasen\cite{Nadasen1981}    &  Parabolic Diffuse-Well\cite{Thomas1959}    &   Berman-Fultz Lorentzian\cite{Berman1975a}
  
\end{tabular}
}
\end{table}

%

The measured cross sections were compared to the predictions by the reaction codes TALYS, EMPIRE, CoH, ALICE, and by the calculations in the TENDL database.
The codes were all run on their default settings, in order to assess their predictive capabilities for the casual user.
The default settings for the optical models and gamma strength function ($\gamma$SF) are listed in  \autoref{tab:fe_defaults}.
The level density models for each are as follows.
For both CoH and TALYS, the default level density model is the Gilbert-Cameron (GC) model \cite{Gilbert2011}, which uses the Constant Temperature model at lower excitation energies and the Fermi Gas model at higher energies.
In EMPIRE, the default level density model is the Enhanced Generalized Superfluid Model (EGSM) \cite{Giardina2002}.
This model uses the Generalized Superfluid Model (GSM) \cite{ignatyuk1979role,Ignatyuk1993} at lower energies and the Fermi Gas model as well at higher energies, and has been normalized to discrete levels.
This normalization is performed in such a way that it only affects the level density below the neutron separation energy.
Finally, the default level density model in ALICE is the Kataria-Rarnamurthy-Kapoor (KRK) model \cite{Kataria1978,Kataria1990}, a semi-empirical nuclear level density formula which provides shell-dependent corrections to the nuclear mass surface, based on a Fourier expansion of the single particle level density of nucleons.

\begin{figure}
 \centering
 \includegraphics[width=0.5\textwidth]{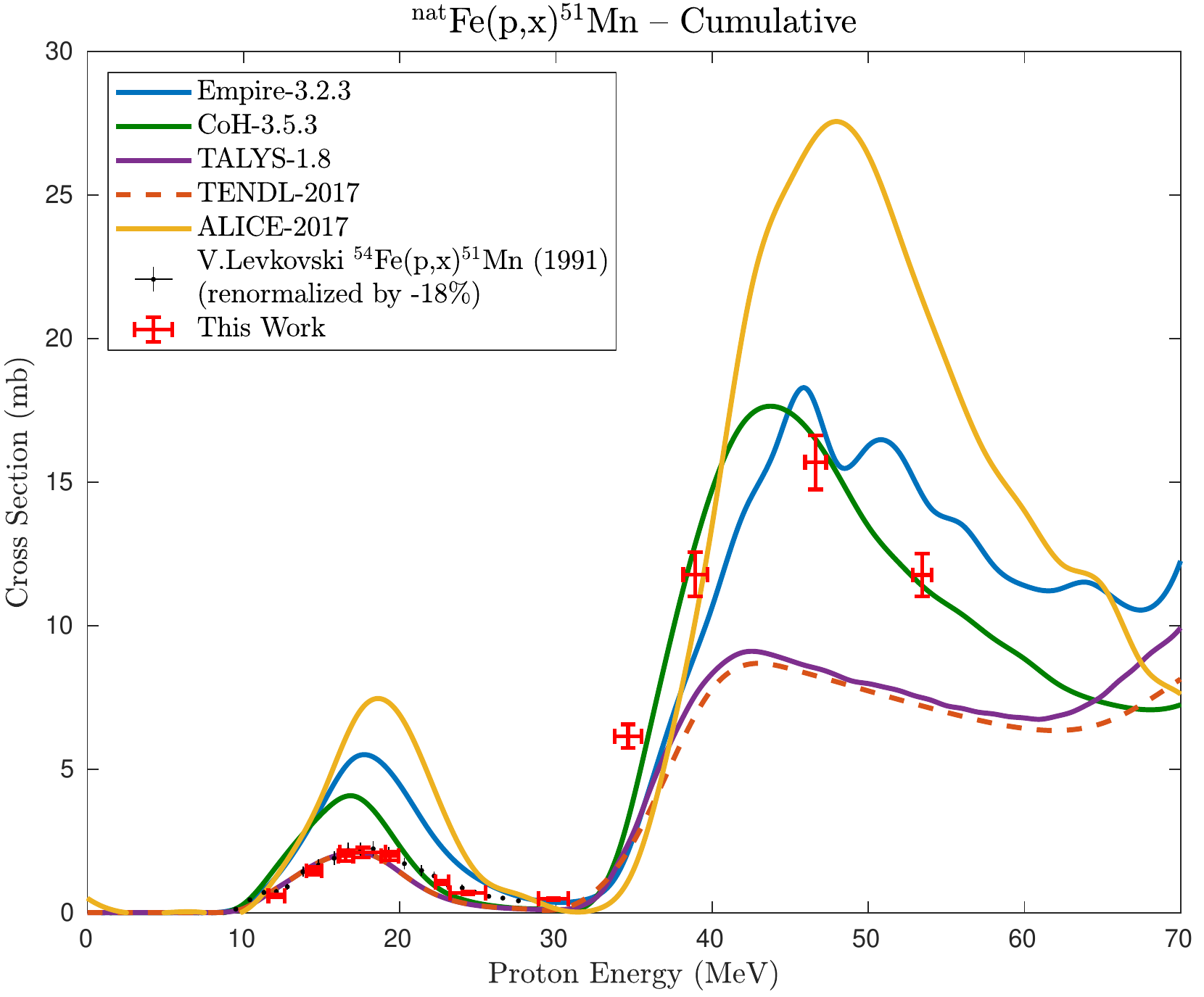}
 \caption{Measured \ce{^{nat}Fe}(p,x)\ce{^{51}Mn} cross section, with the \ce{^{54}Fe}(p,$\alpha$)\ce{^{51}Mn} reaction channel visibly peaking at approximately \mbox{15 MeV}.}
 \label{fig:51Mn}
\end{figure}

For this analysis, the focus will be on the three largest channels  measured -- \ce{^{nat}Fe}(p,x)\ce{^{51}Mn}, \ce{^{nat}Fe}(p,x)\ce{^{52m}Mn}, and \ce{^{nat}Fe}(p,x)\ce{^{52}Mn}.
The use of natural abundance targets in these measurements exacerbates the modeling  challenges seen in our previous work using monoisotopic targets\,\cite{Voyles2018a}, by propagating the difficulties inherent to modeling a wide number of reaction channels, to multiple target isotopes.
As a result, this analysis is useful for a qualitative comparison of predictive capabilities, but no firm conclusions can be drawn about the direct causes of any inaccuracies, as there are many \enquote{moving parts} in these calculations.
For the \ce{^{nat}Fe}(p,x)\ce{^{51}Mn} reactions, seen in \autoref{fig:51Mn}, the lower-energy  \ce{^{54}Fe}(p,$\alpha$)\ce{^{51}Mn} reaction, which peaks around 15\,MeV, is  well-modeled by TALYS and TENDL, and over-predicted by EMPIRE, CoH and ALICE.
The higher-energy reactions on the higher mass Fe isotopes, however, are better matched by  EMPIRE and CoH.
ALICE overpredicts the production of \ce{^{51}Mn} in both peaks, and appears to peak at too high an energy.
As was seen in the modeling of high-energy proton-induced niobium reactions\,\cite{Voyles2018a}, TALYS and TENDL do well with the lower-energy \enquote{compound} reactions but do not accurately predict the higher-energy reactions that have a significant pre-equilibrium component.
EMPIRE and CoH seem to accurately predict the locations of the peaks, but often fail to reliably estimate the magnitude of the cross section.
This is seen again in the  \ce{^{nat}Fe}(p,x)\ce{^{52}Mn} cross section, where the first peak is well-modeled by TALYS and TENDL, but  rises much more rapidly than the data would support above 50\,MeV.
CoH and EMPIRE, again, overpredict the cross sections at all energies, but seem to have the correct shape.
Through measurement of the  \ce{^{nat}Fe}(p,x)\ce{^{52m}Mn} independent cross section, the effect of spin distributions in highly-excited nuclear states can be studied.
The measured data suggest that the independent cross section to the $2^+$ isomer should be a large fraction of the cumulative cross section to \ce{^{52}Mn} (which has a $6^+$ ground state), over 80\% at the peak energy.
All three codes predict about 50\%, which indicates that the isomer feeding is not well-modeled.
It is possible that this is caused by residual nucleus population and/or level density models that are skewed too heavily towards high spin.
In the other isomer-to-ground state ratio measured for Fe,  \ce{^{nat}Fe}(p,x)\ce{^{58}Co}, the opposite is seen -- the codes underpredict the ratio of the isomer-to-ground state, but in this case the isomer is the higher spin state ($5^+$, compared with a  $2^+$ ground state).
Given that the level density model was the same for all of the product nuclei, it points to the spin distribution of the initial population of the residual nucleus as the main problem with modeling.

\begin{figure}
 \centering
 \includegraphics[width=0.5\textwidth]{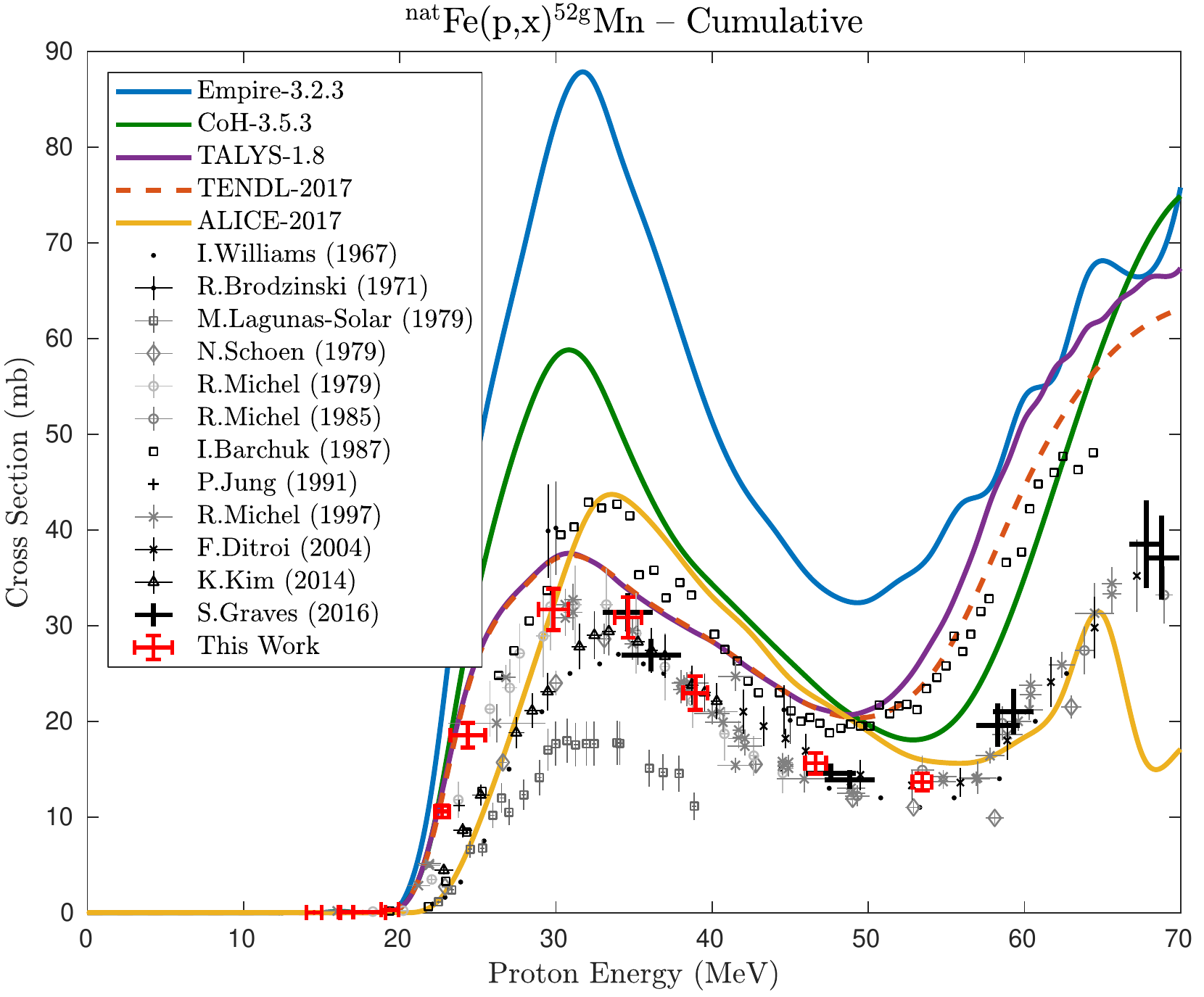}
 \caption{Measured \ce{^{nat}Fe}(p,x)\ce{^{52}Mn} cross section, with \ce{^{56}Fe}(p,$\alpha$n)\ce{^{52}Mn}/\ce{^{54}Fe}(p,2pn)\ce{^{52}Mn} reaction channels visibly peaking at approximately \mbox{30 MeV}.}
 \label{fig:temp_52Mn}
\end{figure}

\begin{figure}
 \centering
 \includegraphics[width=0.5\textwidth]{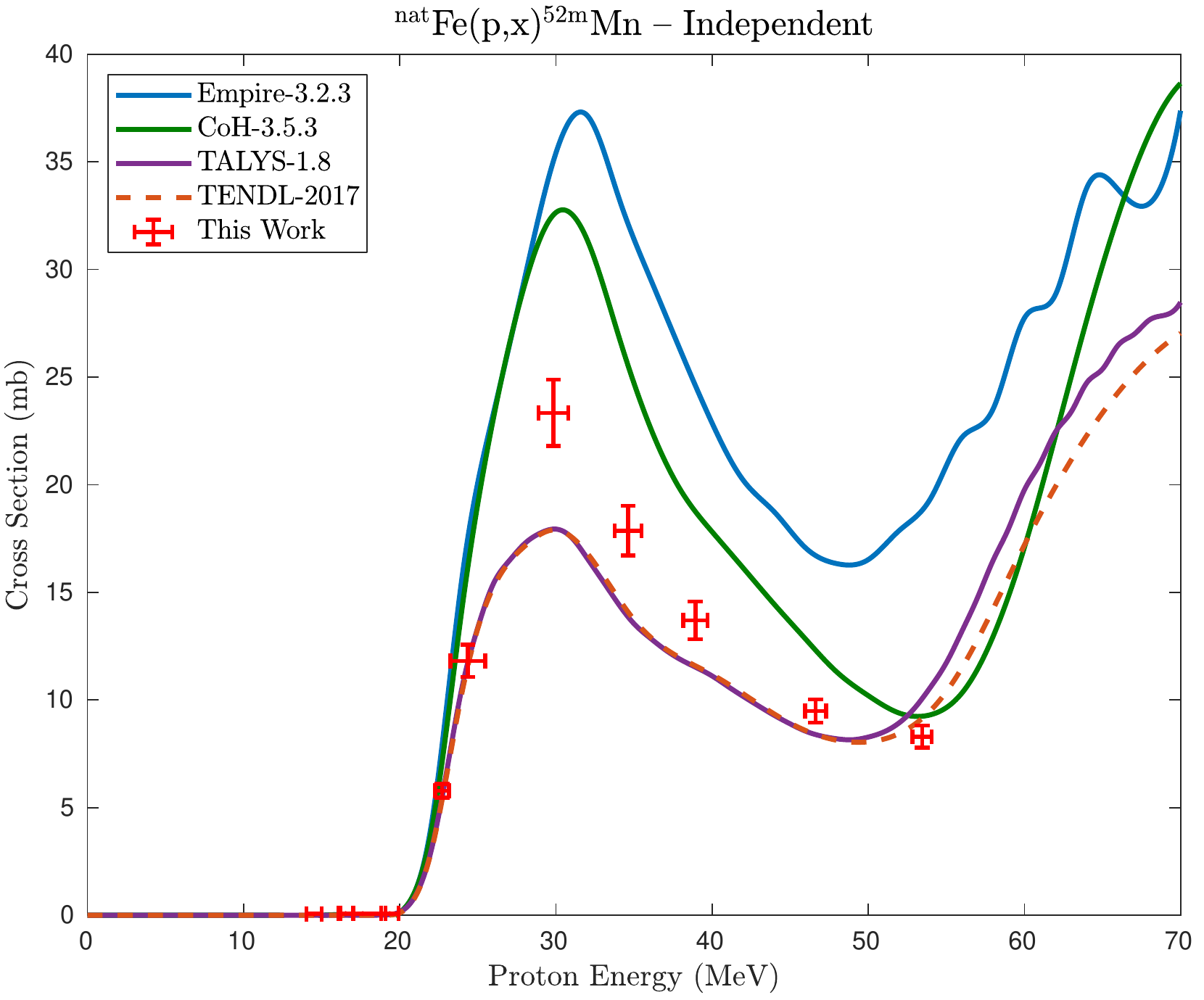}
 \caption{Measured \ce{^{nat}Fe}(p,x)\ce{^{52m}Mn} cross section, with the \ce{^{56}Fe}(p,$\alpha$n)\ce{^{52m}Mn}/\ce{^{54}Fe}(p,2pn)\ce{^{52m}Mn} reaction channels visibly peaking at approximately \mbox{30 MeV}.}
 \label{fig:temp_52mMn}
\end{figure}

%
%
%

 \section{\label{sec:conclusions_fe}Conclusions}

We present here a set of measurements of 34 cross sections for the \ce{^{nat}Fe}(p,x), \ce{^{nat}Cu}(p,x), and  \ce{^{nat}Ti}(p,x) reactions up to 55\,MeV, as well as  independent measurements of three isomer branching ratios.
Nearly all cross sections have been reported with higher precision than previous measurements.
We report the first measurements  for $\leq$70\,MeV protons  of the  \ce{^{nat}Fe}(p,x)\ce{^{49}Cr}, \ce{^{nat}Fe}(p,x)\ce{^{51}Mn}, \ce{^{nat}Fe}(p,x)\ce{^{52m}Mn},  \ce{^{nat}Fe}(p,x)\ce{^{56}Mn}, and \ce{^{nat}Fe}(p,x)\ce{^{58m}Co} reactions, as well as the first measurement of the independent cross sections for    \ce{^{nat}Fe}(p,x)\ce{^{51}Cr}, \ce{^{nat}Fe}(p,x)\ce{^{52g}Mn}, \ce{^{nat}Fe}(p,x)\ce{^{58g}Co}, and the \ce{^{52\text{m}}Mn} ($2^+$) / \ce{^{52\text{g}}Mn}  ($6^+$) and \ce{^{58\text{m}}Co} ($5^+$) / \ce{^{58\text{g}}Co}  ($2^+$)  isomer branching ratios via \ce{^{nat}Fe}(p,x).
We also use these measurements to illustrate the deficiencies in the current state of  reaction modeling up to 55\,MeV for  \ce{^{nat}Fe}(p,x), \ce{^{nat}Cu}(p,x), and  \ce{^{nat}Ti}(p,x) reactions.
Finally, this work provides another example of the current issues with modeling of  stopping power in stacked target charged particle irradiation experiments, corrected using variance minimization techniques.  

 \section{Acknowledgements}
 
 
The authors would like to particularly acknowledge the assistance and support of  Brien Ninemire, Scott Small, Tom Gimpel, and all the rest of the operations, research, and facilities staff of the LBNL 88-Inch Cyclotron.
We also wish to acknowledge Alexander  Springer and Haleema Zaneb, who participated in these experiments.
This research is supported by the U.S. Department of Energy Isotope Program, managed by the Office of Science for  Nuclear Physics.
This work has been carried out  under the auspices of the U.S. Department of Energy by  Lawrence Berkeley National Laboratory and the U.S. Nuclear Data Program under contract \# DE-AC02-05CH11231.
This research was performed under appointment to the Rickover Fellowship Program in Nuclear Engineering, sponsored by the Naval Reactors Division of the U.S. Department of Energy.
Additional support has been provided by the U.S. Nuclear Regulatory Commission.

This research used the Savio computational cluster resource provided by the Berkeley Research Computing program at the University of California, Berkeley (supported by the UC Berkeley Chancellor, Vice Chancellor for Research, and Chief Information Officer).

\appendix

\pagebreak

\section{Stack design } \label{sec:fe_stack_design}


\begin{table*}[h!]
\centering
\caption{Specifications of the 25\,MeV and 55\,MeV target stack designs in the present work. The proton beam enters the stack upstream of the SS-5 and SS-3 profile monitors, respectively, and travels through the stack in the order presented here. The 6061 aluminum degraders have a measured density of approximately 2.68 $\pm$ 0.03 g/cm$^3$. Their areal densities were determined using the variance minimization techniques described  in this work  and an earlier paper\,\cite{Voyles2018a}. 
A 316 stainless steel foil is inserted at both the front and rear of each target stack as a monitor of the beam's spatial profile, by developing radiochromic film (Gafchromic EBT3) after end-of-bombardment (EoB).}
\label{tab:fe_stack_table}
\small
\resizebox{\textwidth}{!}{%
\begin{tabular}{@{}llll|llll@{}}
\toprule\toprule
25\,MeV Target layer            & \begin{tabular}[c]{@{}l@{}}Measured \\ thickness\end{tabular} & \begin{tabular}[c]{@{}l@{}}Measured\\areal density\\(mg/cm$^2$)\end{tabular} & \begin{tabular}[c]{@{}l@{}}Uncertainty \\ in areal\\ density  (\%)\end{tabular} & 55\,MeV Target layer            & \begin{tabular}[c]{@{}l@{}}Measured \\ thickness\end{tabular} & \begin{tabular}[c]{@{}l@{}}Measured\\areal density\\(mg/cm$^2$)\end{tabular} & \begin{tabular}[c]{@{}l@{}}Uncertainty \\ in areal\\ density  (\%)\end{tabular} \\
\midrule
SS profile monitor SS-5 & 130.94 \mmicro m                                              & 100.57                                                                      & 0.17                                                                      & SS profile monitor SS-3 & 130.9 \mmicro m                                               & 100.48                                                                      & 0.17                                                                      \\
Fe-08                   & 26.25 \mmicro m                                               & 19.69                                                                       & 0.17                                                                      & Fe-01                   & 25.75 \mmicro m                                               & 20.22                                                                       & 0.21                                                                      \\
Ti-14                   & 25.01 \mmicro m                                               & 10.87                                                                       & 0.36                                                                      & Ti-01                   & 25.88 \mmicro m                                               & 11.09                                                                       & 0.16                                                                      \\
Cu-14                   & 24.01 \mmicro m                                               & 17.49                                                                       & 0.40                                                                      & Cu-01                   & 28.81 \mmicro m                                               & 22.40                                                                       & 0.11                                                                      \\
Al Degrader E-09        & 256.5 \mmicro m                                               & --                                                                          & --                                                                        & Al Degrader A-1         & 2.24 mm                                                       & --                                                                          & --                                                                        \\
Fe-09                   & 26.5 \mmicro m                                                & 19.90                                                                       & 0.09                                                                      & Fe-02                   & 25.5 \mmicro m                                                & 19.91                                                                       & 0.13                                                                      \\
Ti-15                   & 23.81 \mmicro m                                               & 10.97                                                                       & 0.11                                                                      & Ti-02                   & 25.74 \mmicro m                                               & 10.94                                                                       & 0.24                                                                      \\
Cu-15                   & 21.81 \mmicro m                                               & 17.63                                                                       & 0.46                                                                      & Cu-02                   & 28.75 \mmicro m                                               & 22.32                                                                       & 0.40                                                                      \\
Al Degrader H-01        & 127.09 \mmicro m                                              & --                                                                          & --                                                                        & Al Degrader A-2         & 2.24 mm                                                       & --                                                                          & --                                                                        \\
Fe-10                   & 26.5 \mmicro m                                                & 19.84                                                                       & 0.11                                                                      & Fe-03                   & 25.25 \mmicro m                                               & 20.00                                                                       & 0.27                                                                      \\
Ti-16                   & 24.6 \mmicro m                                                & 10.96                                                                       & 0.32                                                                      & Ti-03                   & 25.91 \mmicro m                                               & 11.25                                                                       & 0.15                                                                      \\
Cu-16                   & 22.01 \mmicro m                                               & 17.22                                                                       & 0.25                                                                      & Cu-03                   & 28.86 \mmicro m                                               & 22.49                                                                       & 0.20                                                                      \\
Fe-11                   & 27.26 \mmicro m                                               & 19.96                                                                       & 0.17                                                                      & Al Degrader C-1         & 0.97 mm                                                       & --                                                                          & --                                                                        \\
Ti-17                   & 25.01 \mmicro m                                               & 10.88                                                                       & 0.25                                                                      & Fe-04                   & 25.25 \mmicro m                                               & 19.93                                                                       & 0.33                                                                      \\
Cu-17                   & 29 \mmicro m                                                  & 21.91                                                                       & 0.33                                                                      & Ti-04                   & 25.84 \mmicro m                                               & 10.91                                                                       & 0.18                                                                      \\
Fe-12                   & 27.01 \mmicro m                                               & 20.03                                                                       & 0.12                                                                      & Cu-04                   & 28.78 \mmicro m                                               & 22.38                                                                       & 0.29                                                                      \\
Ti-18                   & 25.01 \mmicro m                                               & 11.00                                                                       & 0.87                                                                      & Al Degrader C-2         & 0.97 mm                                                       & --                                                                          & --                                                                        \\
Cu-18                   & 28.75 \mmicro m                                               & 22.33                                                                       & 0.14                                                                      & Fe-05                   & 25.64 \mmicro m                                               & 20.02                                                                       & 0.24                                                                      \\
Fe-13                  & 26.25 \mmicro m                                               & 20.05                                                                       & 0.16                                                                      & Ti-05                   & 25.86 \mmicro m                                               & 10.99                                                                       & 0.30                                                                      \\
Ti-19                   & 26.6 \mmicro m                                                & 11.01                                                                       & 0.22                                                                      & Cu-05                   & 28.77 \mmicro m                                               & 22.35                                                                       & 0.12                                                                      \\
Cu-19                   & 28.75 \mmicro m                                               & 22.32                                                                       & 0.19                                                                      & Al Degrader C-3         & 0.97 mm                                                       & --                                                                          & --                                                                        \\
Fe-14                   & 25.75 \mmicro m                                               & 20.11                                                                       & 0.19                                                                      & Fe-06                   & 25.75 \mmicro m                                               & 20.21                                                                       & 0.26                                                                      \\
Ti-20                   & 27.01 \mmicro m                                               & 11.06                                                                       & 0.35                                                                      & Ti-06                   & 25.5 \mmicro m                                                & 11.15                                                                       & 0.23                                                                      \\
Cu-20                   & 28.26 \mmicro m                                               & 22.34                                                                       & 0.28                                                                      & Cu-06                   & 28.83 \mmicro m                                               & 22.43                                                                       & 0.10                                                                      \\
SS profile monitor SS-6 & 131.5 \mmicro m                                               & 100.99                                                                      & 0.17                                                                      & Al Degrader C-4         & 0.97 mm                                                       & --                                                                          & --                                                                        \\
                        &                                                               &                                                                             &                                                                           & Fe-07                   & 25.76 \mmicro m                                               & 19.93                                                                       & 0.19                                                                      \\
                        &                                                               &                                                                             &                                                                           & Ti-07                   & 25.75 \mmicro m                                               & 11.17                                                                       & 0.33                                                                      \\
                        &                                                               &                                                                             &                                                                           & Cu-07                   & 28.76 \mmicro m                                               & 22.34                                                                       & 0.24                                                                      \\
                        &                                                               &                                                                             &                                                                           & Al Degrader H-02        & 127.04 \mmicro m                                              & --\cmmnt{34.51}                                                                       & --\cmmnt{0.20}                                                                      \\
                        &                                                               &                                                                             &                                                                           & SS profile monitor SS-4 & 131.21 \mmicro m                                              & 101.25                                                                      & 0.16                                                                     \\
 \bottomrule\bottomrule
\end{tabular}
}
\end{table*}

\section{Measured excitation functions} \label{sec:fe_xs_figures}

Figures of the cross sections measured in this work are presented here, in comparison with literature data
\cite{Al-Abyad2009,A2006,Aleksandrov1987,barchuk1987excitation,Barrandon1975,Belhout2007,Brodzinski1971,Brodzinski1971a,daum1997investigation,Ditroi2005,Fink1990,Garrido2016,Graves2016,Greenwood1984,Grutter1982,Jung1987,Khandaker2009,Kim2014,Kopecky1993,Lagunas-Solar1979a,levkovski1991cross,MICHEL1997153,MICHEL1979a,Michel1980,Michel1978,Michel1985,Mills1992,Neumann1999a,Schoen1979a,Shahid2015,Sudar1994,Takacs1994a,Voyles2018a,PhysRev.162.1055,YashimaH2003,Zarie2006a,zhao1993measurement}.



\begin{figure*}
    \sloppy
    \centering
    \subfloat{
        \centering
        \subfigimg[width=0.496\textwidth]{}{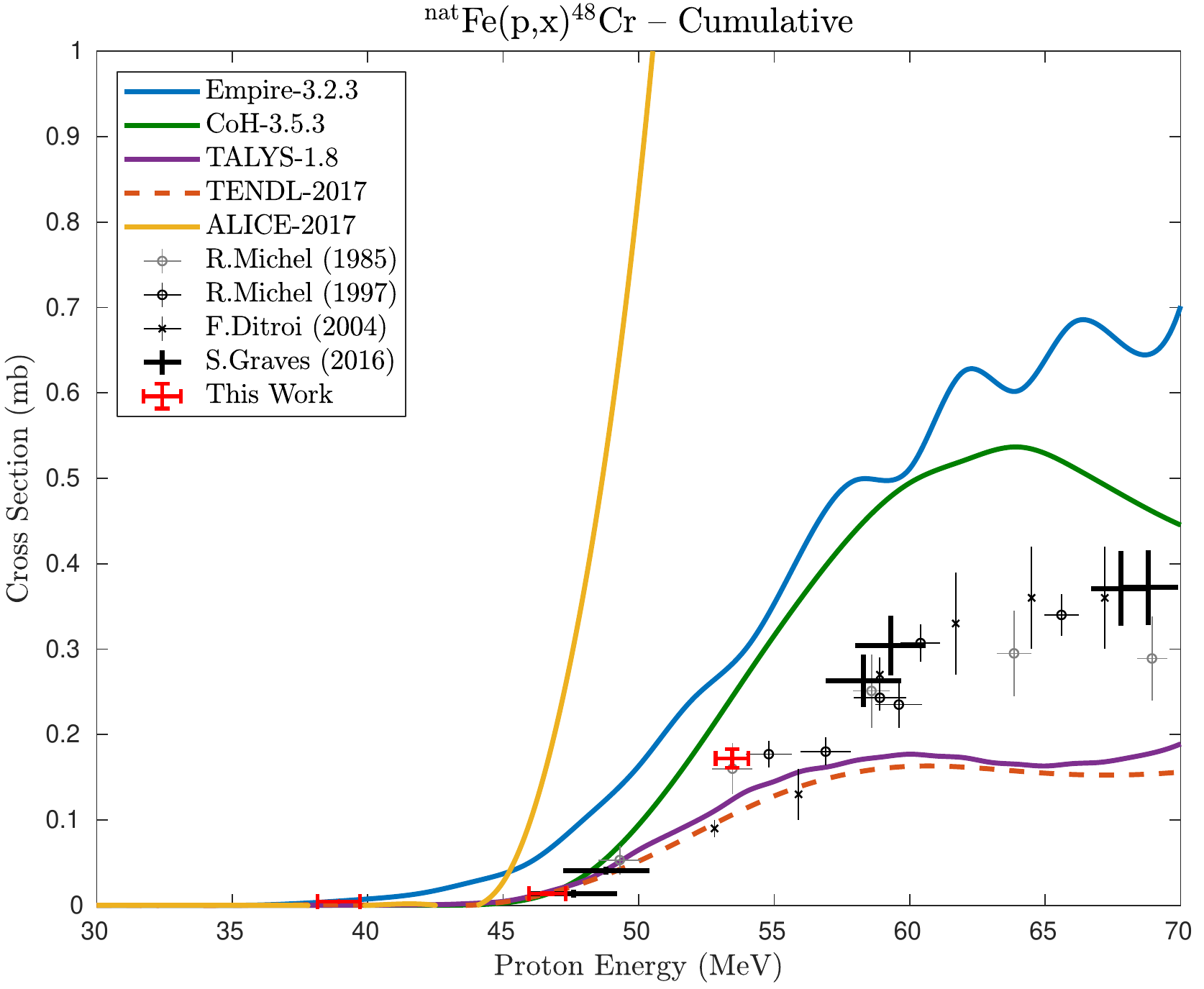}{50}
        \subfigimg[width=0.496\textwidth]{}{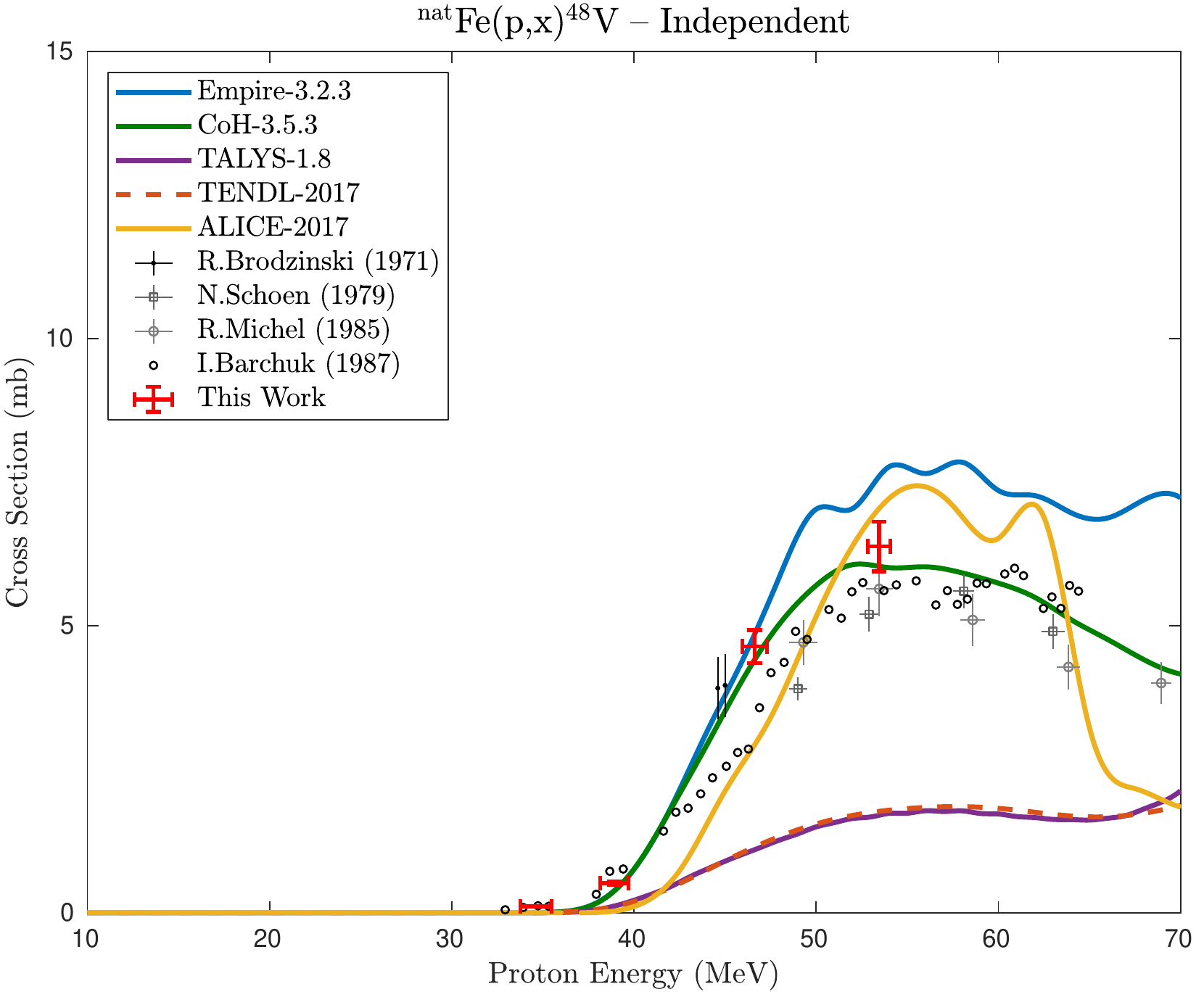}{50}
   \hspace{-10pt}}%
    \\
    \subfloat{
        \centering
        \subfigimg[width=0.496\textwidth]{}{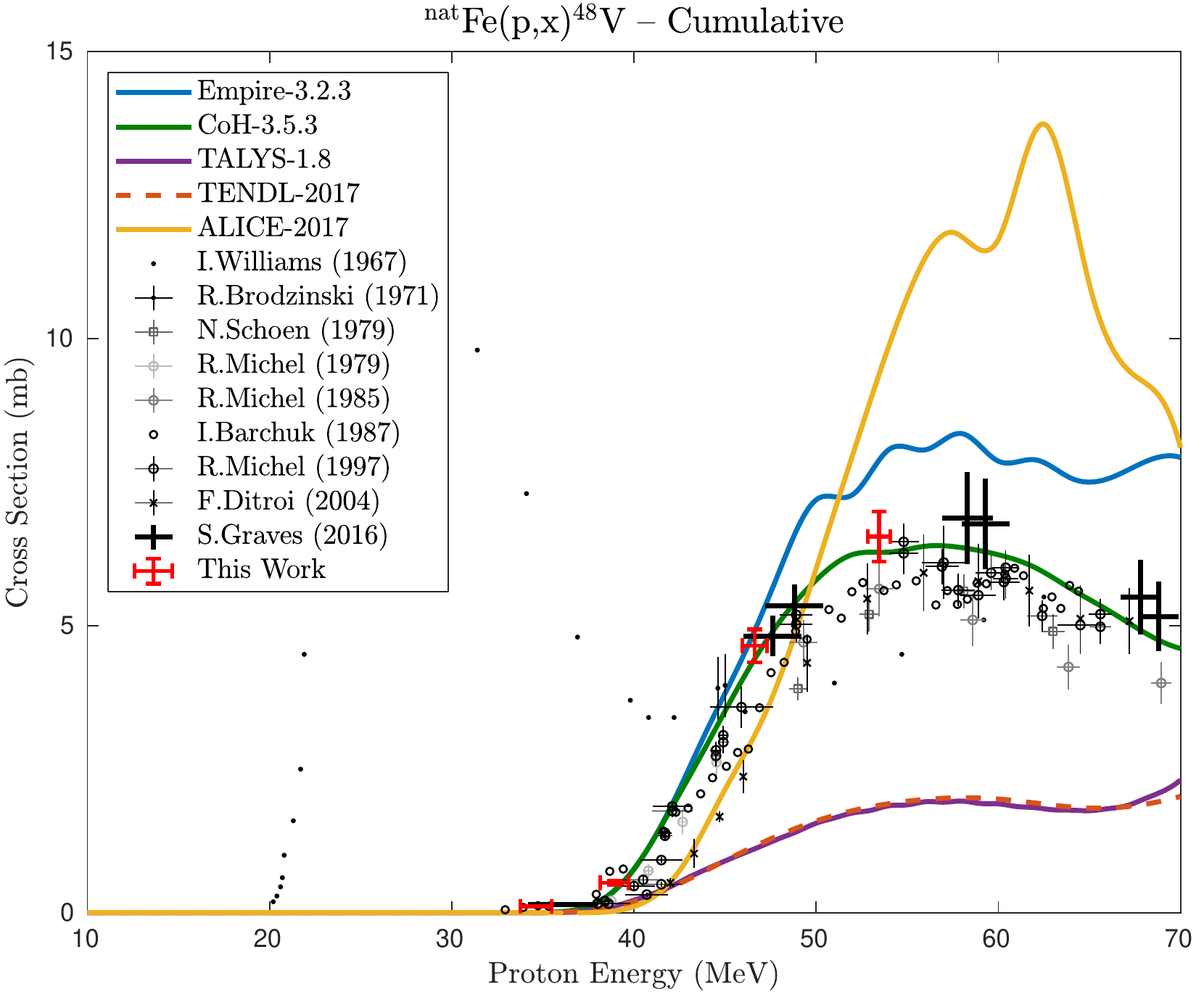}{50}
        \subfigimg[width=0.496\textwidth]{}{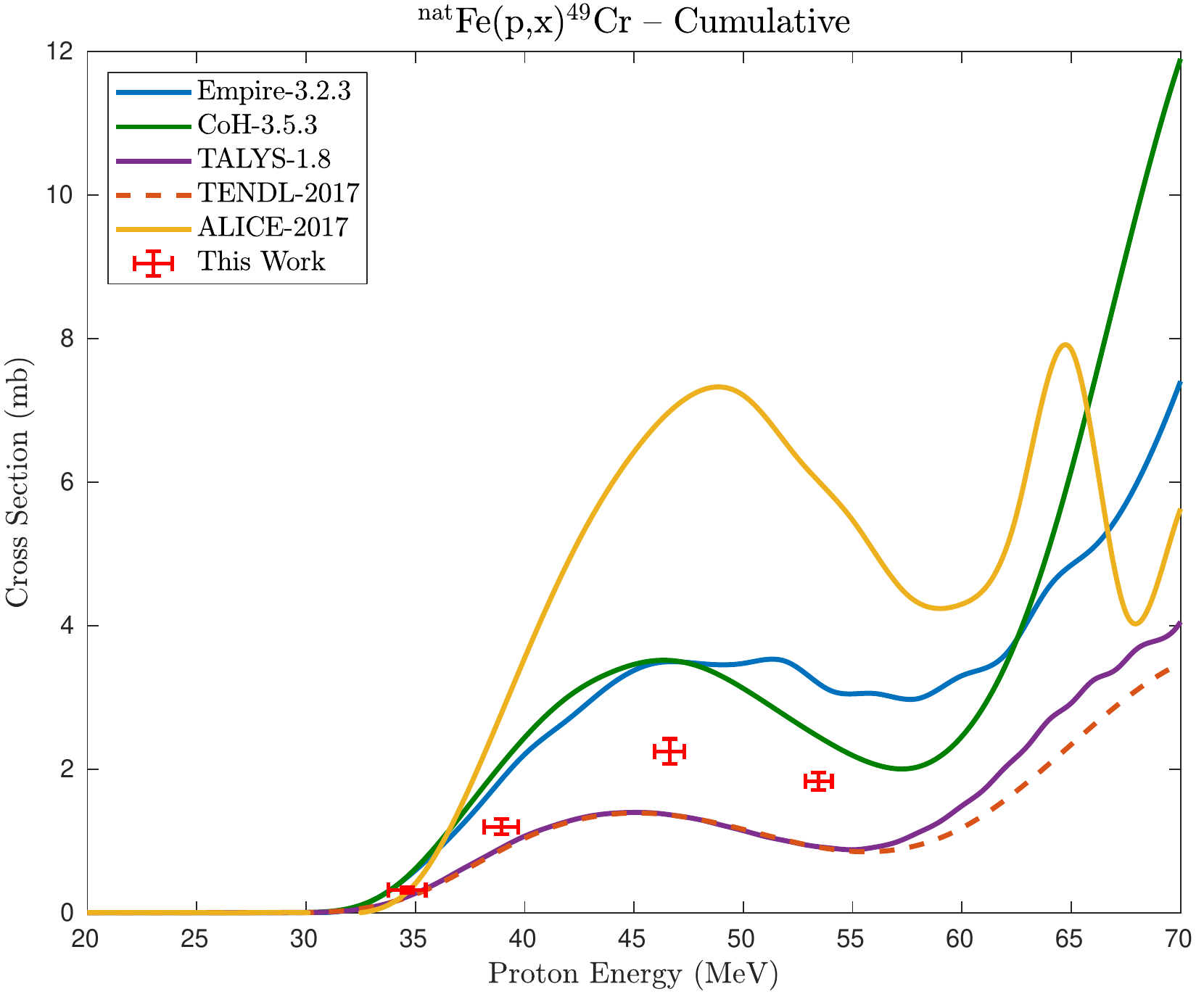}{50}
   \hspace{-10pt}}%
    \\
    \subfloat{
        \centering
        \subfigimg[width=0.496\textwidth]{}{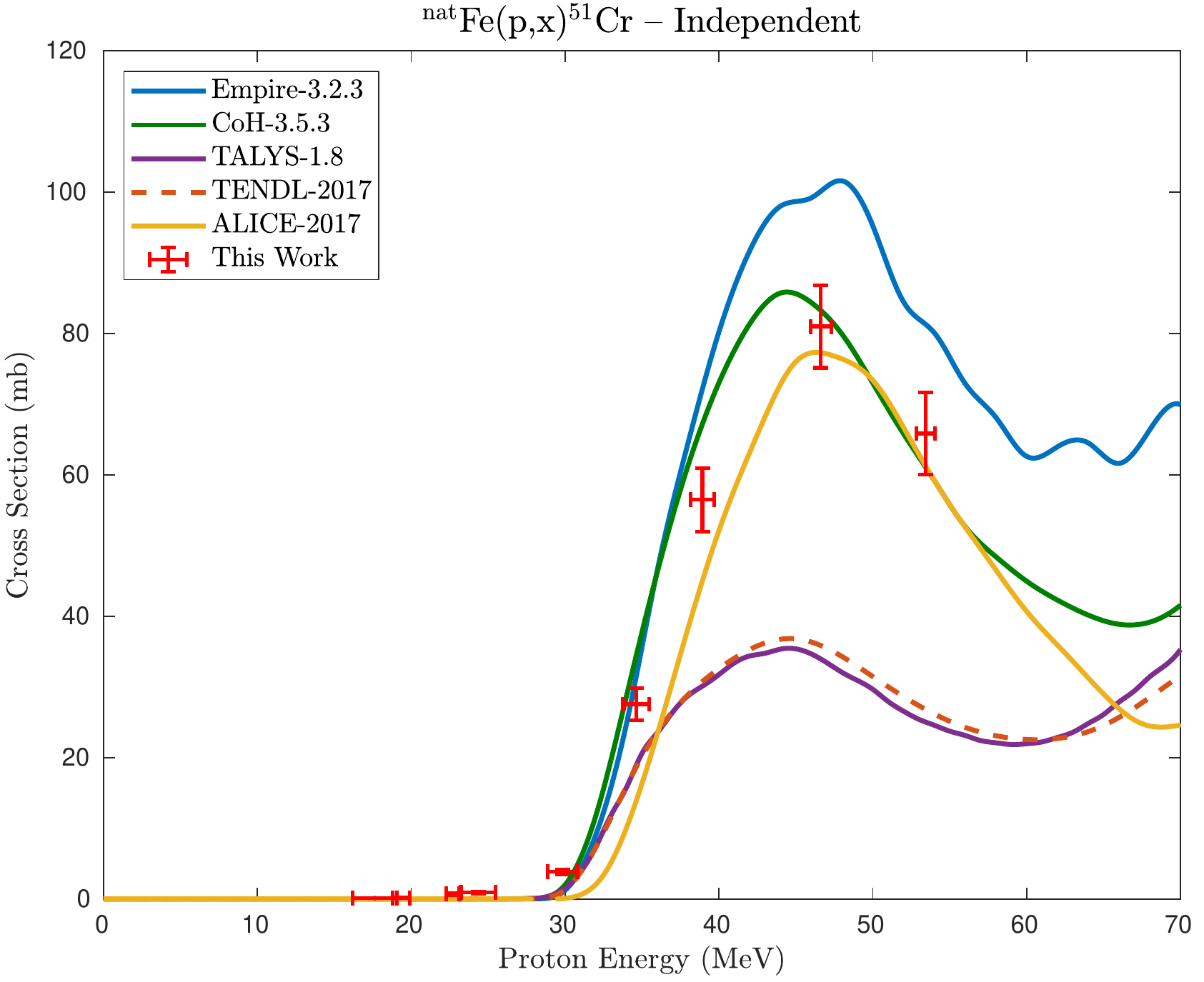}{50}
%
        \subfigimg[width=0.496\textwidth]{}{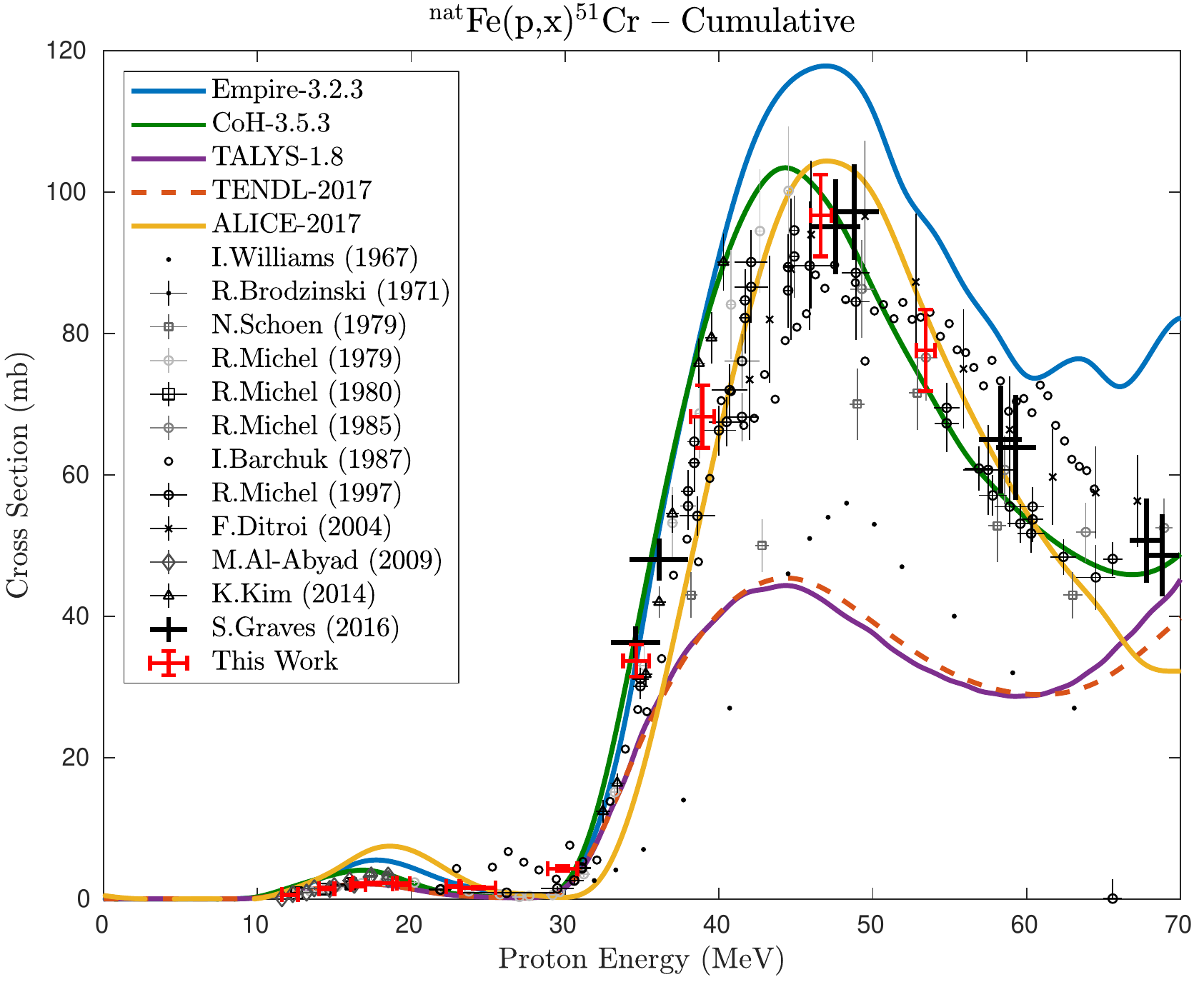}{50}
   \hspace{-10pt}}%
     \phantomcaption{}\label{fig:xs_curves_p1}
\end{figure*}

\begin{figure*}
    \centering
    \subfloat{
        \centering
        \subfigimg[width=0.496\textwidth]{}{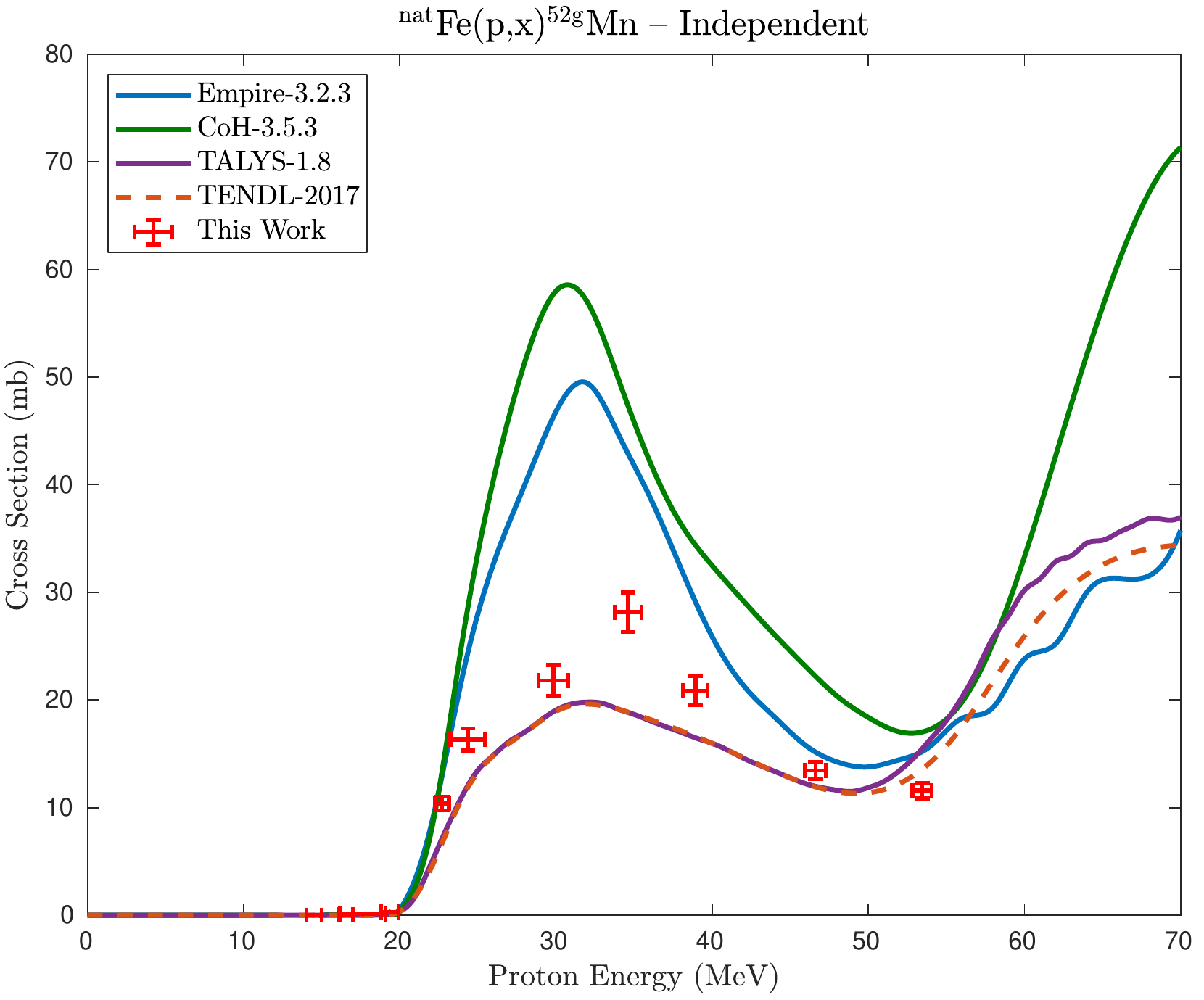}{50}
%
        \subfigimg[width=0.496\textwidth]{}{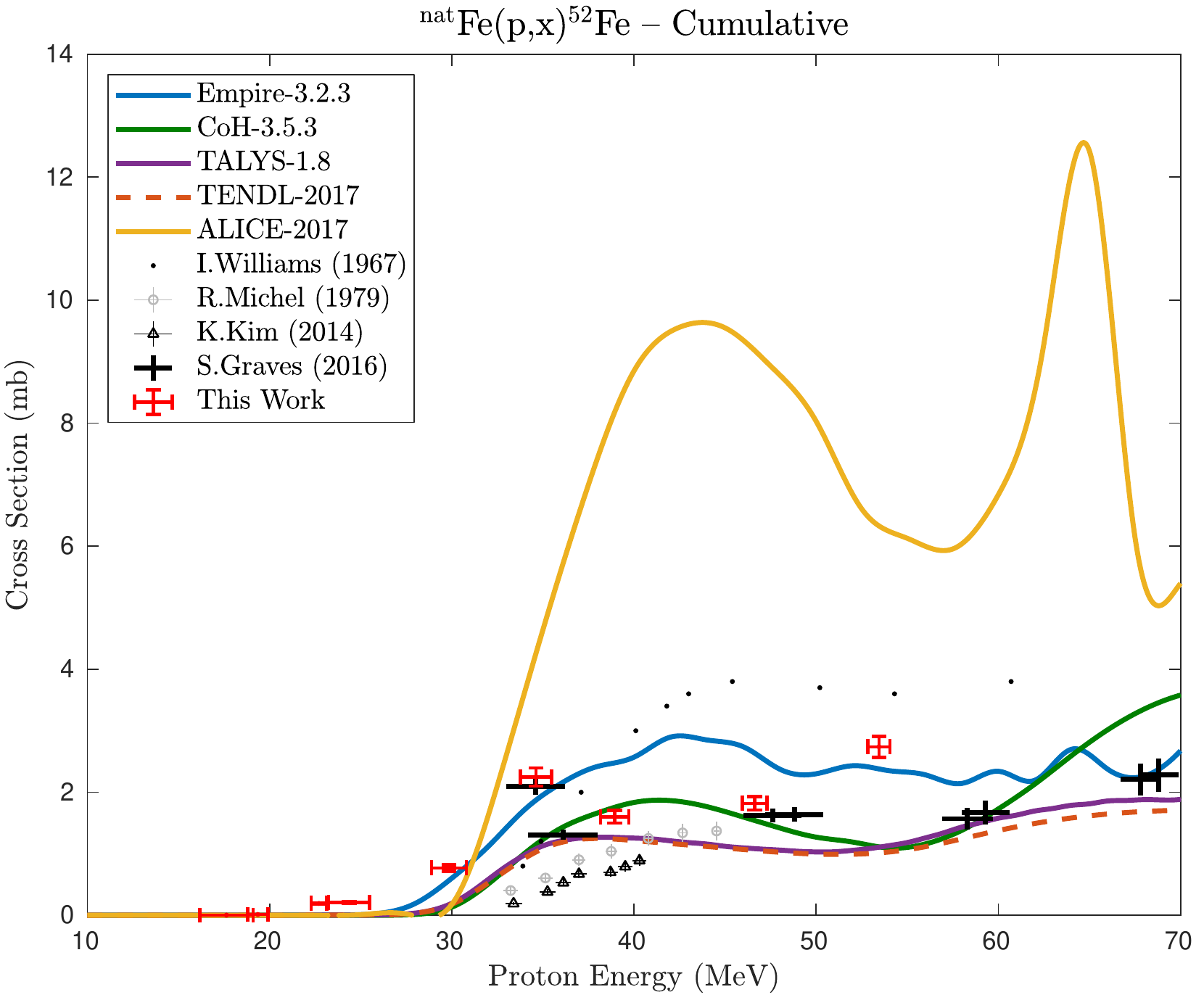}{50}
   \hspace{-10pt}}%
    \\
    \subfloat{
        \centering
        \subfigimg[width=0.496\textwidth]{}{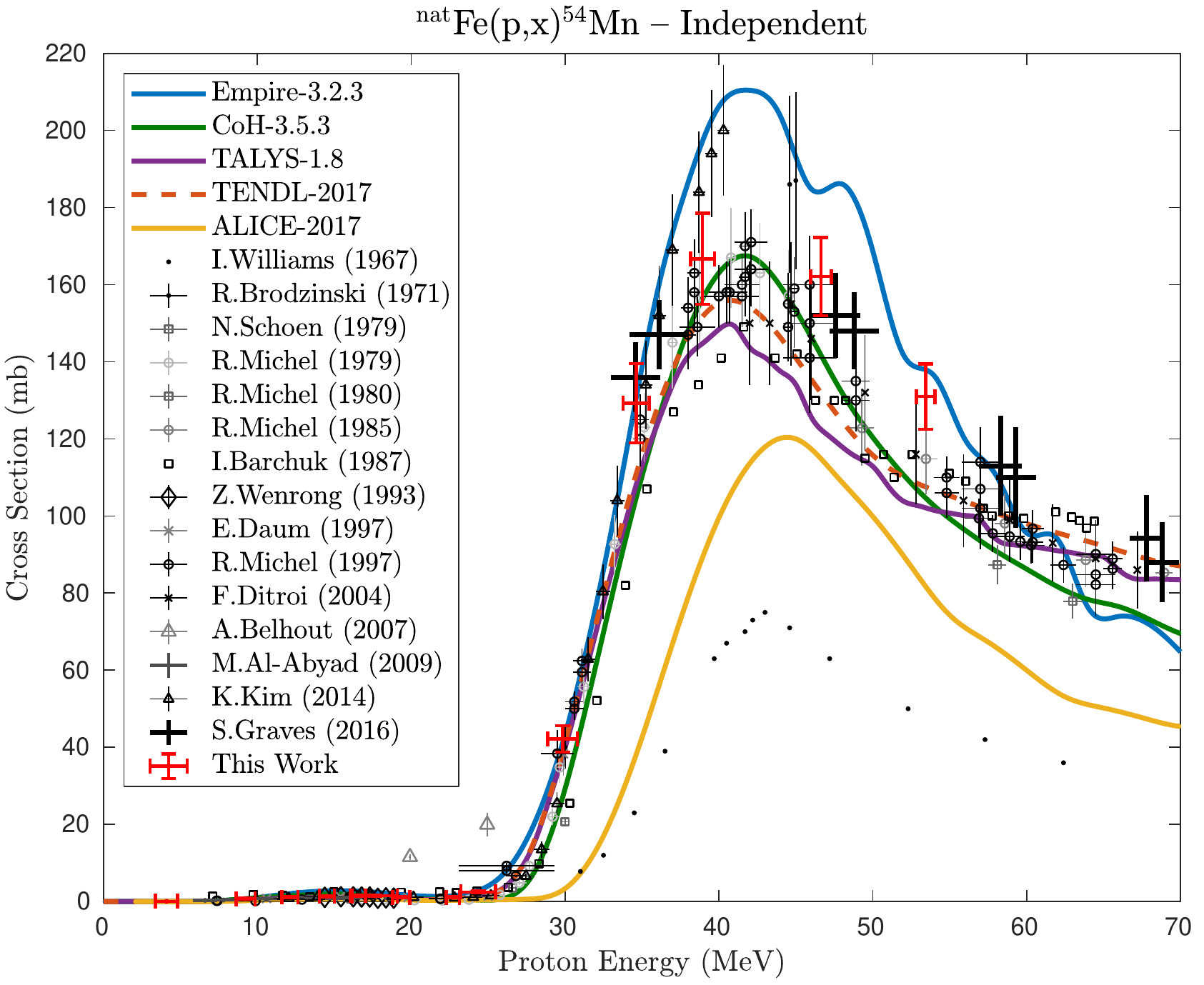}{50}
        \subfigimg[width=0.496\textwidth]{}{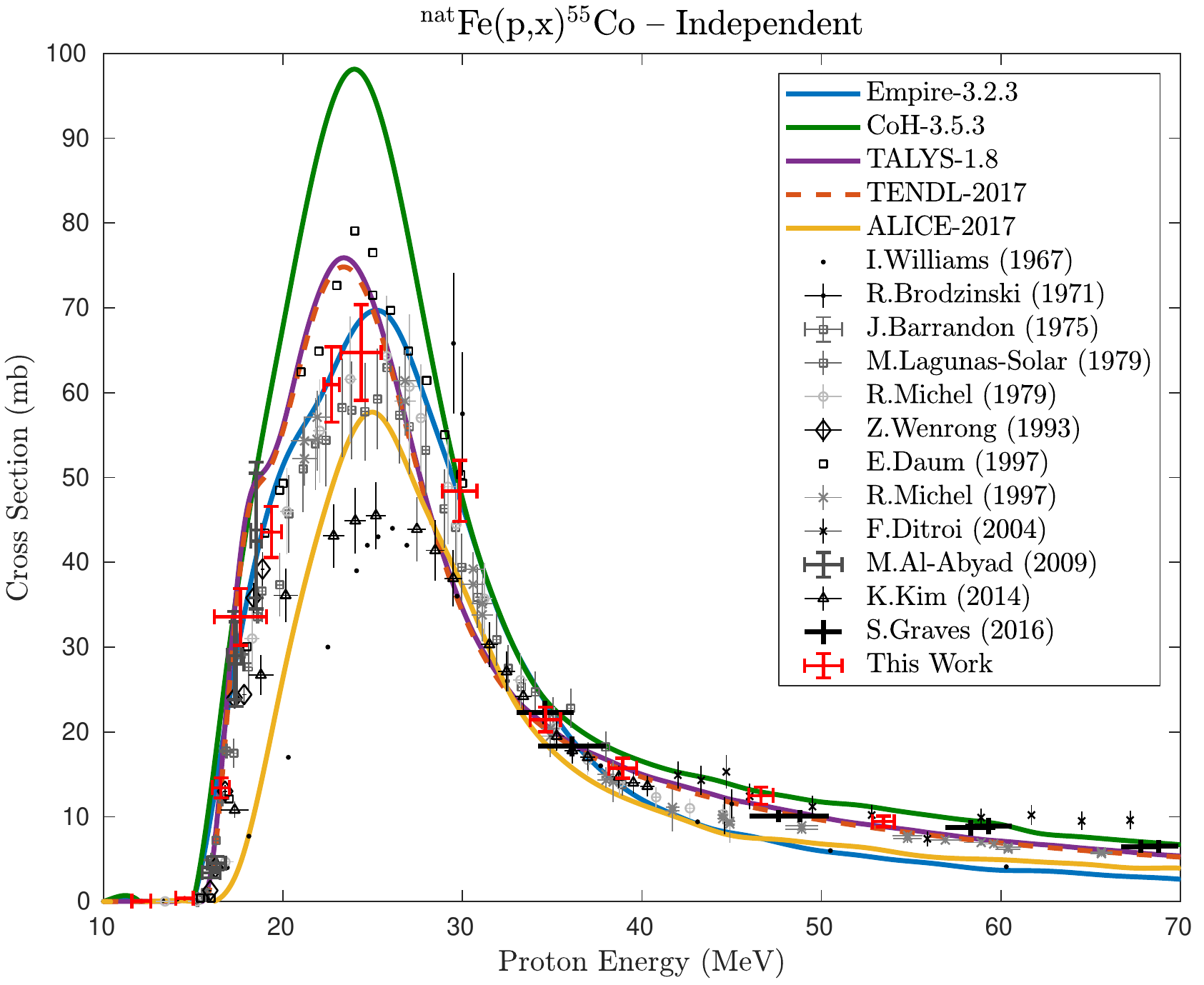}{50}
   \hspace{-10pt}}%
    \\
    \subfloat{
        \centering
        \subfigimg[width=0.496\textwidth]{}{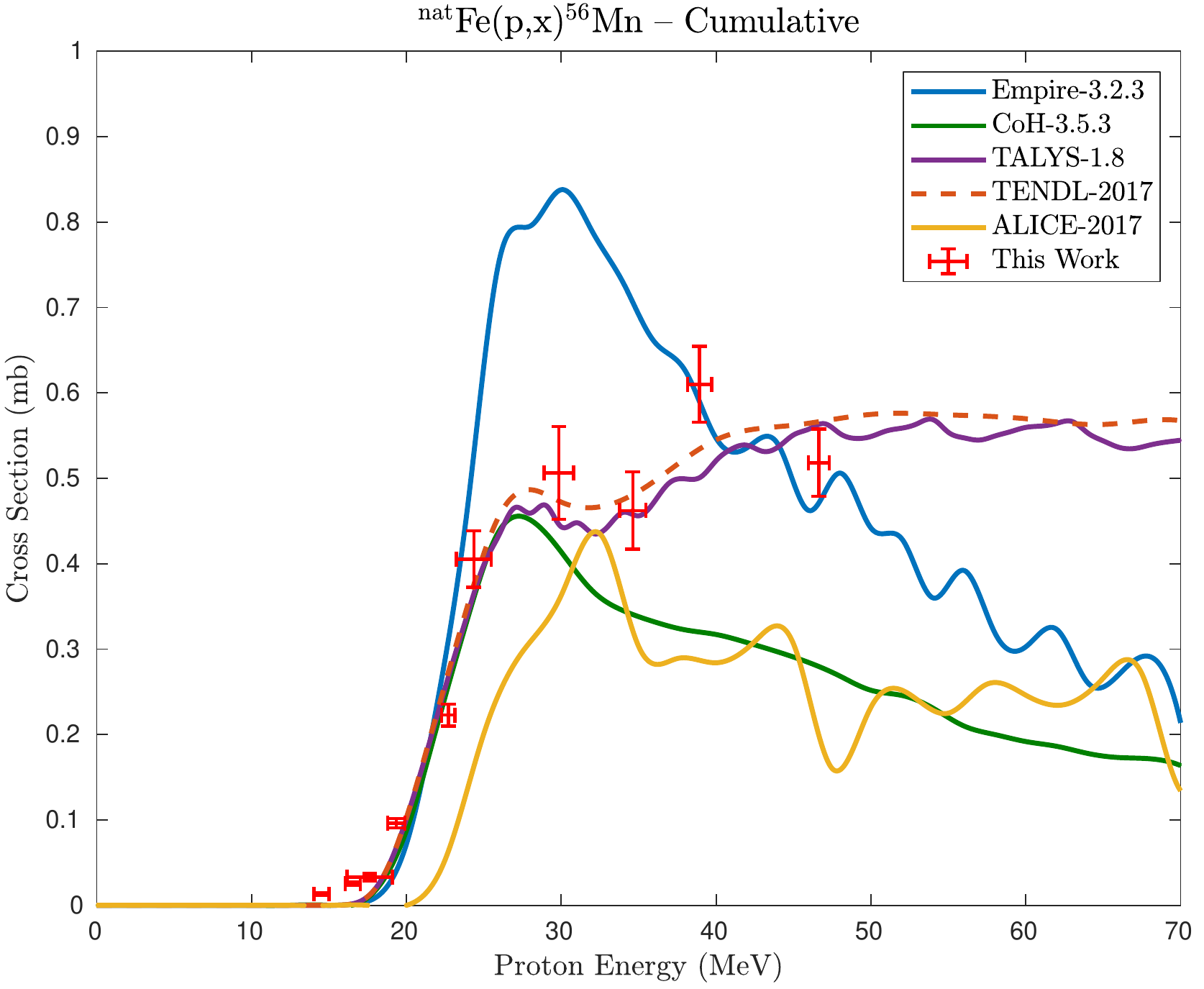}{50}
%
        \subfigimg[width=0.496\textwidth]{}{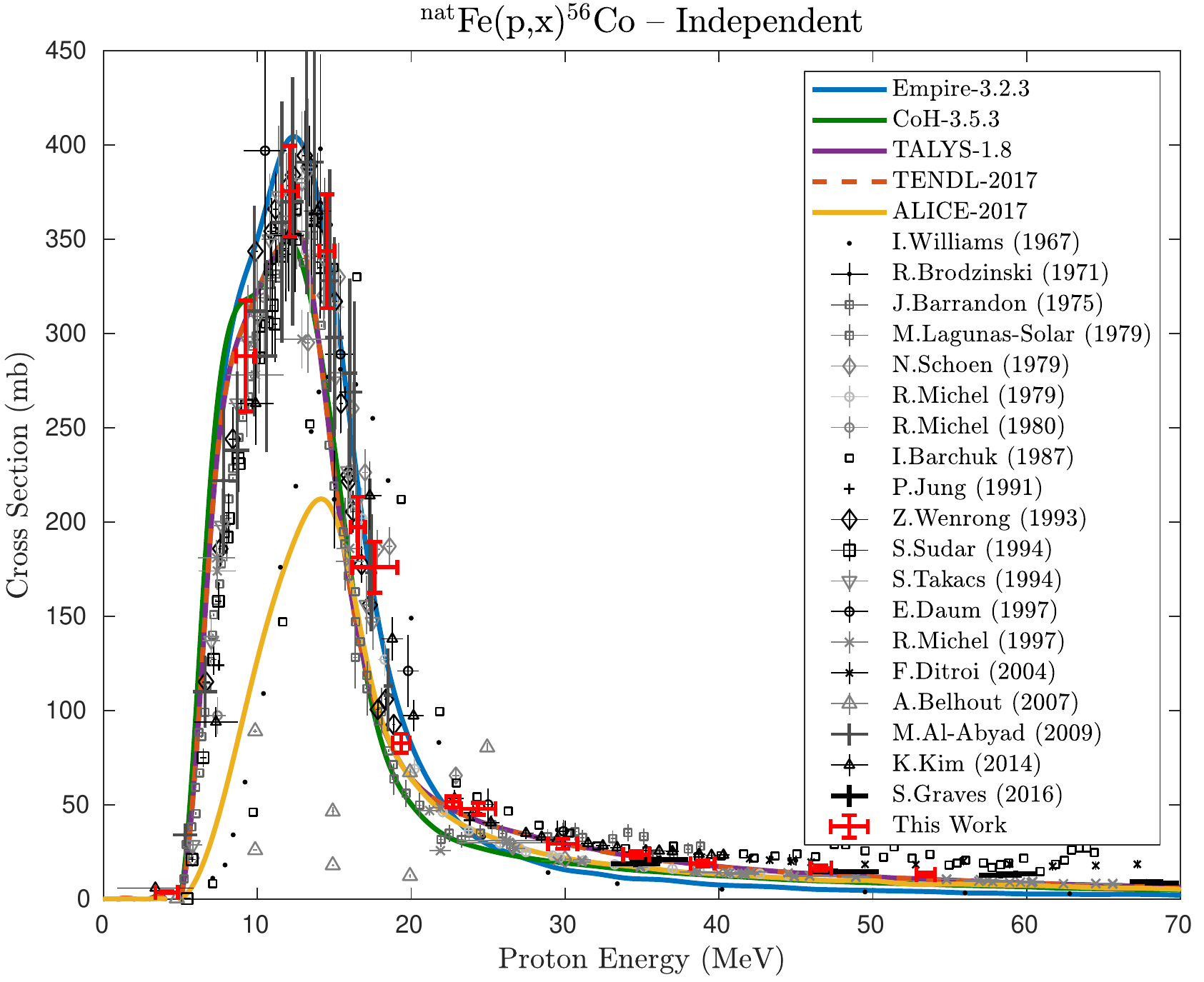}{50}
   \hspace{-10pt}}%
     \phantomcaption{}\label{fig:xs_curves_p2}
\end{figure*}

\begin{figure*}
    \centering
    \subfloat{
        \centering
        \subfigimg[width=0.496\textwidth]{}{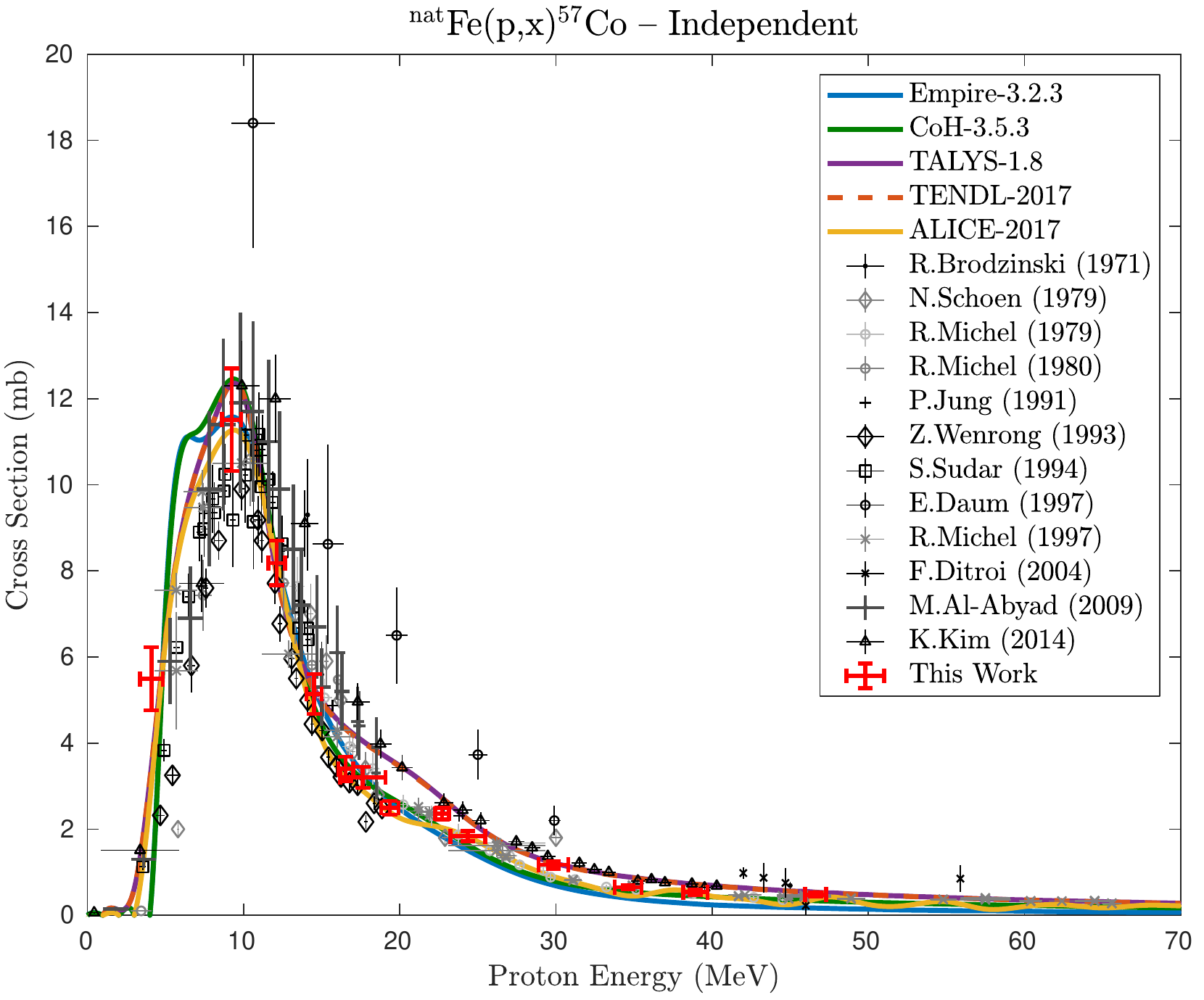}{50}
%
        \subfigimg[width=0.496\textwidth]{}{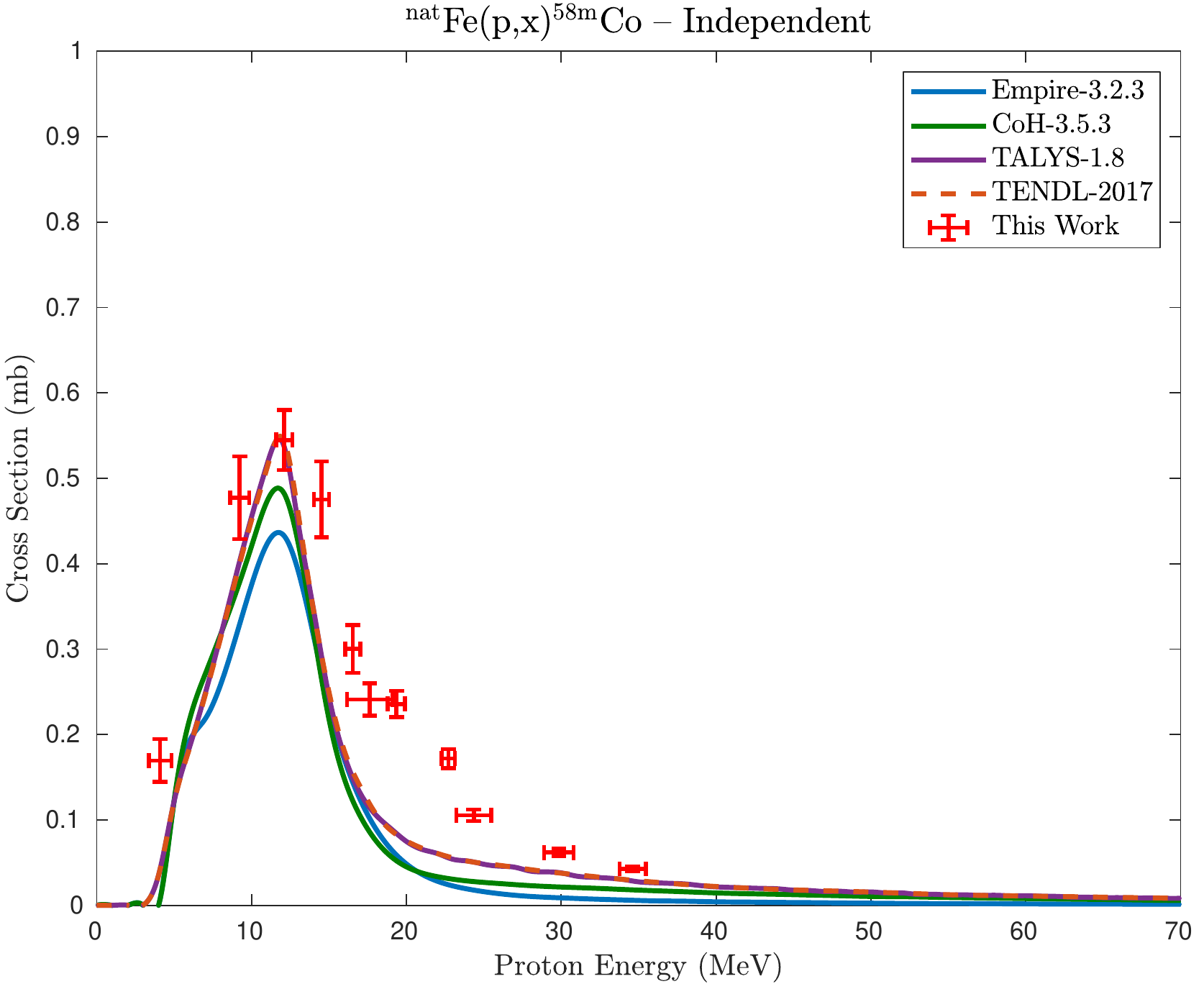}{50}
   \hspace{-10pt}}%
    \\
    \subfloat{
        \centering
        \subfigimg[width=0.496\textwidth]{}{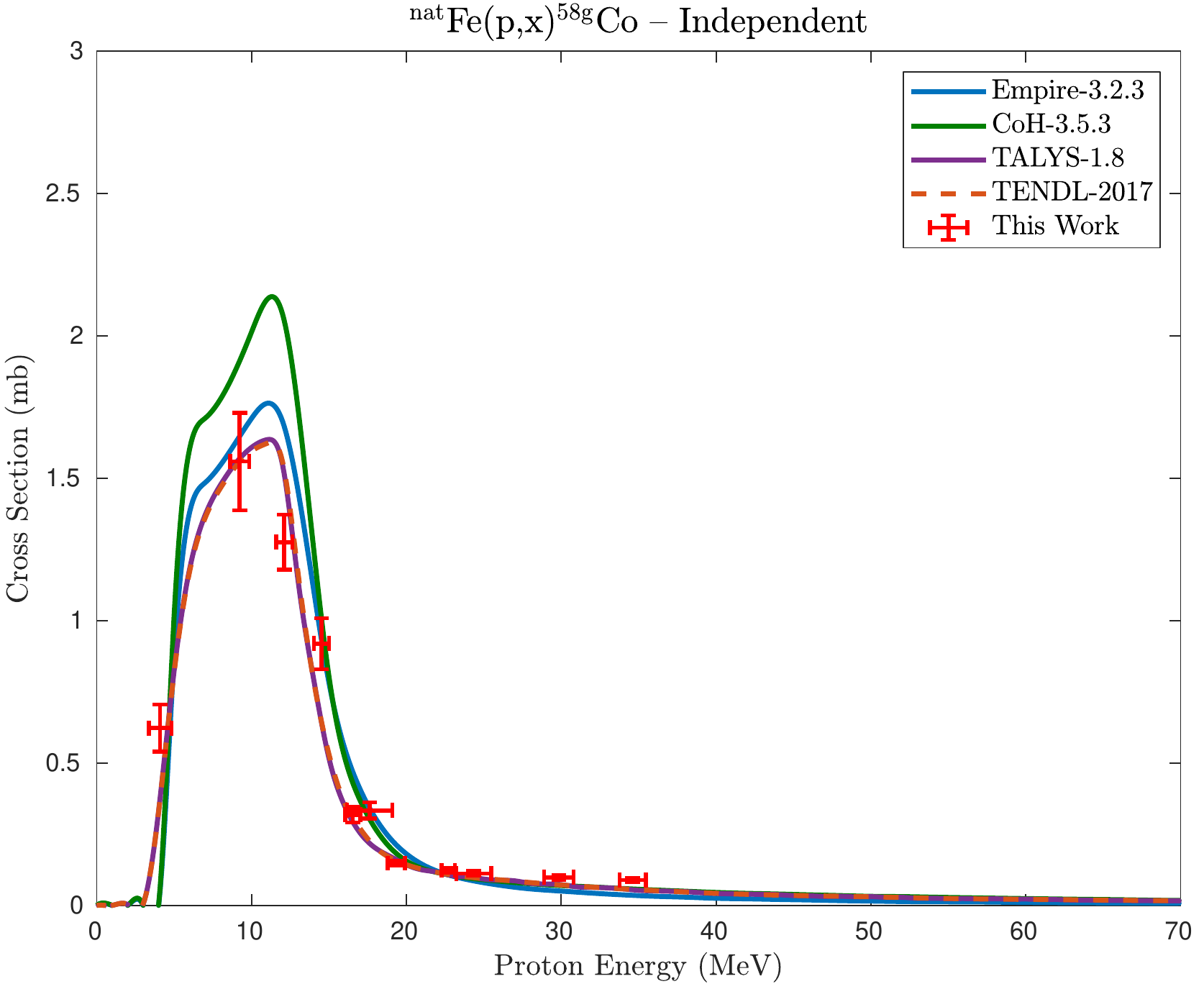}{50}
%
        \subfigimg[width=0.496\textwidth]{}{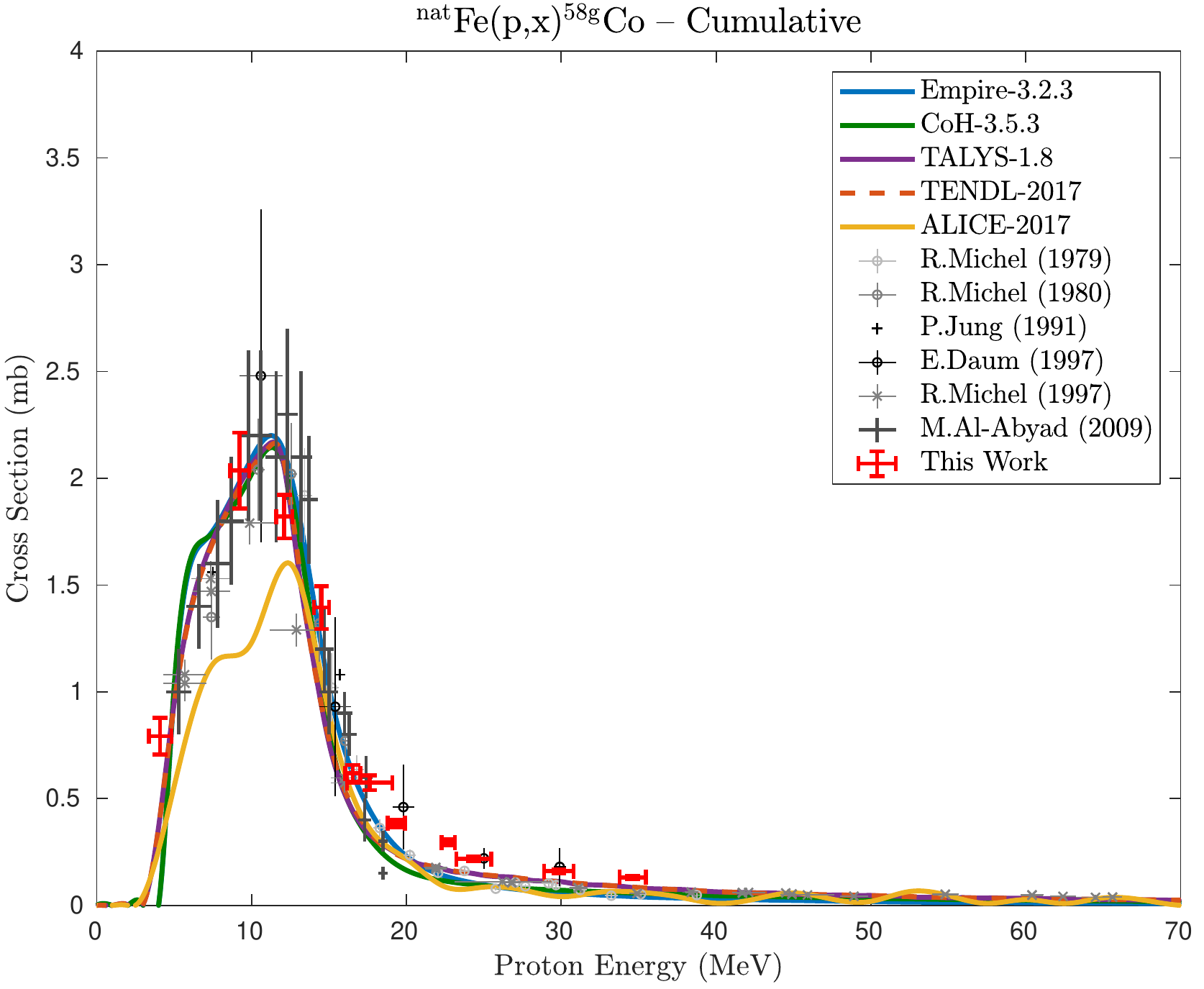}{50}
   \hspace{-10pt}}%
    \\
    \subfloat{
        \centering
        \subfigimg[width=0.496\textwidth]{}{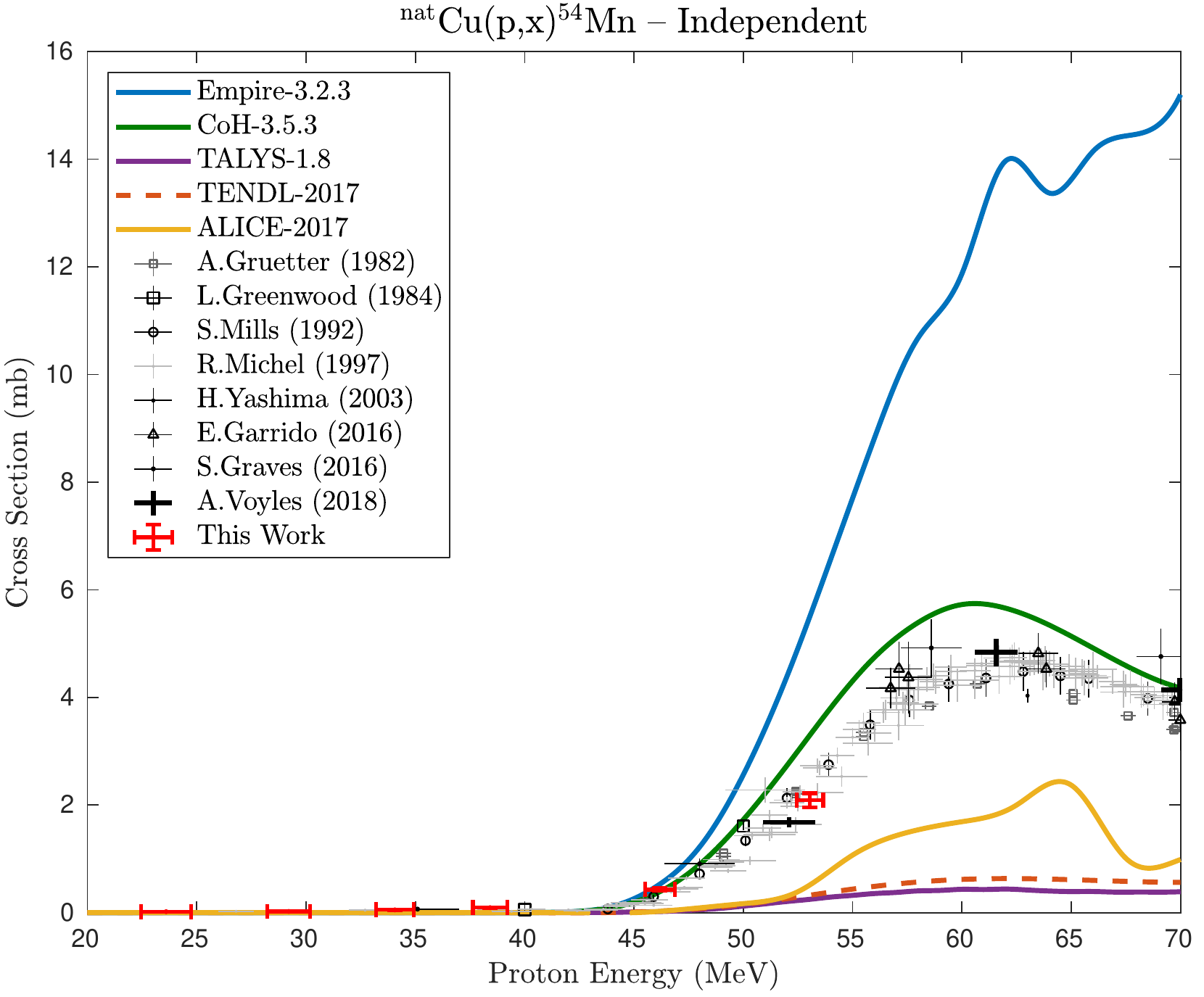}{50}
        \subfigimg[width=0.496\textwidth]{}{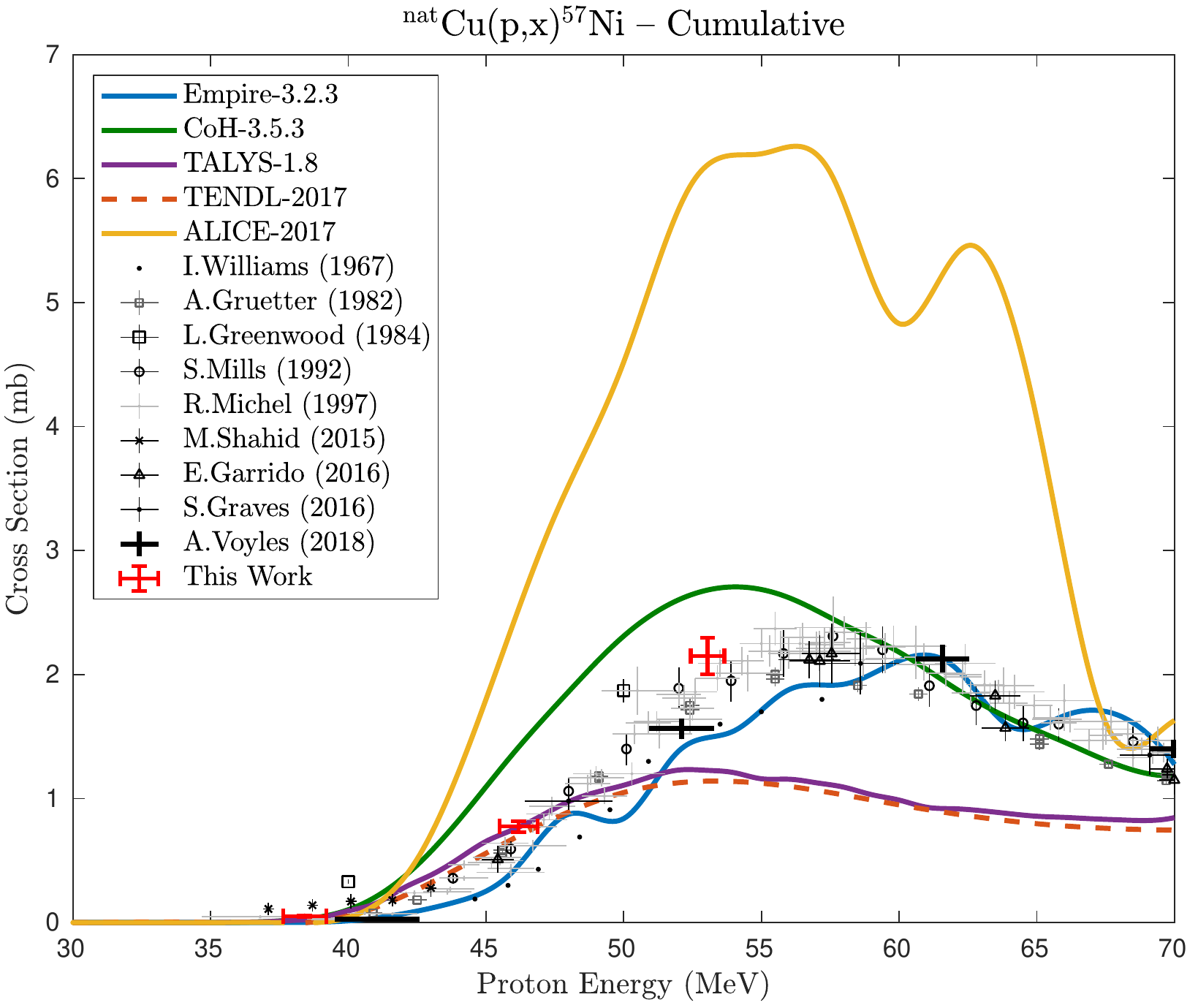}{50}
   \hspace{-10pt}}%
     \phantomcaption{}\label{fig:xs_curves_p3}
\end{figure*}

\begin{figure*}
    \centering
    \subfloat{
        \centering
        \subfigimg[width=0.496\textwidth]{}{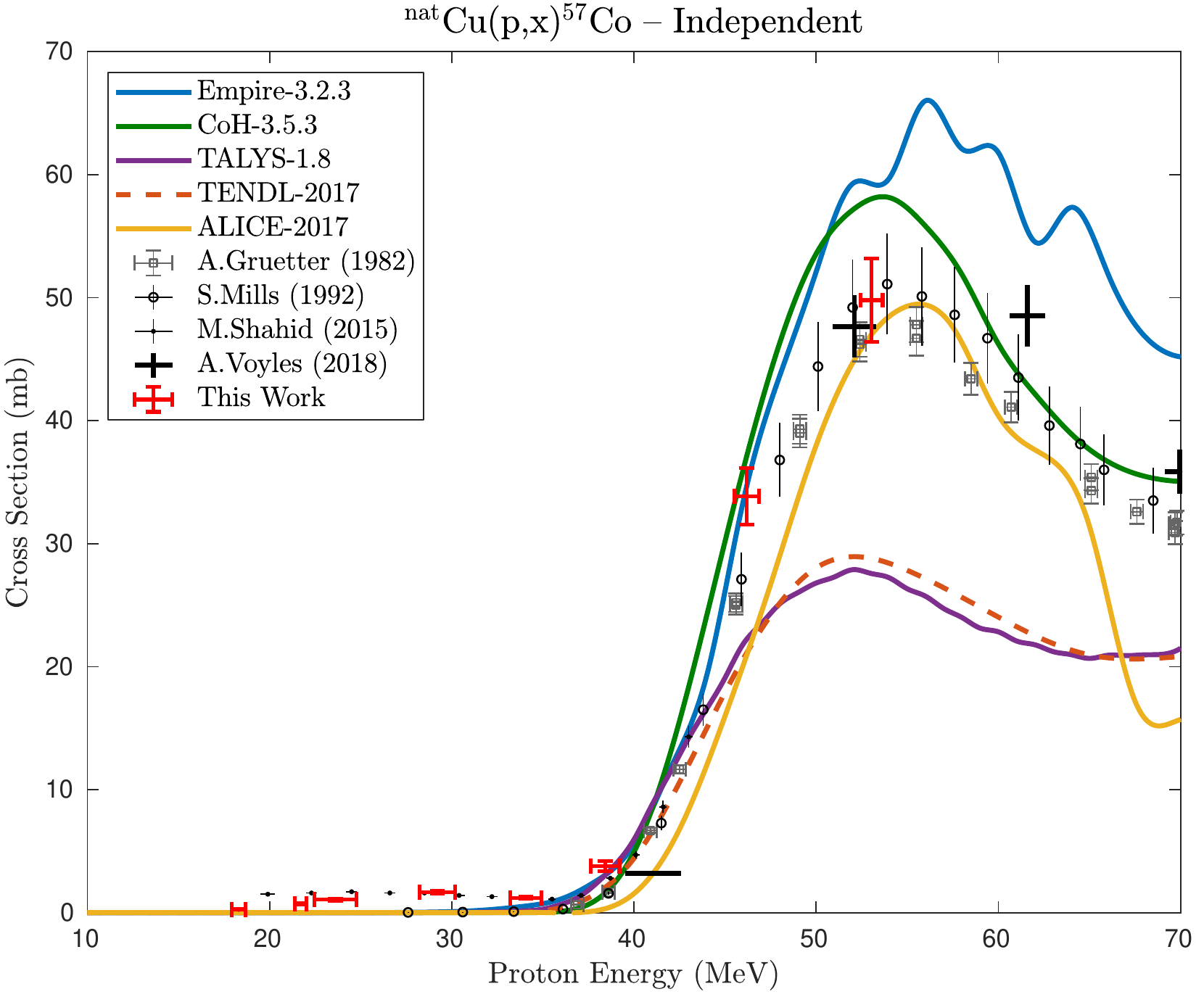}{50}
%
        \subfigimg[width=0.496\textwidth]{}{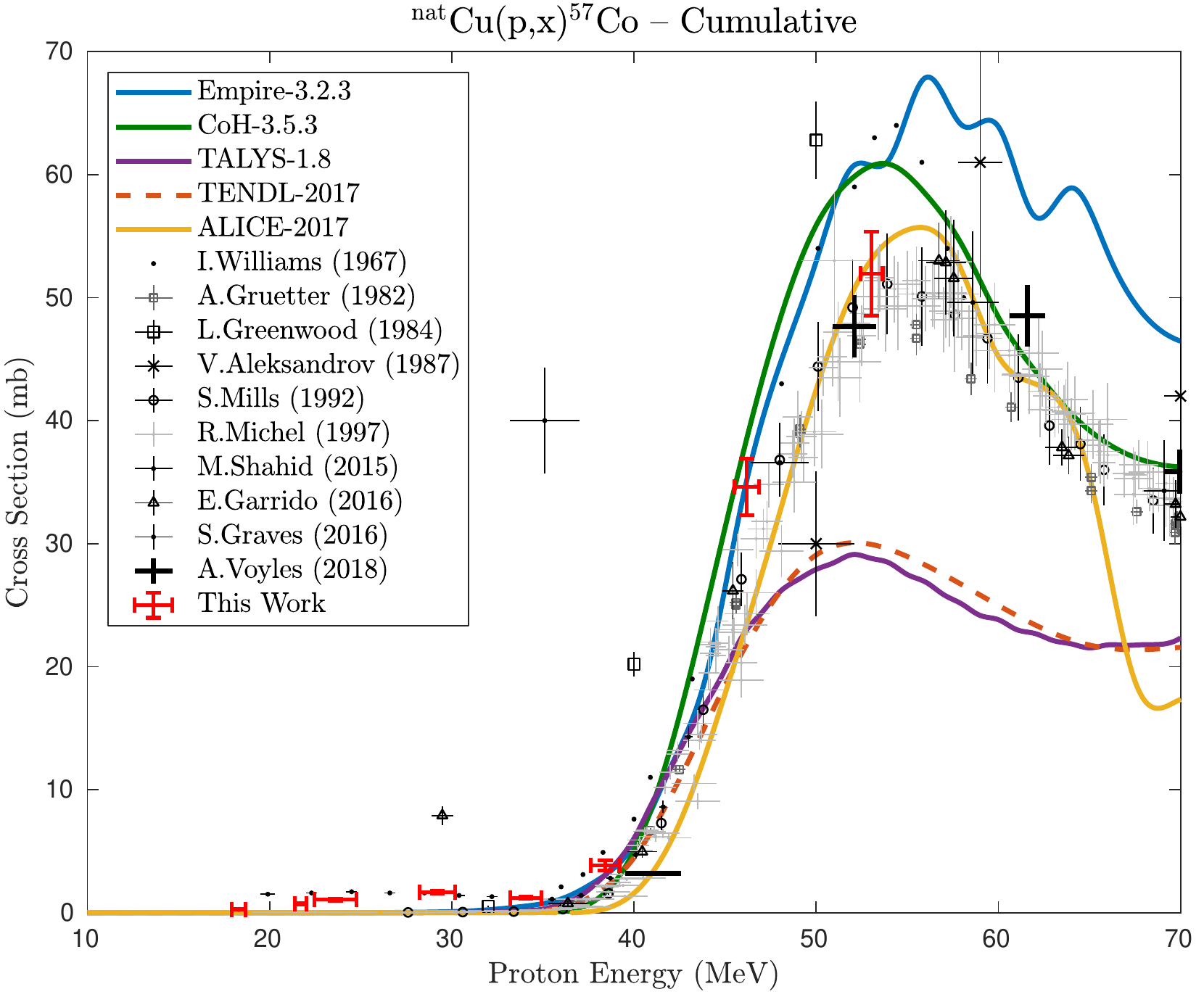}{50}
   \hspace{-10pt}}%
    \\
    \subfloat{
        \centering
        \subfigimg[width=0.496\textwidth]{}{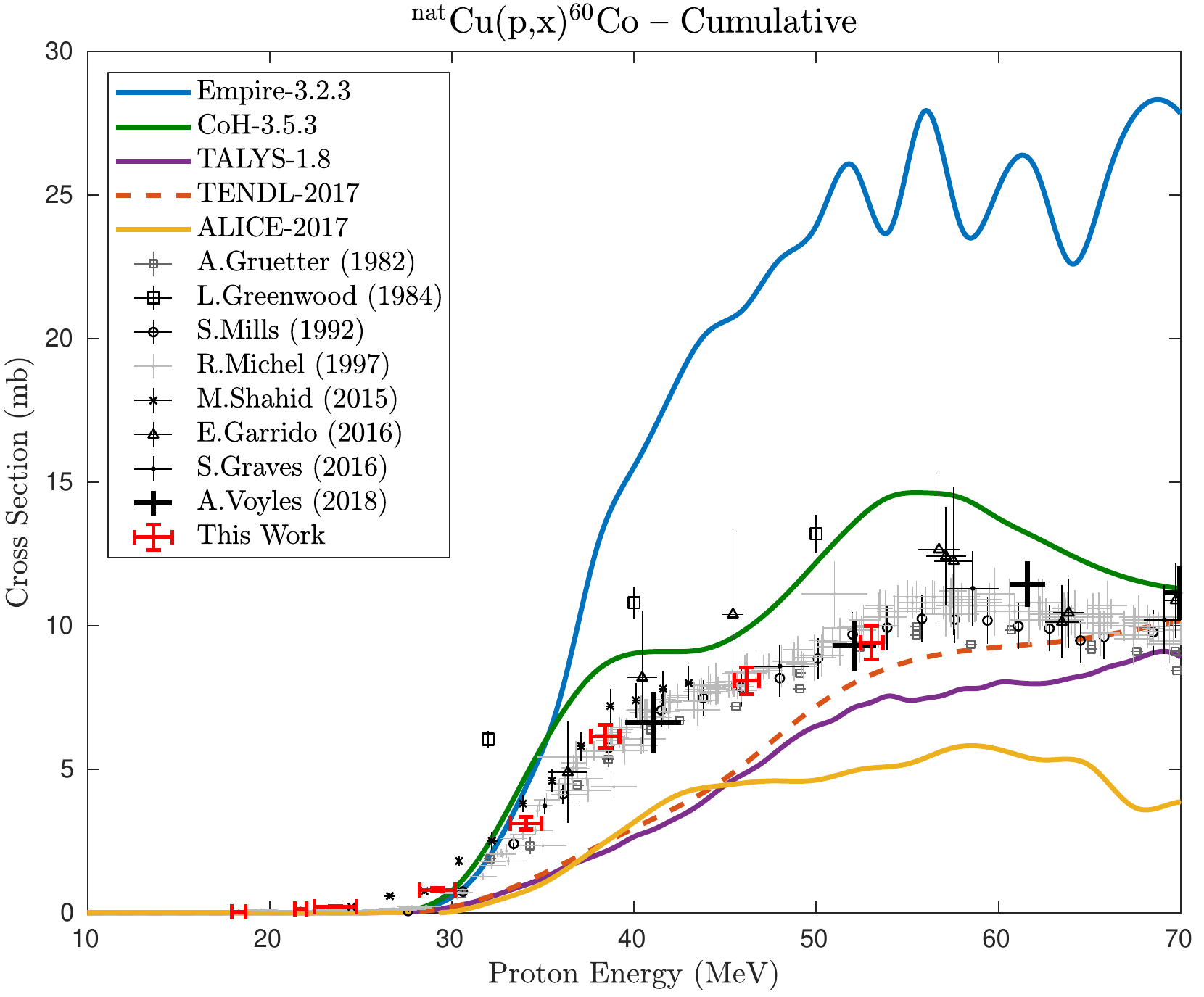}{50}
        \subfigimg[width=0.496\textwidth]{}{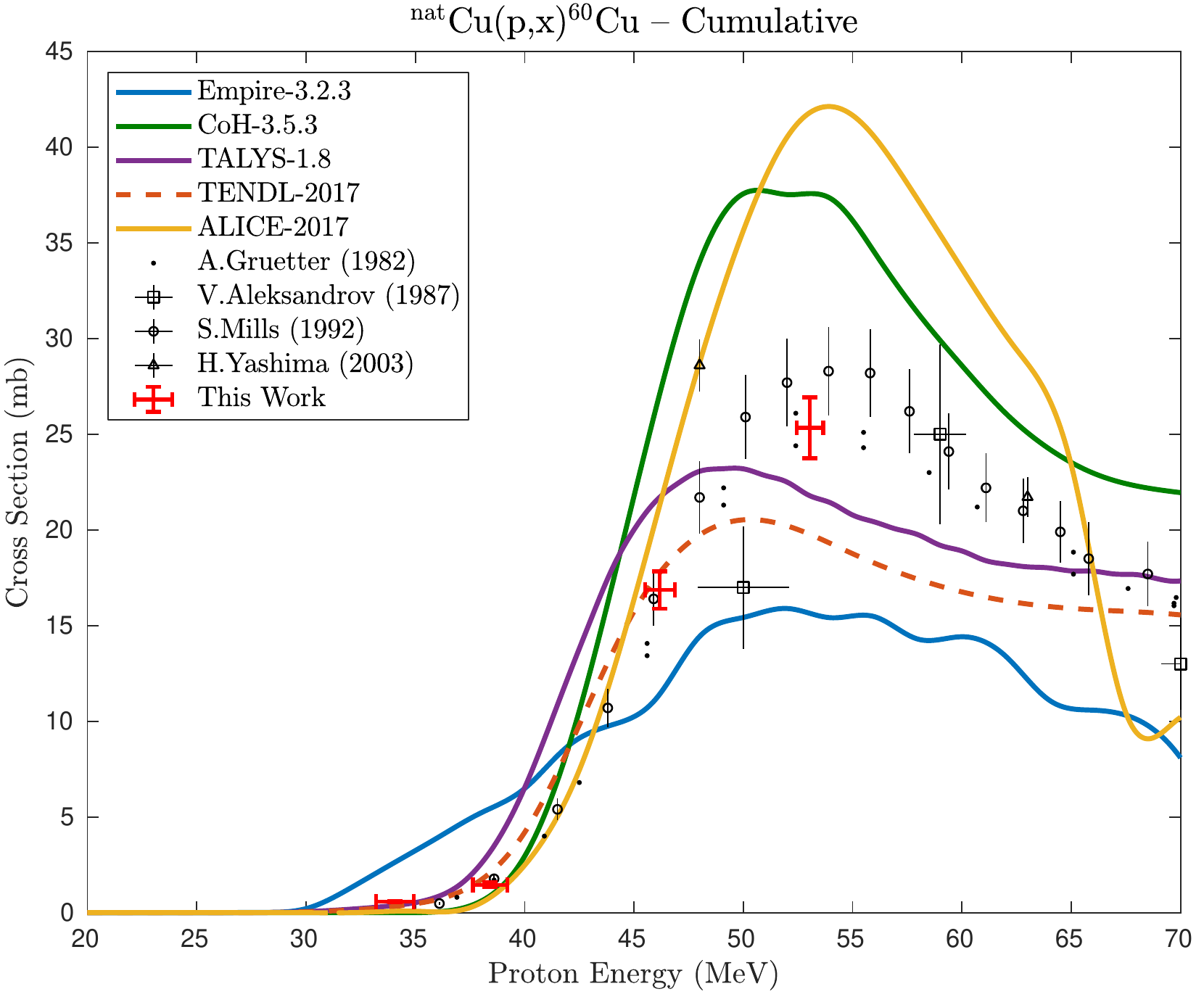}{50}
   \hspace{-10pt}}%
    \\
    \subfloat{
        \centering
        \subfigimg[width=0.496\textwidth]{}{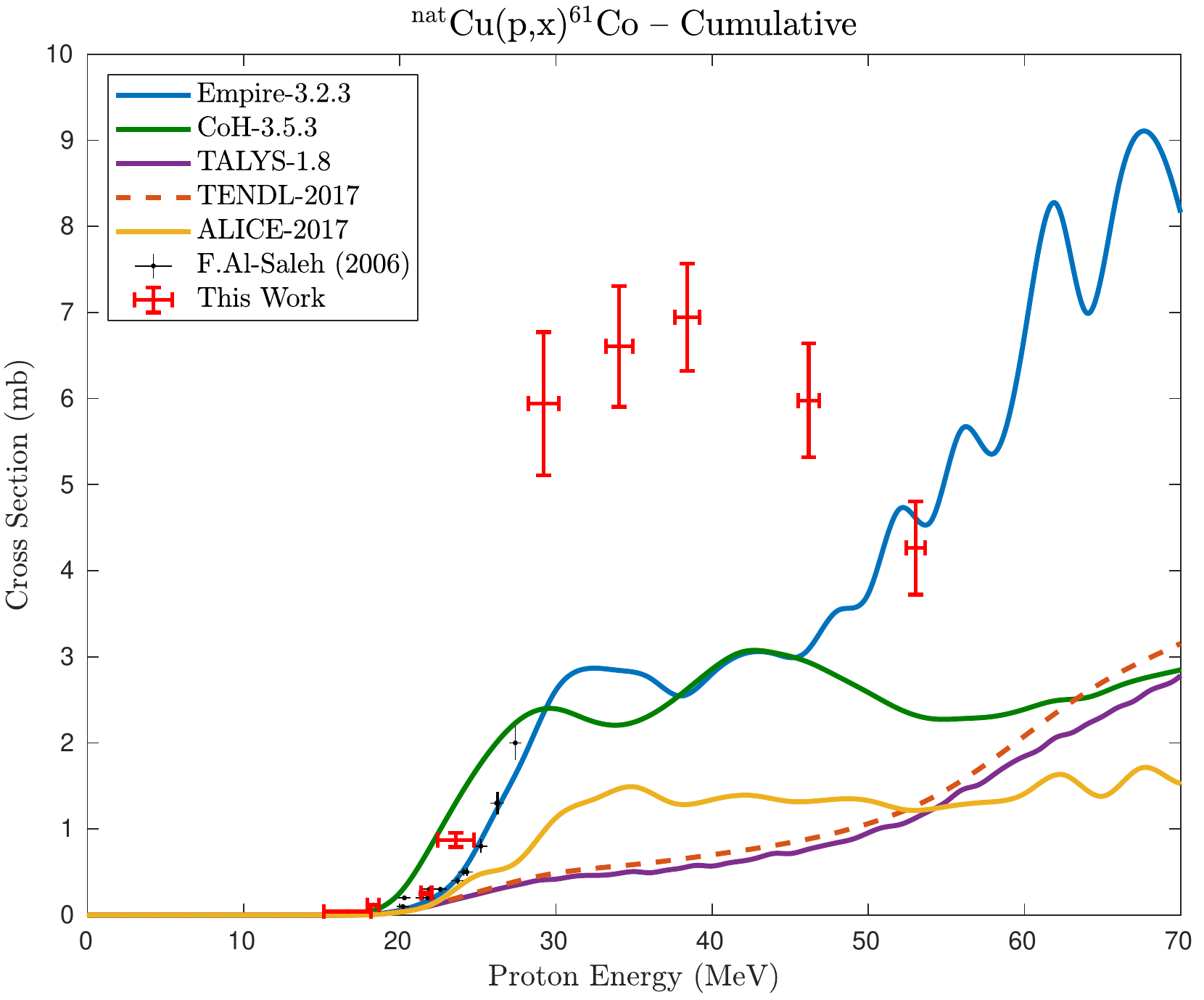}{50}
        \subfigimg[width=0.496\textwidth]{}{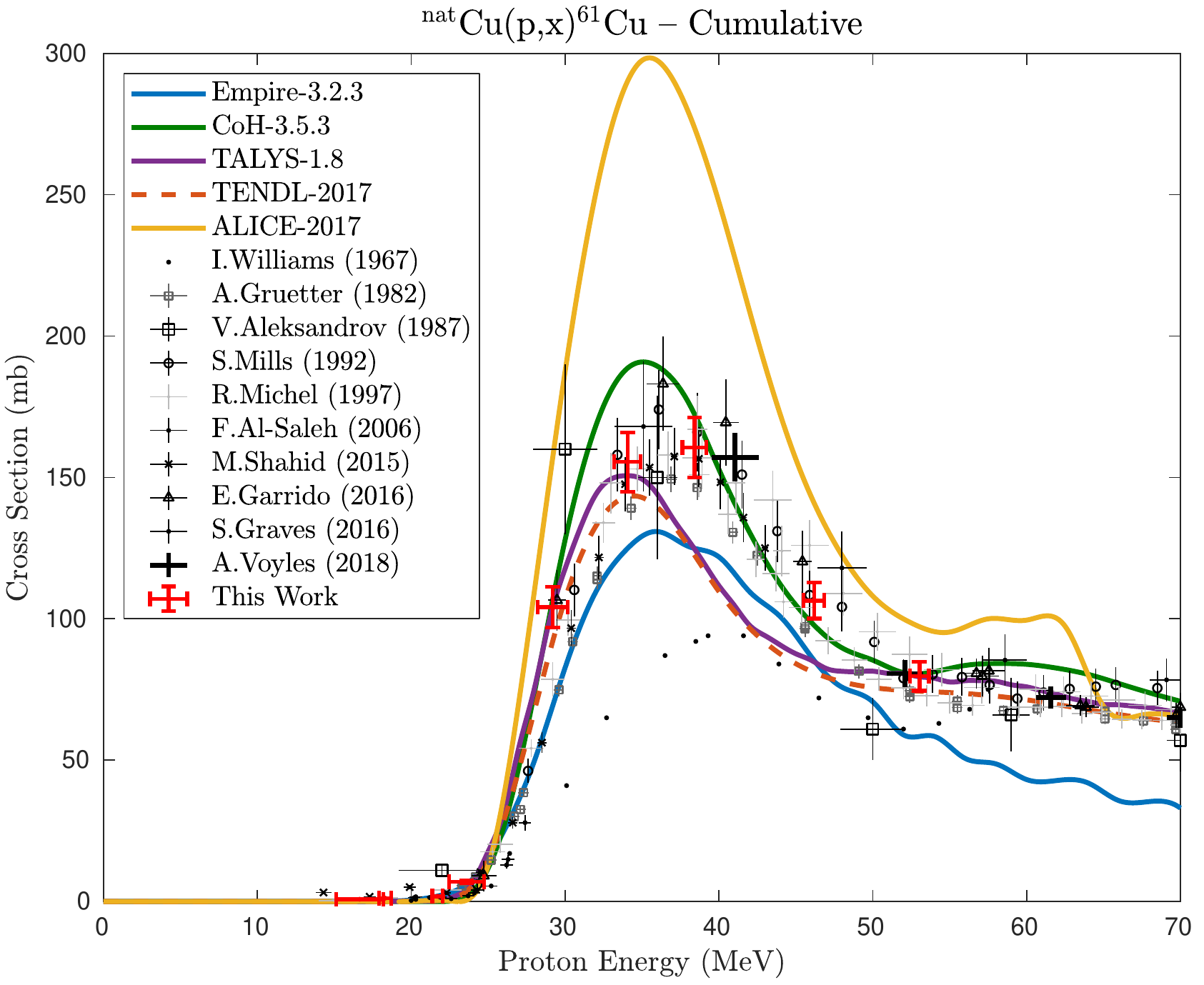}{50}
   \hspace{-10pt}}
     \phantomcaption{}\label{fig:xs_curves_p4}
\end{figure*}

\begin{figure*}
    \centering
     \subfloat{
        \centering
        \subfigimg[width=0.496\textwidth]{}{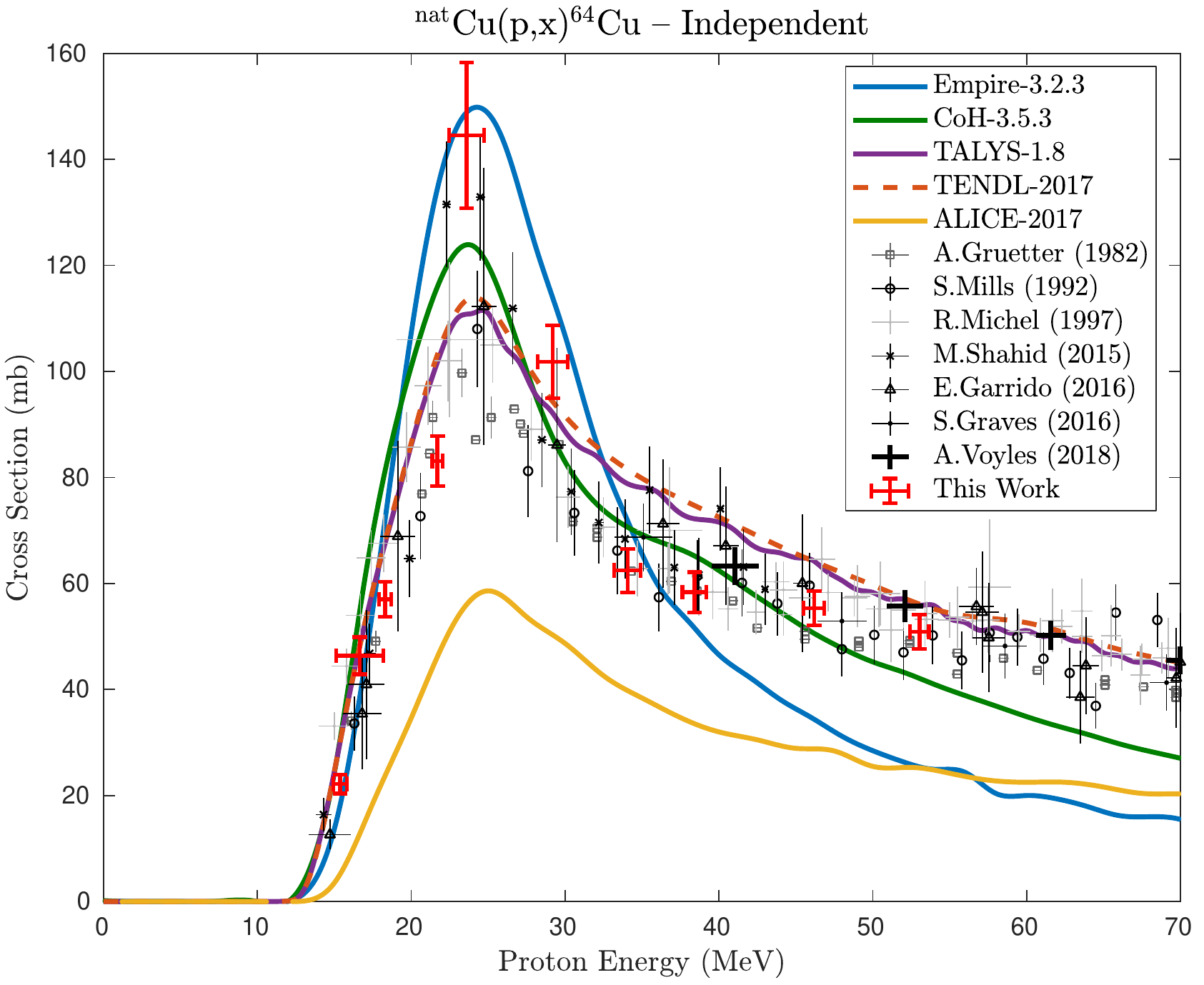}{50}
        \subfigimg[width=0.496\textwidth]{}{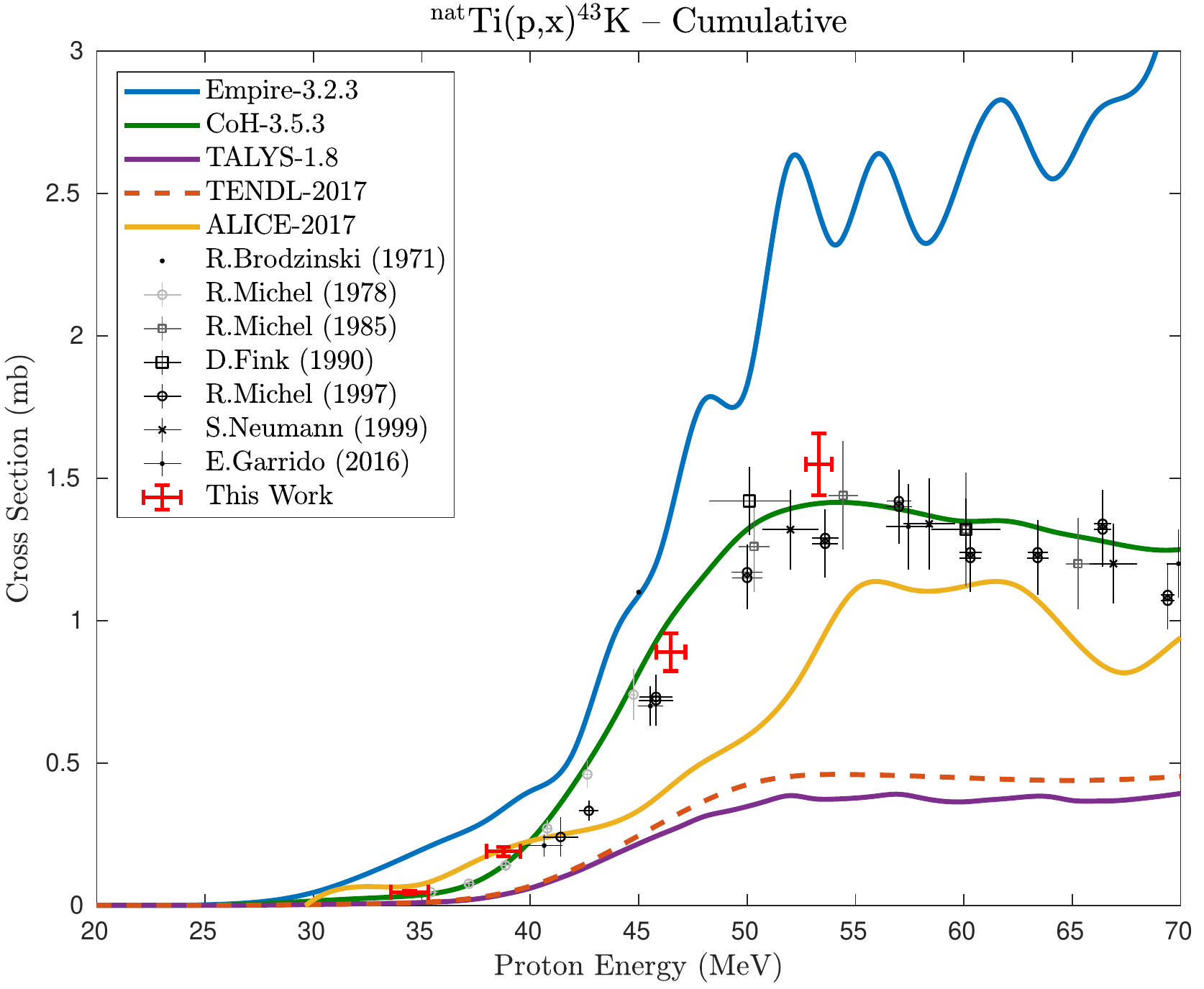}{50}
   \hspace{-10pt}}%
    \\
    \subfloat{
        \centering
        \subfigimg[width=0.496\textwidth]{}{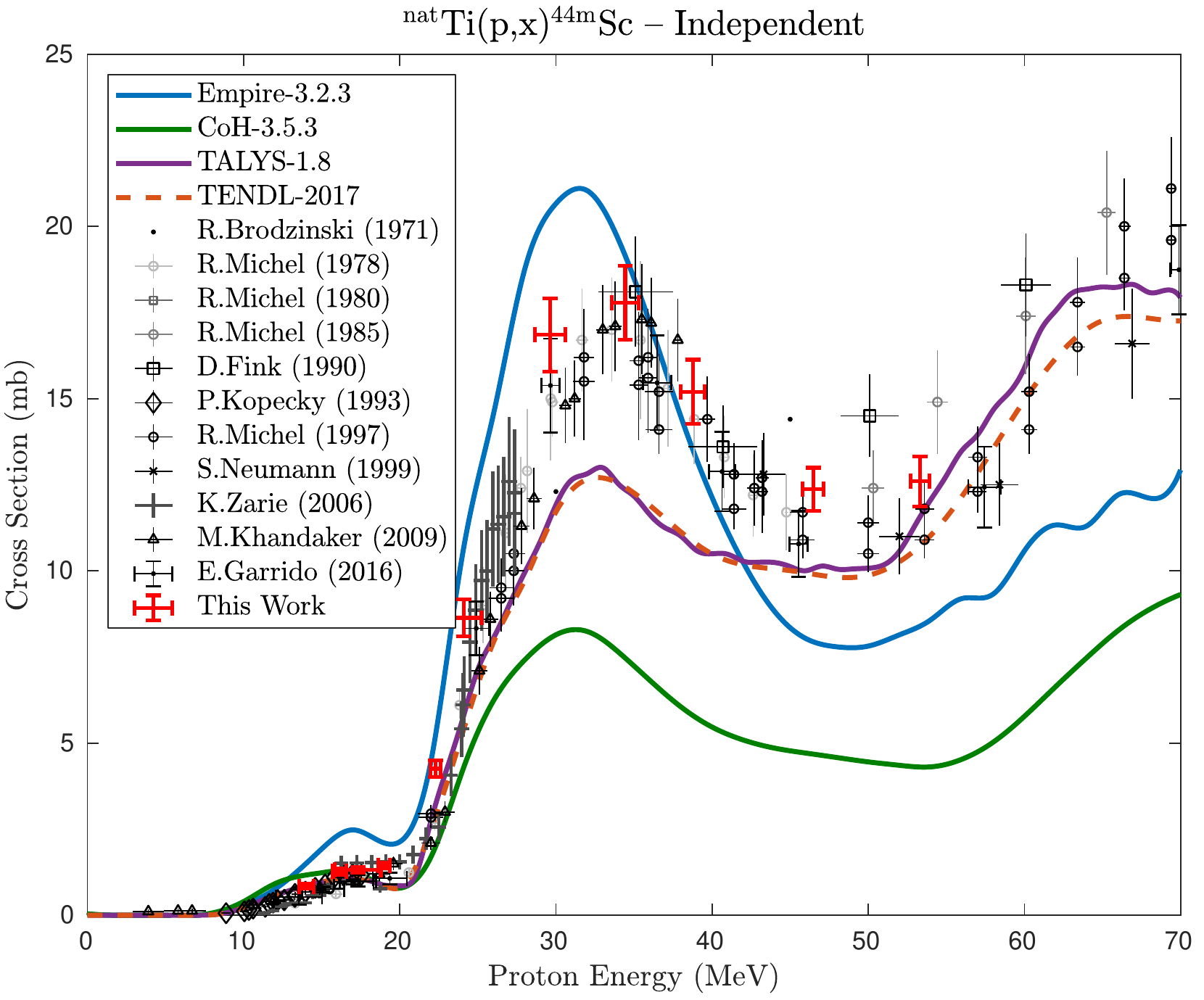}{50}
        \subfigimg[width=0.496\textwidth]{}{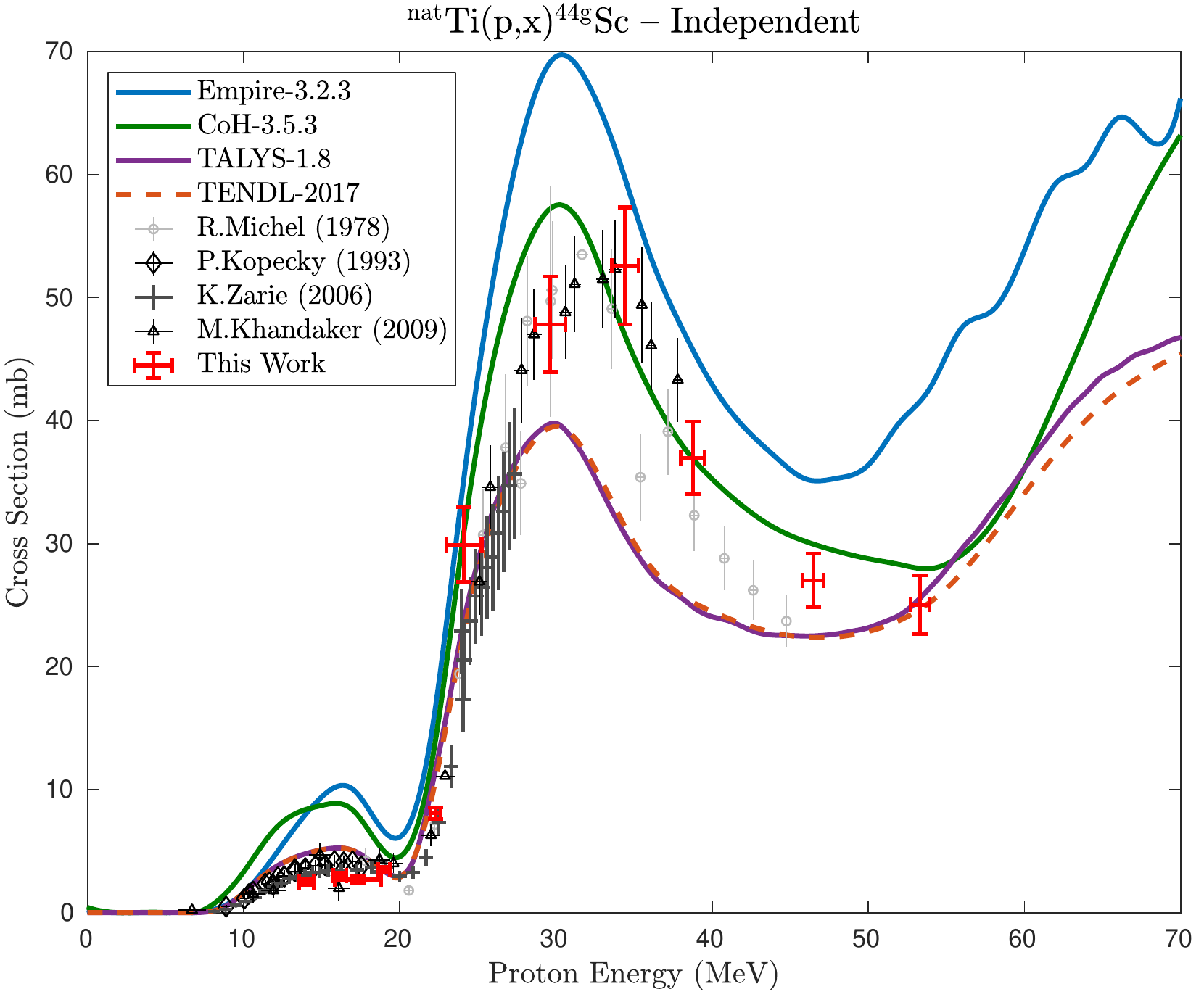}{50}
   \hspace{-10pt}}%
    \\
         \subfloat{
        \centering
        \subfigimg[width=0.496\textwidth]{}{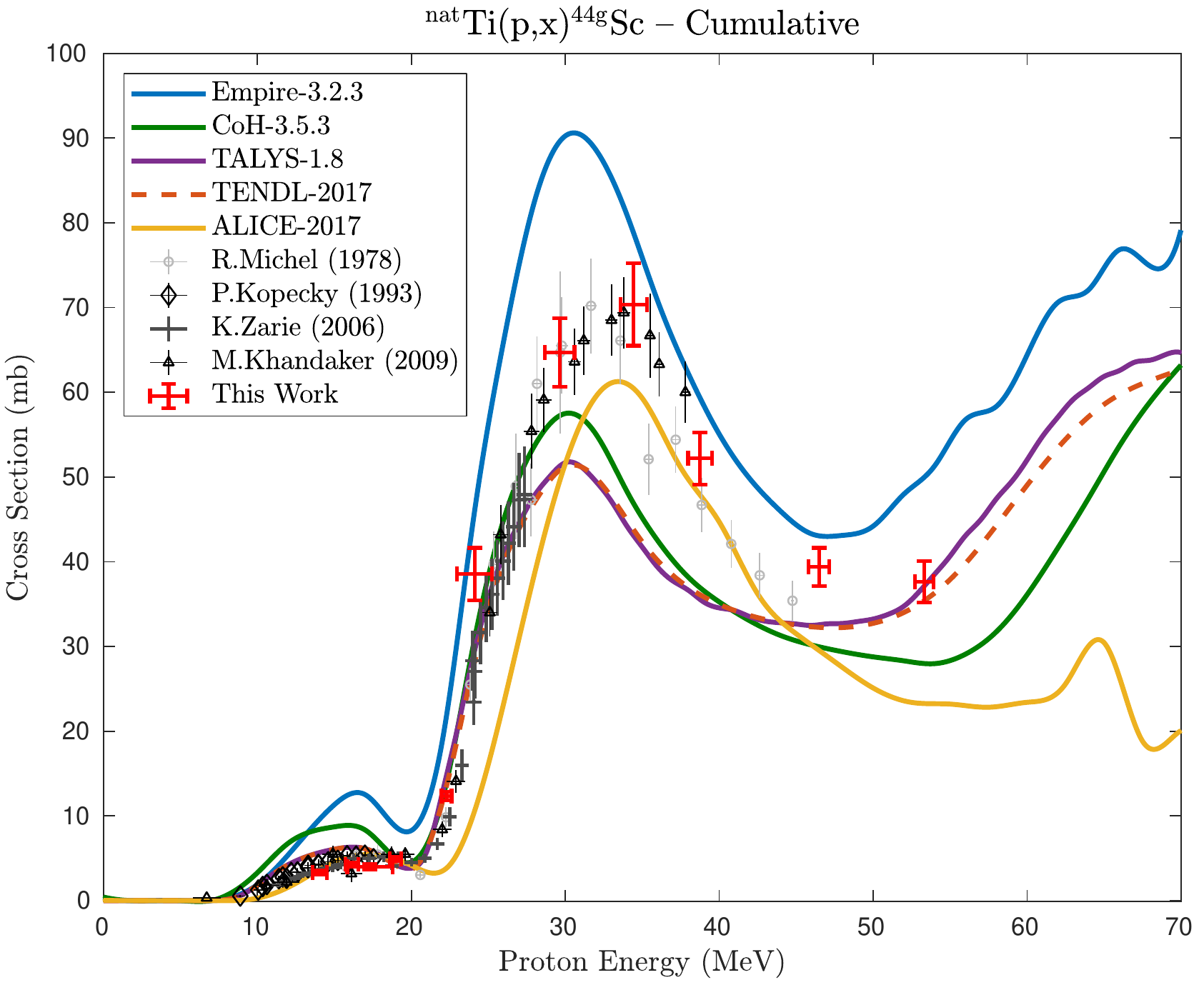}{50}
        \subfigimg[width=0.496\textwidth]{}{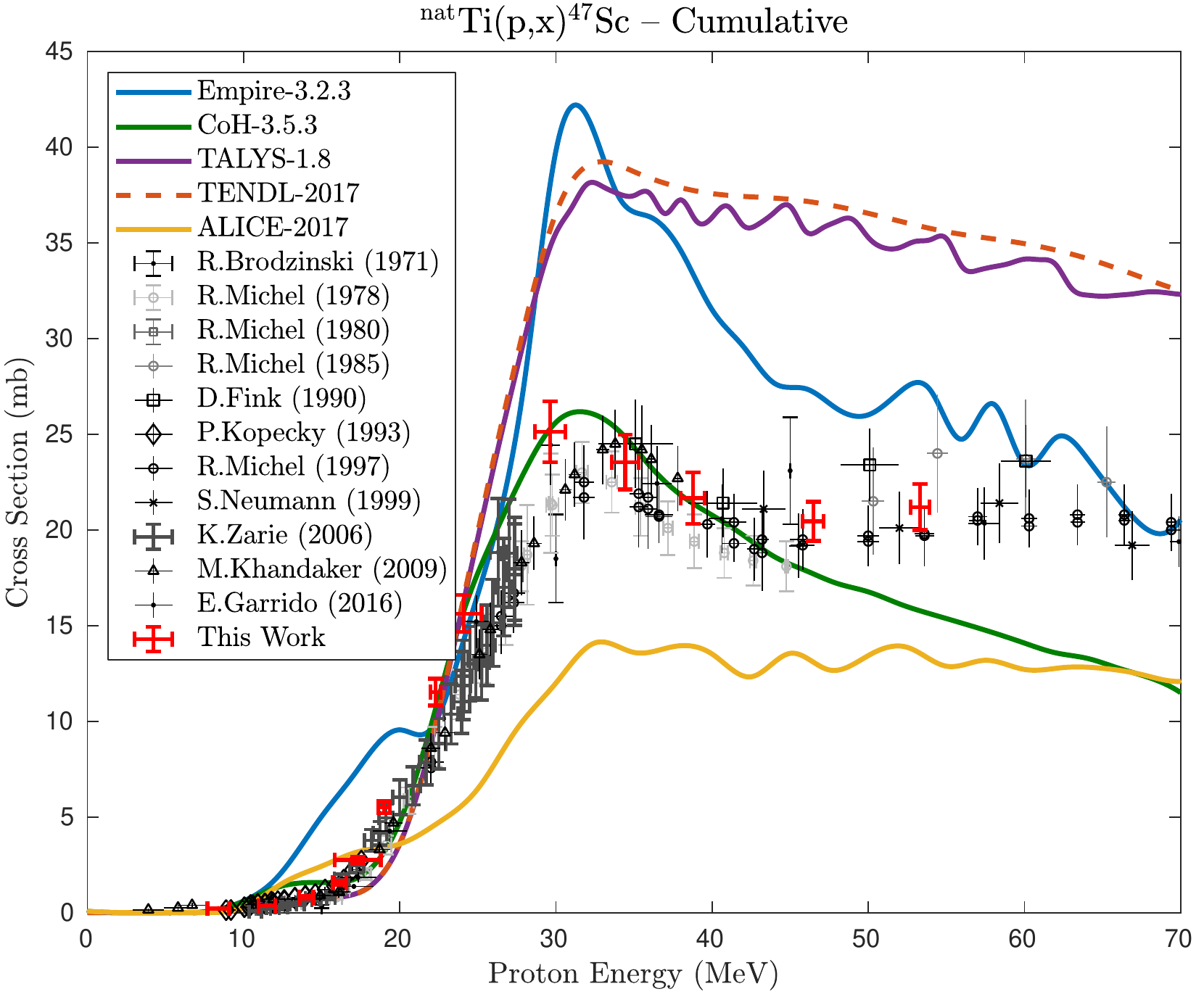}{50}
   \hspace{-10pt}}%
     \phantomcaption{}\label{fig:xs_curves_p8}
\end{figure*}
\begin{figure*}
    \centering    
         \subfloat{
        \centering
        \subfigimg[width=0.497\textwidth]{}{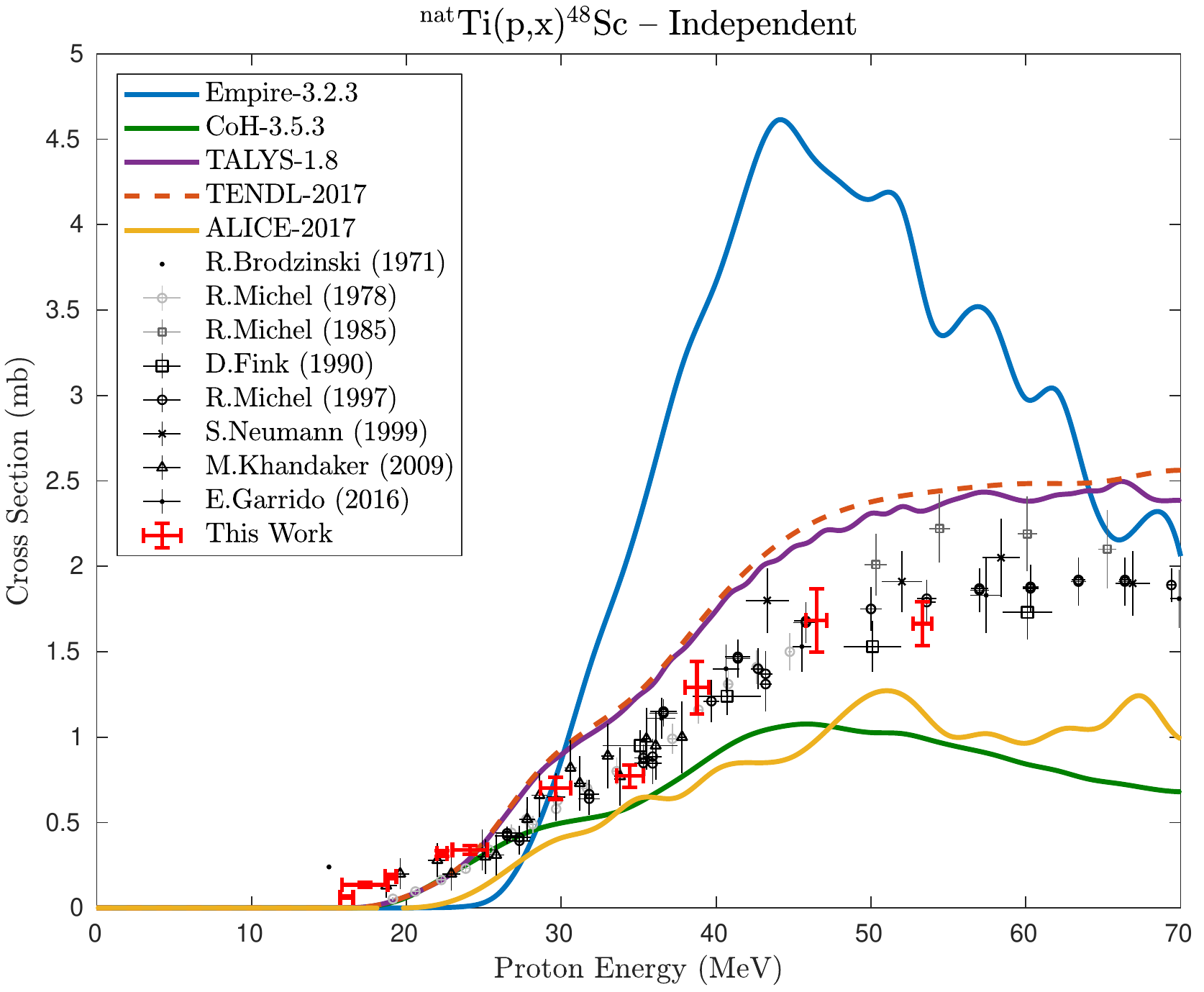}{50}
   }%
     \phantomcaption{}\label{fig:xs_curves_p7}
\end{figure*}

\pagebreak

\vspace{-20cm}

\section{Measured isomer-to-ground state branching ratios } \label{sec:fe_ibr_figures}

Plots of the isomer-to-ground state ratios measured in this work are presented here, in comparison with literature data and reaction modeling codes 
\cite{Michel1978,Kopecky1993,Zarie2006a,Khandaker2009}.

\begin{figure*}
    \centering
    \subfloat{
        \centering
        \subfigimg[width=0.497\textwidth]{}{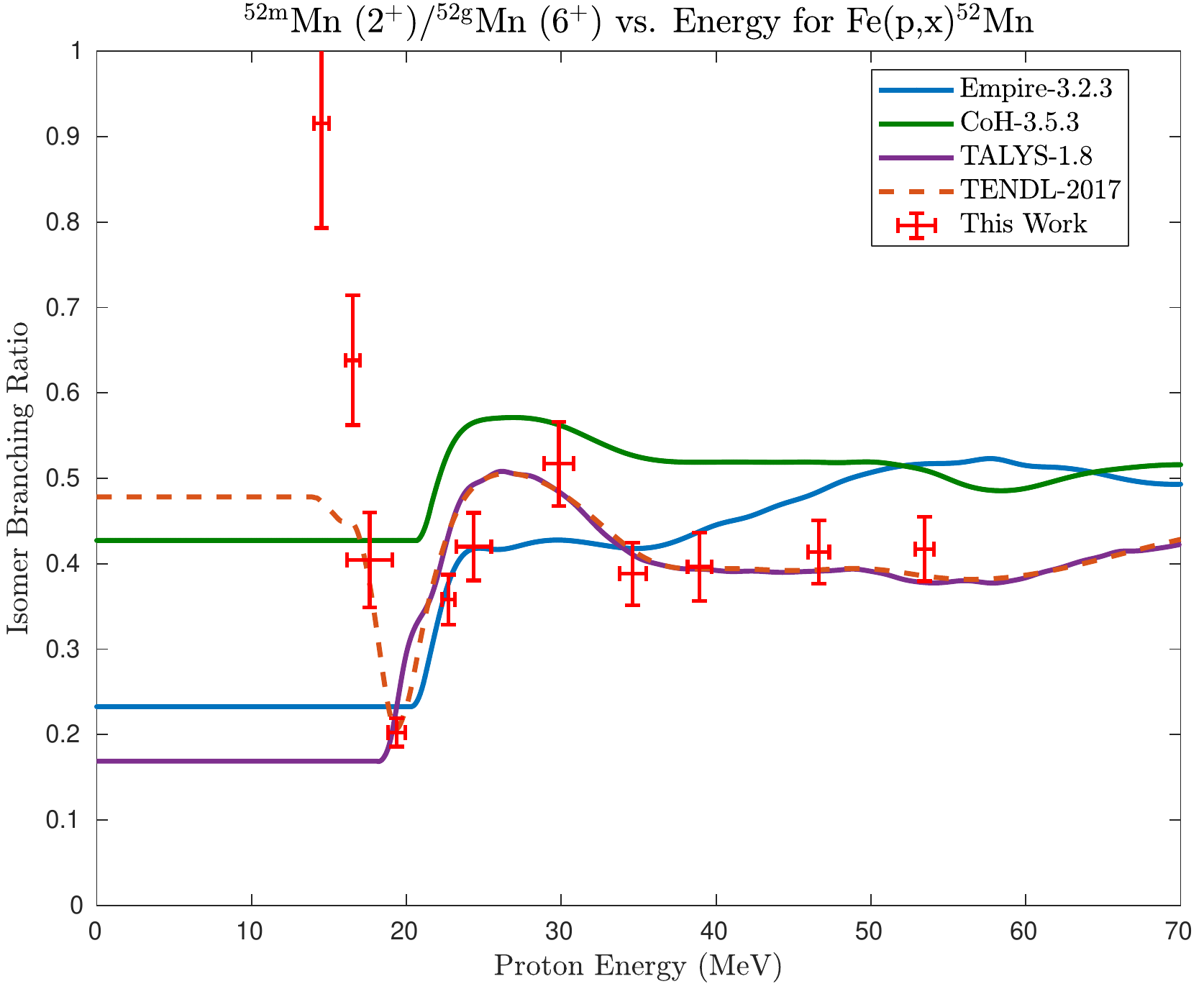}{50}
        \subfigimg[width=0.497\textwidth]{}{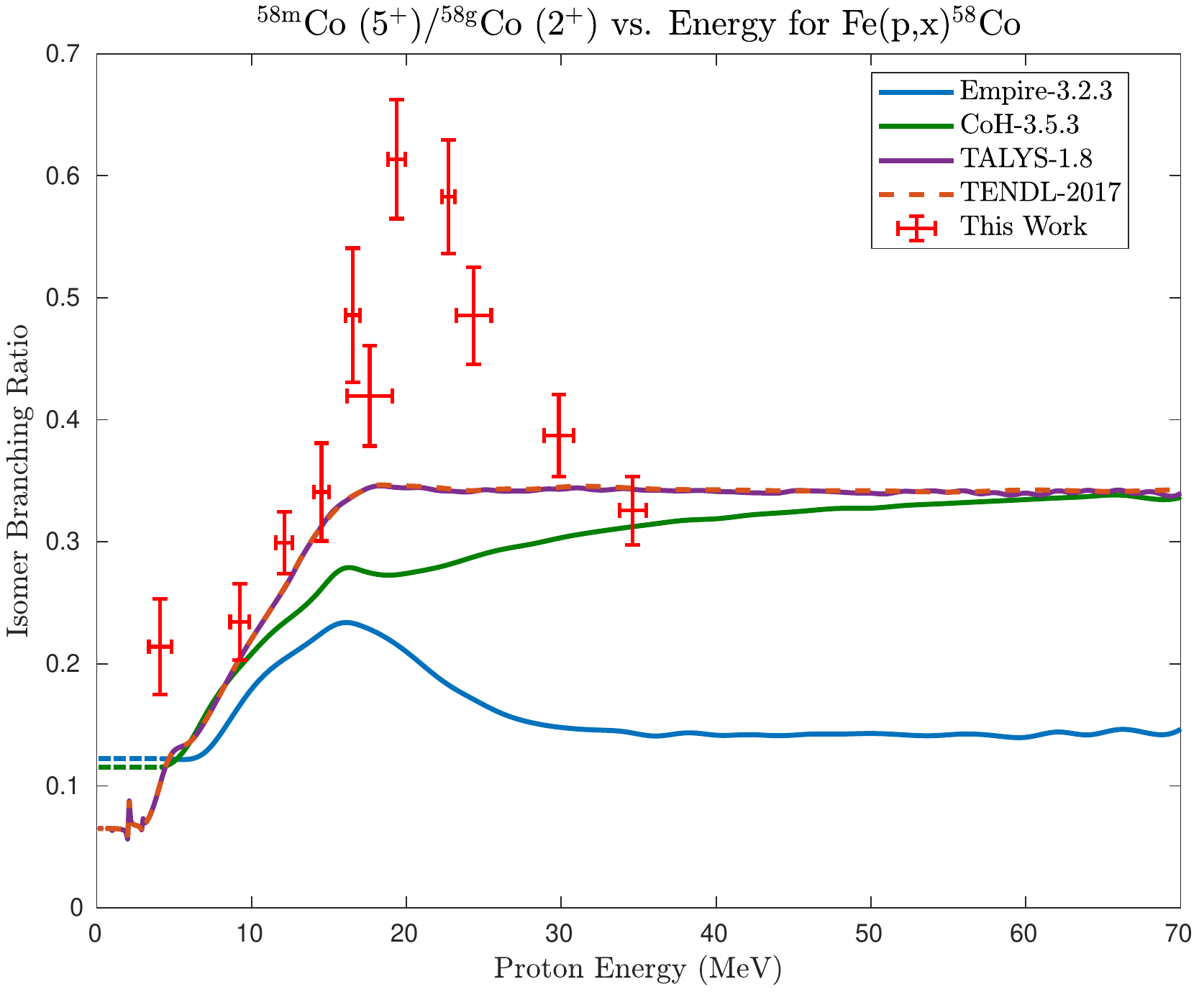}{50}
   \hspace{-10pt}}%
    \\
    \subfloat{
        \centering
        \subfigimg[width=0.497\textwidth]{}{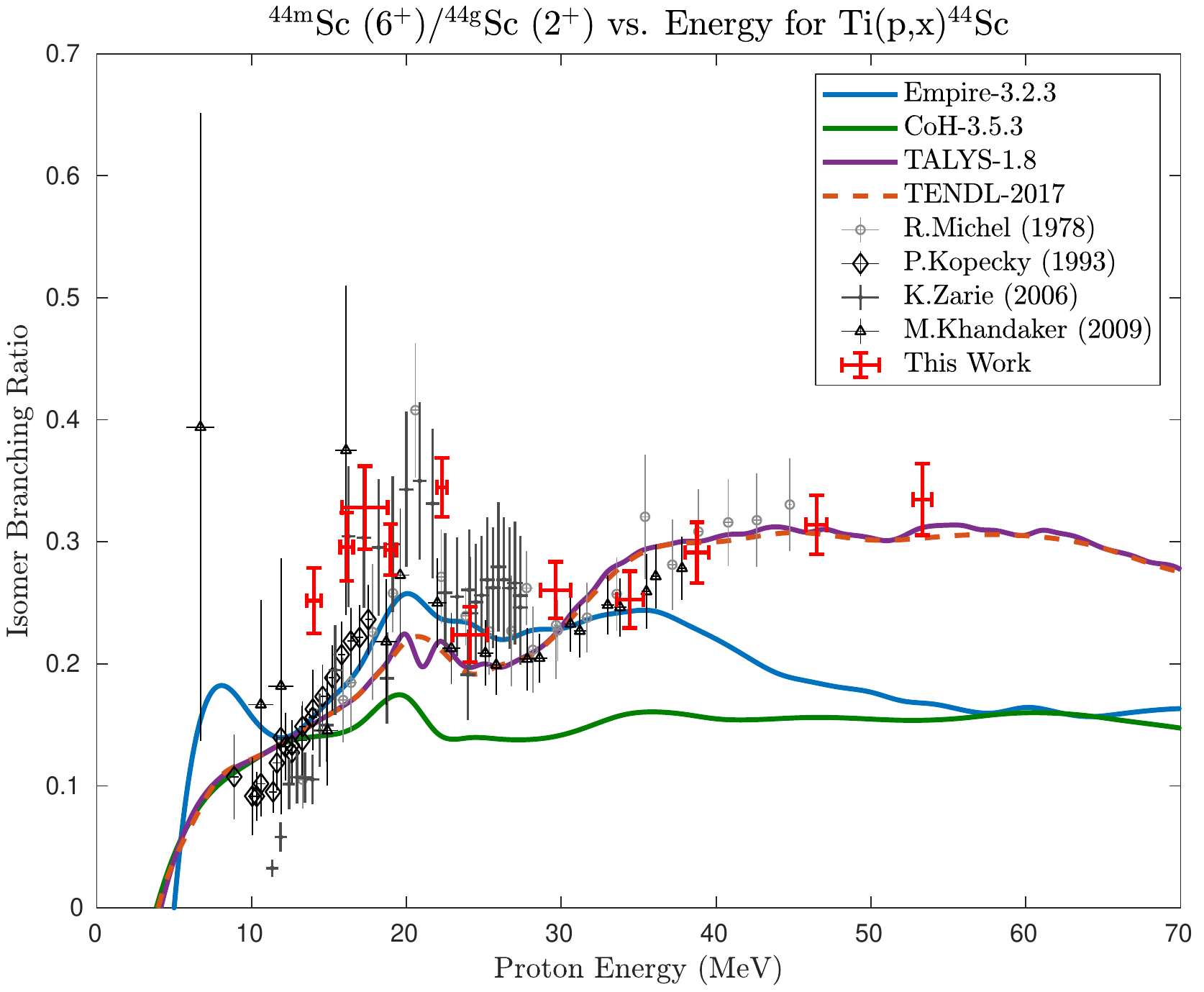}{50}
   }
     \phantomcaption{}\label{fig:ibr_curves}
\end{figure*}

\pagebreak

\IfFileExists{../../library.bib}{\bibliography{../../library}}{\bibliography{library}}

See Supplemental Material at [URL will be inserted by publisher] for a tabulation of the relevant nuclear data used in the analysis for the present work.

\pagebreak



\section{Supplemental Material} \label{fe_supp_material}
%
The   half-lives and gamma-ray branching ratios  listed in these tables were used for all calculations of measured cross sections reported in this work, and have been taken from the most recent edition of  Nuclear Data Sheets for each  mass chain  [11, 12, 20--35].




\begin{table}[ht]
\centering
\small
\begin{adjustbox}{totalheight=\textheight}
\begin{tabular}{@{}llll@{}}
\toprule\toprule
Nuclide & Half-life & E$_\gamma$ (keV) & I$_\gamma$ (\%)\\
\midrule
\ce{^{43}K}   & 22.3(1) h     & 372.760   & 86.8(2)\\
   &      & 617.490   & 79.2(6)\\
\ce{^{44g}Sc} & 3.97(4) h     & 1157.020  & 99.9(4)\\
 &      & 1499.46   & 0.908(15)\\
\ce{^{44m}Sc} & 58.61(10) h   & 271.241   & 86.7(3)\\
 &    & 1001.83   & 1.20(7)\\
 &    & 1126.06   & 1.20(7)\\
\ce{^{46}Sc}  & 83.79(4) d    & 889.277   & 99.9840(10)\\
  &     & 1120.545  & 99.9870(10)\\
\ce{^{47}Sc}  & 3.3492(6) d   & 159.381   & 68.3(4)\\
\ce{^{48}Cr}  & 21.56(3) h    & 112.31    & 96.0(20)\\
  &     & 308.24    & 100(2)\\
\ce{^{48}Sc}  & 43.67(9) h    & 175.361   & 7.48(10)\\
  &     & 1037.522  & 97.6(7)\\
\ce{^{48}V}   & 15.9735(25) d & 928.327   & 0.783(3)\\
   &  & 944.130   & 7.870(7)\\
   &  & 2240.396  & 2.333(13)\\
\ce{^{49}Cr}  & 42.3(1) m     & 62.289    & 16.4(6)\\
  &      & 90.639    & 53.2(19)\\
  &    & 152.928   & 30.3(11)\\
\ce{^{51}Cr}  & 27.704(3) d   & 320.0824  & 9.910(10)\\
\ce{^{51}Mn}  & 46.2(1) m     & 749.07    & 0.265(7)\\
\ce{^{52}Fe}  & 45.9(6) s     & --        & --     \\
\ce{^{52g}Mn} & 5.591(3) d    & 346.02    & 0.980(14)\\
 &  & 600.16    & 0.390(11)\\
 &  & 647.47    & 0.400(20)\\
 &  & 744.233   & 90.0(12)\\
 &  & 935.544   & 94.5(13)\\
 &  & 1246.278  & 4.21(7)\\
 &  & 1333.649  & 5.07(7)\\
 &  & 1434.092  & 100.0(14)\\
\ce{^{52m}Mn} & 21.1(2) m     & 377.738   & 1.68(3)\\
\ce{^{54}Mn}  & 312.20(20) d  & 834.848   & 99.9760(10)\\
\ce{^{55}Co}  & 17.53(3) h    & 91.9      & 1.16(9)\\
  &  & 385.4     & 0.54(5)\\
  &  & 477.2     & 20.2(17)\\
  &  & 520.0     & 0.83(8)\\
  &  & 803.7     & 1.87(15)\\
  &  & 827.0     & 0.21(6)\\
  &  & 931.1     & 75.0(35)\\
  &  & 984.6     & 0.52(10)\\
  &  & 1212.8    & 0.26(3)\\
  &  & 1316.6    & 7.1(3)\\
  &  & 1408.5    & 16.9(8)\\
  &  & 2177.6    & 0.29(4)\\
\ce{^{56}Co}  & 77.236(26) d  & 733.514   & 0.191(3)\\
  &  & 787.743   & 0.311(3)\\
  &  & 847.770   & 99.9399(23)\\
  &  & 977.372   & 1.421(6)\\
  &  & 996.948   & 0.111(4)\\
  &  & 1037.843  & 14.05(4)\\
  &  & 1140.368  & 0.132(3)\\
  &  & 1175.101  & 2.252(6)\\
  &  & 1238.288  & 66.46(12)\\
  &  & 1335.40   & 0.1224(12)\\
  &  & 1360.212  & 4.283(12)\\
  &  & 1442.746  & 0.180(4)\\
  &  & 1771.357  & 15.41(6)\\
  &  & 1810.757  & 0.640(3)\\
  &  & 1963.741  & 0.707(4)\\
  &  & 2015.215  & 3.016(12)\\
\bottomrule\bottomrule
\end{tabular}
\end{adjustbox}
\end{table}

\begin{table}[ht]
\centering
\small
\begin{adjustbox}{totalheight=\textheight}
\begin{tabular}{@{}llll@{}}
\toprule\toprule
Nuclide & Half-life & E$_\gamma$ (keV) & I$_\gamma$ (\%)\\
\midrule
\ce{^{56}Co}  & 77.236(26) d  & 2034.791  & 7.77(3)\\
  &  & 2113.135  & 0.377(3)\\
  &  & 2212.944  & 0.388(4)\\
  &  & 2276.131  & 0.118(4)\\
  &  & 2598.500  & 16.97(4)\\
  &  & 3009.645  & 1.036(13)\\
\ce{^{56}Mn}  & 2.5789(1) h   & 846.7638  & 98.85(3)\\
\ce{^{57}Co}  & 271.74(6) d   & 122.06065 & 85.60(17)\\
  &    & 136.47356 & 10.68(8)\\
  &    & 352.33    & 0.0030(3)\\
  &    & 692.41    & 0.149(10)\\
\ce{^{57}Ni}  & 35.60(6) h    & 1377.63   & 81.7(24)\\
  &     & 1919.52   & 12.3(4)\\
\ce{^{58g}Co} & 70.86(6) d    & 810.7593  & 99.45(1)\\
 &     & 863.951   & 0.686(10)\\
 &     & 1674.725  & 0.517(10)\\
\ce{^{58m}Co} & 9.10(9) h     & --        & --     \\
\ce{^{60}Co}  & 1925.28(14) d & 1173.228  & 99.85(3)\\
  &  & 1332.492  & 99.9826(6)\\
\ce{^{60}Cu}  & 23.7(4) m     & 467.3     & 3.52(18)\\
  &  & 826.4     & 21.7(11)\\
  &  & 952.4     & 2.73(18)\\
  &  & 1035.2    & 3.70(18)\\
  &  & 1173.2    & 0.26(9)\\
  &  & 1293.7    & 1.85(18)\\
  &  & 1332.5    & 88.0(1)\\
  &  & 1791.6    & 45.4(23)\\
  &  & 1861.6    & 4.8(3)\\
  &  & 1936.9    & 2.20(9)\\
  &  & 2158.9    & 3.34(18)\\
  &  & 2403.3    & 0.77(8)\\
  &  & 3124.1    & 4.8(3)\\
\ce{^{61}Cu}  & 3.339(8) h    & 67.412    & 4.2(8)\\
  &  & 282.956   & 12.2(22)\\
  &  & 373.050   & 2.1(4)\\
  &  & 529.169   & 0.38(7)\\
  &  & 588.605   & 1.17(21)\\
  &  & 656.008   & 10.8(20)\\
  &  & 816.692   & 0.31(6)\\
  &  & 841.211   & 0.21(4)\\
  &  & 1099.560  & 0.25(4)\\
  &  & 1132.351  & 0.090(17)\\
  &  & 1185.234  & 3.7(7)\\
  &  & 1446.492  & 0.045(8)\\
\ce{^{62}Zn}  & 9.193(15) h   & 40.85     & 25.5(24)\\
  &  & 243.36    & 2.52(23)\\
  &  & 246.95    & 1.90(18)\\
  &  & 260.43    & 1.35(13)\\
  &  & 304.88    & 0.29(3)\\
  &  & 349.60    & 0.45(4)\\
  &  & 394.03    & 2.24(17)\\
  &  & 548.35    & 15.3(14)\\
  &  & 596.56    & 26(2)\\
  &  & 637.41    & 0.25(3)\\
\ce{^{63}Zn}  & 38.47(5) m    & 669.62    & 8.2(3)\\
  &  & 962.06    & 6.5(4)\\
  &  & 1412..08  & 0.75(4)\\
  &  & 1547.04   & 0.122(7)\\
  &  & 2336.5    & 0.075(6)\\
  &  & 2536.0    & 0.066(7)\\
\ce{^{64}Cu}  & 12.701(2) h   & 1345.77   & 0.475(11)\\   
\bottomrule\bottomrule
\end{tabular}
\end{adjustbox}
\end{table}

\end{document}